\documentclass{ws-ijmpe}
\usepackage[super,compress]{cite}
\usepackage{graphicx}
\usepackage[english]{babel}
\usepackage{epsfig}

\def\gsim{ \,\, \vcenter{\hbox{$\buildrel{\displaystyle >}\over\sim$}}
 \,\,}
\def\lsim{ \,\, \vcenter{\hbox{$\buildrel{\displaystyle <}\over\sim$}}
 \,\,}
\def\be{\begin{equation}}
\def\ee{\end{equation}}
\def\bea{\begin{eqnarray}}
\def\eea{\end{eqnarray}}

\def\beq{\begin{equation}}
\def\eeq{\end{equation}}
\def\bea{\begin{eqnarray}}
\def\eea{\end{eqnarray}}
\def\eq#1{{Eq.~(\ref{#1})}}
\def\fig#1{{Fig.~\ref{#1}}}

\newcommand{\bsp}{\begin{split}}
\newcommand{\esp}{\end{split}}

\def\lsim{\raise0.3ex\hbox{$<$\kern-0.75em\raise-1.1ex\hbox{$\sim$}}}
\def\gsim{\raise0.3ex\hbox{$>$\kern-0.75em\raise-1.1ex\hbox{$\sim$}}}

\def\d+Au  {$d$Au}
\def\d+Aum  {d\mathrm{Au}}

\def\jpsi    {\mbox{$J/\psi$}}

\def\beq     {\begin{equation}}
\def\eeq     {\end{equation}}

\long\def\symbolfootnote[#1]#2{\begingroup%
\def\thefootnote{\fnsymbol{footnote}}\footnote[#1]{#2}\endgroup}

\newcommand{\Ep}{E_\mathrm{p}}
\newcommand{\mT}{m_T}
\newcommand{\dd}{{d}}
\newcommand{\qzero}{\hat{q}_0}
\newcommand{\gevsqfm}{GeV$^2$/fm}
\newcommand{\ellb}{L}
\newcommand{\qhat}{\hat{q}}
\newcommand{\Ea}{E}

\newcommand{\epsa}{\varepsilon}
\newcommand{\epsamax}{\varepsilon^{\rm max}}
\newcommand{\sqrts}{\sqrt{s}}

\begin{document}

\markboth{Albacete {\it et al}}{Predictions for $p+$Pb Collisions at 
$\sqrt{s_{_{NN}}} = 5$ TeV: Comparison with Data}

\catchline{}{}{}{}{}

\title{Predictions for $p+$Pb Collisions at $\sqrt{s_{_{NN}}} = 5$ TeV: \\
Comparison with Data}

\author{JAVIER L. ALBACETE} 
\address{IPNO, Universit\'e Paris-Sud 11, CNRS/IN2P3, 91406 Orsay,
  France} 

\author{FRAN\c{C}OIS ARLEO}
\address{Laboratoire Leprince-Ringuet, \'Ecole Polytechnique,
CNRS/IN2P3, Universit\'e Paris-Saclay, 91128, Palaiseau, France}

\author{GERGELY G. BARNAF\"OLDI} 
\address{Institute for Particle and Nuclear Physics, Wigner Research Centre for
Physics, Hungarian Academy of Sciences, P.O.Box 49, Budapest, 1525, Hungary}

\author{JEAN BARRETTE} 
\address{McGill University, Montreal, H3A 2T8, Canada}

\author{WEI-TIAN DENG}
\address{Theory Center, IPNS, KEK, 1-1 Oho, Tsukuba, Ibaraki 305-0801, 
Japan}

\author{ADRIAN DUMITRU}
\address{Department of
    Natural Sciences, Baruch College, CUNY, 17 Lexington Avenue, New
    York, NY 10010, USA}
\address{RIKEN BNL Research Center,
    Brookhaven National Laboratory, Upton, NY 11973, USA}

\author{KARI J. ESKOLA}
\address{University of Jyvaskyla, Department of Physics, P.O. Box 35, 
FI-40014 University of Jyvaskyla, Finland}
\address{Helsinki Institute of Physics, P.O. Box 64, FI-00014 University 
of Helsinki, Finland}

\author{ELENA G.~FERREIRO}
\address{Departamento de F{\'\i}sica de Part{\'\i}culas, Universidade de 
Santiago de Compostela, 15782 Santiago de Compostela, Spain}

\author{FREDERIC FLEURET}
\address{Laboratoire Leprince-Ringuet, Ecole polytechnique, CNRS/IN2P3, 
Universit\'e Paris-Saclay, 91128 Palaiseau, France}

\author{HIROTSUGU FUJII}
\address{Institute of Physics, University of Tokyo, Komaba,
    Tokyo 153-8902, Japan}

\author{MIKLOS GYULASSY} 
\address{Department of Physics, Columbia University, New York, NY 10027, USA}
\address{Institute for Particle and Nuclear Physics, Wigner Research Centre for
Physics, Hungarian Academy of Sciences, P.O.Box 49, Budapest, 1525, Hungary}

\author{SZILVESZTER MIKL\'OS HARANGOZ\'O}
\address{Institute for Particle and Nuclear Physics, Wigner Research Centre for
Physics, Hungarian Academy of Sciences, P.O.Box 49, Budapest, 1525, Hungary}
\address{E\"otv\"os Lor\'and University, P\'azm\'any P\'eter s\'et\'any 1/A,
H-1117, Budapest, Hungary}

\author{ILKKA HELENIUS}
\address{Department of Astronomy and Theoretical Physics, Lund University, 
S\"{o}lvegatan 14A, SE-223 62 Lund, Sweden}

\author{ZHONG-BO KANG}  
\address{Theoretical Division, MS B283, Los Alamos National Laboratory, 
Los Alamos, NM 87545, USA}

\author{PIOTR KOTKO}
\address{Department of Physics, Penn State University, University Park
PA 16803, USA}

\author{KRZYSZTOF KUTAK}
\address{Instytut Fizyki Jadrowej 
im. Henryka Niewodnicza\'nskiego,
Radzikowskiego 152, 31-342 Krak\'ow, Poland}

\author{JEAN-PHILIPPE LANSBERG}
\address{IPNO, Universit\'e Paris-Sud, CNRS/IN2P3, Universit\'e Paris-Saclay,
91406 Orsay Cedex, France}

\author{PETER LEVAI} 
\address{Institute for Particle and Nuclear Physics, Wigner Research Centre for
Physics, Hungarian Academy of Sciences, P.O.Box 49, Budapest, 1525, Hungary}

\author{ZI-WEI LIN}
\address{C-209 Howell Science Complex, Department of Physics, East
  Carolina University, Greenville, NC 27858, USA} 

\author{YASUSHI NARA}
\address{Akita International
    University, Yuwa, Akita-city 010-1292, Japan}

\author{ANDRY RAKOTOZAFINDRABE}
\address{IRFU/SPhN, CEA Saclay, 91191 Gif-sur-Yvette Cedex, France}

\author{G\' ABOR PAPP}
\address{E\"otv\"os Lor\'and University, P\'azm\'any P\'eter s\'et\'any 1/A,
H-1117, Budapest, Hungary}

\author{HANNU PAUKKUNEN}
\address{University of Jyvaskyla, Department of Physics, P.O. Box 35, 
FI-40014 University of Jyvaskyla, Finland}
\address{Departamento de F\'\ isica de Part\'\i culas and IGFAE, Universidade 
de Santiago de Compostela, E-15782 Galicia, Spain}
\address{Helsinki Institute of Physics, P.O. Box 64, FI-00014 University 
of Helsinki, Finland}

\author{ST\'EPHANE PEIGN\'E}
\address{SUBATECH, Universit\'e de Nantes, Ecole des Mines de Nantes,
CNRS/IN2P3, 4 rue Alfred Kastler, 44307 Nantes cedex 3, France}

\author{MIHAI PETROVICI} 
\address{National Institute for Physics and Nuclear Engineering,
Horia~Hulubei, R-077125, Bucharest, Romania}

\author{JIAN-WEI QIU}
\address{Physics Department, Brookhaven National Laboratory, Upton, NY 
11973, USA} 
\address{C.N. Yang Institute for Theoretical Physics, Stony Brook 
University, Stony Brook, NY 11794, USA}
  
\author{AMIR H. REZAEIAN}
\address{Departamento de F\'\i sica, Universidad T\'ecnica
Federico Santa Mar\'\i a, Avda. Espa\~na 1680,
Casilla 110-V, Valpara\'iso, Chile}
\address{Centro Cient\'\i fico Tecnol\'ogico de Valpara\'\i so (CCTVal), 
Universidad T\'ecnica
Federico Santa Mar\'\i a, Casilla 110-V, Valpara\'\i so, Chile }

\author{PENG RU}
\address{School of Physics $\&$ Optoelectronic Technology,Dalian
University of Technology,Dalian,116024, China}
\address{Key Laboratory of Quark \& Lepton Physics (MOE) and 
Institute of Particle Physics,
Central China Normal University, Wuhan 430079, China}

\author{SEBASTIAN SAPETA}
\address{The H.\ Niewodnicza\'nski Institute of Nuclear Physics PAN, 
Radzikowskiego 152, 31-342 Krak\'ow, Poland}
\address{CERN PH-TH, CH-1211, Geneva 23, Switzerland}

\author{VASILE TOPOR POP} 
\address{McGill University, Montreal, H3A 2T8, Canada}

\author{IVAN VITEV}  
\address{Theoretical Division, MS B283, Los Alamos National Laboratory, 
Los Alamos, NM 87545, USA}

\author{RAMONA VOGT}
\address{Nuclear and Chemical Sciences 
Division, Lawrence Livermore National Laboratory, 
Livermore, CA 94551, USA}
\address{Physics Department, University of California at Davis, 
Davis, CA 95616, USA}

\author{ENKE WANG}
\address{Key Laboratory of Quark \& Lepton Physics (MOE) 
and Institute of Particle Physics,
Central China Normal University, Wuhan 430079, China}

\author{XIN-NIAN WANG}
\address{Key Laboratory of Quark and Lepton Physics (MOE) and Institute of 
Particle Physics, Central China  Normal University, Wuhan 430079, China}
\address{Nuclear Science Division, MS 70R0319, Lawrence Berkeley National 
Laboratory, Berkeley, CA 94720, USA}

\author{HONGXI XING}
\address{Theoretical Division, MS B283, Los Alamos National Laboratory, 
Los Alamos, NM 87545, USA}

\author{RONG XU}
\address{Key Laboratory of Quark and Lepton Physics (MOE) and Institute of 
Particle Physics, Central China Normal University, Wuhan 430079, China}

\author{BEN-WEI ZHANG}
\address{Key Laboratory of Quark $\&$ Lepton Physics (MOE) and 
Institute of Particle Physics, Central China Normal University, Wuhan 430079, 
China}

\author{WEI-NING ZHANG}
\address{School of Physics $\&$ Optoelectronic Technology,Dalian
University of Technology, Dalian 116024, China}

\maketitle

\begin{abstract}
Predictions made in Albacete {\it et al.} \cite{Albacete:2013ei} prior to the
LHC $p+$Pb run at $\sqrt{s_{_{NN}}} = 5$ TeV are
compared to currently available data.  Some predictions shown here
have been updated
by including the same experimental cuts as the data.  Some additional
predictions are also presented, especially for quarkonia, that were provided
to the experiments before the data were made public but were too late for the
original publication are also shown here.  
\end{abstract}

\ccode{12.38.Bx, 25.75.Bh, 25.75.Cj, 13.87.-a}

\section{Introduction}

Members and friends of the JET Collaboration made predictions for 
the $\sqrt{s_{_{NN}}} = 5.02$ TeV $p+$Pb run at the LHC in the winter of 2013. 
Predictions were collected for charged
hadrons; identified particles such as $\pi^0$, $K^\pm$, and
$p/\overline p$; photons; jets; $J/\psi$; and gauge bosons.  The
observables
included individual distributions, ratios such as $R_{p{\rm Pb}}$, and
correlation
functions.   The paper in which these predictions were compiled 
\cite{Albacete:2013ei} was submitted to this journal and to arXiv.org 
before the $p+$Pb run began in 2013.  This paper presents the confrontation of
the predictions with data currently available. 

The test beam results for $dN_{\rm ch}/d\eta$ published by the ALICE 
Collaboration \cite{ALICE:2012xs} 
were presented for the case where the lead beam
moved to the right, in the direction of positive rapidity, in 
Ref.~[\refcite{Albacete:2013ei}] because this was the accelerator 
configuration employed for the test run.  Therefore,
all predictions for $R_{p {\rm Pb}}(y)$ were
reflected to conform to that convention.  However, for the full 2013 run, since
some of the detectors, ALICE and LHCb in particular, are asymmetric around
midrapidity, some of the data were taken in a $p+$Pb configuration
(the proton beam moving toward forward rapidity) and the rest were taken in a 
Pb$+p$ configuration (the lead beam moving toward forward rapidity, as in the
case of the test run).  Thus, further publications have generally employed
the typical convention, from fixed-target facilities, where the proton 
beam moves in the direction of positive rapidity.  Thus, in this paper, all 
results are presented assuming this convention unless explicitly stated 
otherwise.

\section{Charged particles}

In Ref.~[\refcite{Albacete:2013ei}], 
detailed descriptions of the approaches used to
calculate the charged particle multiplicities, $p_T$ distributions and 
nuclear modification factors, $R_{p{\rm Pb}}$, in $p+$Pb collisions  were given.  
Therefore the model descriptions will not
be repeated here and instead only a brief summary of the various approaches
is presented in this section.
Note that almost all approaches involve some parameters
tuned at a specific energy to predict results for other energies.  
For details, consult Ref.~[\refcite{Albacete:2013ei}] 
and the original references included therein.

Event generators determine multiplicities from their models 
of soft particle production followed by fragmentation and hadronization.
Hard particle production is typically based on a $p+p$ generator such as
$\mathtt{PYTHIA}$ \cite{Bengtsson:1987kr}.  Examples employed here include 
$\mathtt{HIJING}$ 
\cite{Wang:1991hta,Gyulassy:1994ew,Deng:2010mv,Deng:2010xg,Xu:2012au}, 
$\mathtt{HIJINGB\overline B}$
\cite{ToporPop:2011wk,ToporPop:2010qz,Barnafoldi:2011px,Pop:2012ug} and 
$\mathtt{AMPT}$ \cite{Lin:2004en}.  See Secs.~2.3-2.5 in 
Ref.~[\refcite{Albacete:2013ei}].

Perturbative QCD approaches involving collinear factorization at leading and
next-to-leading order (LO and NLO)
typically require a minimum $p_T$ for validity, making
estimates of total multiplicity difficult. However, above this minimum $p_T$,
they can calculate the $p_T$ distributions and modification factors.  These
calculations differ in the cold nuclear matter effects employed and the 
parameters used.  Nuclear shadowing is generally included, as is isospin, 
differences due to the proton and neutron number of the target nucleus (most
important for Drell-Yan and gauge boson production). Broadening of
the $p_T$ distributions in cold matter and medium-induced energy loss are
also often included.  See Secs.~2.6-2.7, based on 
Refs.~[\refcite{Kang:2011bp,Zhang:2001ce,Papp:2002ub}], in 
Ref.~[\refcite{Albacete:2013ei}].

A more first-principles QCD approach that can provide an estimate of the total
multiplicity is the
color glass condensate (CGC).  This provides a saturation-based description
of the initial state in which nuclei in a high-energy
nuclear or proton-nucleus 
collision appear to be sheets of high-density gluon matter. In this 
approach, gluon production can be described by $k_{\rm T}$-factorization which
assumes an ordering in intrinsic transverse momentum rather than momentum
fraction $x$, as in collinear factorization. The unintegrated gluon density 
associated with $k_T$ factorization is related to the color dipole forward 
scattering amplitude which satisfies the JIMWLK evolution equations
\cite{JalilianMarian:1997jx,JalilianMarian:1997gr,Iancu:2000hn,Ferreiro:2001qy}.
In the large $N_c$ limit, the JIMWLK equations simplify to the 
Balitsky-Kovchegov (BK) equation, a closed-form result for the rapidity
evolution of the dipole amplitude
\cite{Balitsky:1995ub,Kovchegov:1999yj,Kovchegov:1999ua,Balitsky:2006wa}.  
The running coupling corrections to the 
leading log BK equation, rcBK, have been phenomenologically successful in 
describing the rapidity/energy evolution of the dipole 
\cite{Balitsky:1995ub,Kovchegov:1999yj,Kovchegov:1999ua,Balitsky:2006wa,Albacete:2007yr,Albacete:2012xq,Rezaeian:2012ye,Rezaeian:2011ia}.  
The initial condition still needs to be modeled, generally employing the 
McLerran-Venugopalan model 
\cite{McLerran:1993ni,McLerran:1993ka,McLerran:1994vd}
with parameters constrained by data. 
The impact parameter dependent dipole saturation
model (IP-Sat) 
\cite{Kowalski:2003hm,Tribedy:2010ab,Tribedy:2011aa}
is a refinement of the dipole
saturation model that reproduces the correct limit when the dipole radius
$r_T \rightarrow 0$.  It includes power corrections to the collinear DGLAP 
evolution and should be valid where logs in $Q^2$ dominate logs of $x$.
See Secs.~2.1-2.2 in 
Ref.~[\refcite{Albacete:2013ei}] for a more thorough description.

In this update, we do not show all the calculations for the minimum bias charged
particle distributions or the $p_T$-dependent nuclear suppression factor from
ALICE, available from the test beam data, again.  Here we only show calculations
that have been updated or are shown against data taken during the full $p+$Pb
run and were thus not previously available for 
comparison.  In particular, we show updates of the CGC minimum bias charged
particle multiplicity distributions, $dN_{\rm ch}/d\eta$; comparisons of the
centrality dependence of $dN_{\rm ch}/d\eta$ to the ATLAS data calculated
in the same centrality bins; comparison of the $p_T$ distributions to the 
ALICE and CMS midrapidity data; calculations of the average $p_T$ as a function
of the charge particle multiplicity; and comparisons of the nuclear suppression
factor, $R_{p{\rm Pb}}(p_T)$, at midrapidity for ALICE and CMS.

\subsection[Multiplicity distribution]{Multiplicity distribution (J. Albacete, A. Drumitru and A. Rezaeian)}

In the original compilation \cite{Albacete:2013ei}, 
it was shown that the charged particle
pseudorapidity distributions, $dN_{\rm ch}/d\eta$, particularly in the CGC
approach, exhibited a considerably 
steeper slope than the data, especially for $\eta$
in the direction of the lead nucleus.  
Since then, the CGC calculations have
been adjusted, as described below.  

\begin{figure}[t]    
\begin{center}
\includegraphics[width=7 cm] {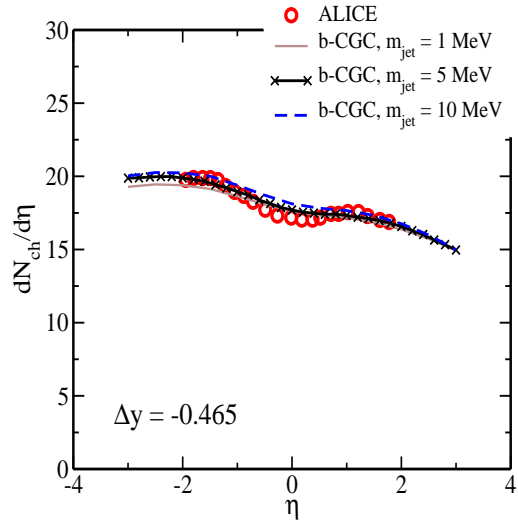}
\end{center}             
\caption{(Color online)
The ALICE charged particle pseudorapidity distribution in minimum-bias 
$p+$Pb collisions at $\sqrt{s_{_{NN}}}=5.02$ TeV \protect\cite{ALICE:2012xs}. 
The b-CGC curves are based on leading log $k_T$-factorization 
and the b-CGC saturation model.  The results are obtained for  
$m_{\rm jet} =1$, 5, and 10~MeV.  The plot is taken from
Ref.~[\protect\refcite{Rezaeian:2012ye}].}
\label{f-alice1a}
\end{figure}

In \fig{f-alice1a}, results are shown for the charged-particle pseudorapidity 
density for non-single diffractive $p+$Pb collisions at $\sqrt{s_{_{NN}}}=5.02$ 
TeV.  In order to compare to the ALICE data \cite{ALICE:2012xs}, the boost of 
the $\eta=0$ laboratory frame is accounted for by adding a rapidity shift of 
$\Delta y=-0.465$. The details of calculation can be found in 
Ref.~[\refcite{Rezaeian:2012ye}]. 
The results are based on $k_T$-factorization and 
the impact-parameter Color Glass Condensate saturation model (b-CGC). 
The parameters of the b-CGC model were determined from a fit to the small-$x$ 
HERA data, including data from diffractive vector meson production 
\cite{Rezaeian:2013tka,Armesto:2014sma}. 

When employing $k_T$ factorization, the rapidity distribution has to be
recast in terms of pseudorapidity:
\begin{eqnarray} 
y(h)=\frac{1}{2}\log
\frac{\sqrt{\cosh^2 \eta+\mu^2}+\sinh \eta}{\sqrt{\cosh^2 \eta+\mu^2} -
\sinh \eta} \, \, .
\label{eq:ytoeta}
\end{eqnarray}
The Jacobian of the rapidity to pseudorapidity transformation is
$h=\partial y/\partial \eta$. 
The scale $\mu$ is determined from the typical transverse mini-jet mass, 
$m_{\rm jet}$, and transverse momentum, $p_T$
\cite{Levin:2010dw,Levin:2010br,Levin:2011hr,Levin:2010zy,Rezaeian:2011ia}.
Different definitions of $\mu$ can be found in the description of 
$k_T$ factorization \cite{Rezaeian:2012ye}.  Here 
$\mu^2=m_{\rm jet}^2/p_T^2$  is employed. 
The main theoretical uncertainty in this 
approach is the choice of the mini-jet mass.  The value of mini-jet mass 
changes the overall $K$-factor in the $k_T$-factorization approach, indicating 
that $m_{\rm jet}$ may mimic some higher-order corrections.
Unfortunately, the 
value of $m_{\rm jet}$ is connected to both soft and hard physics and its true 
value cannot be determined at the current level of calculational accuracy. 

Variations in the choice of $\mu$ and $m_{\rm jet}$ may result in uncertainties 
as large as $\sim 15- 20$\% at the LHC. The RHIC data alone, previously used
to fix $m_{\rm jet}$ and $K$ is not enough to uniquely fix 
$m_{\rm jet}$\footnote{The $K$ factor
is not calculable but is absorbed into an overall factor determined from fits to
lower energy midrapidity data.  This factor includes contributions from
fragmentation and the effective interaction area.  Note that $m_{\rm jet}$ changes
the shape of $dN_{\rm ch}/d\eta$ while $K$ does not.  While $K$ and $m_{\rm jet}$
are correlated, fixing $m_{\rm jet}$ to the ALICE data can put better limits on
$K$.}. 

In \fig{f-alice1a} results on the minimum bias $dN_{\rm ch}/d\eta$ are shown for 
different values of $m_{\rm jet}$.  Values of $m_{\rm jet}$ in the range 
$1 \leq m_{\rm jet} < 30$ MeV provide equally good descriptions of the RHIC data 
on charged hadron multiplicity. It appears that 
$m_{\rm jet} \approx 5$~MeV gives the best description of the ALICE data with 
an uncertainty of less than $4\%$ \cite{Rezaeian:2012ye}. This value of 
$m_{\rm jet}$ is remarkably similar to the up and down current quark masses. 


\begin{figure}[htbp]
\begin{center}
\includegraphics[width=0.795\textwidth]{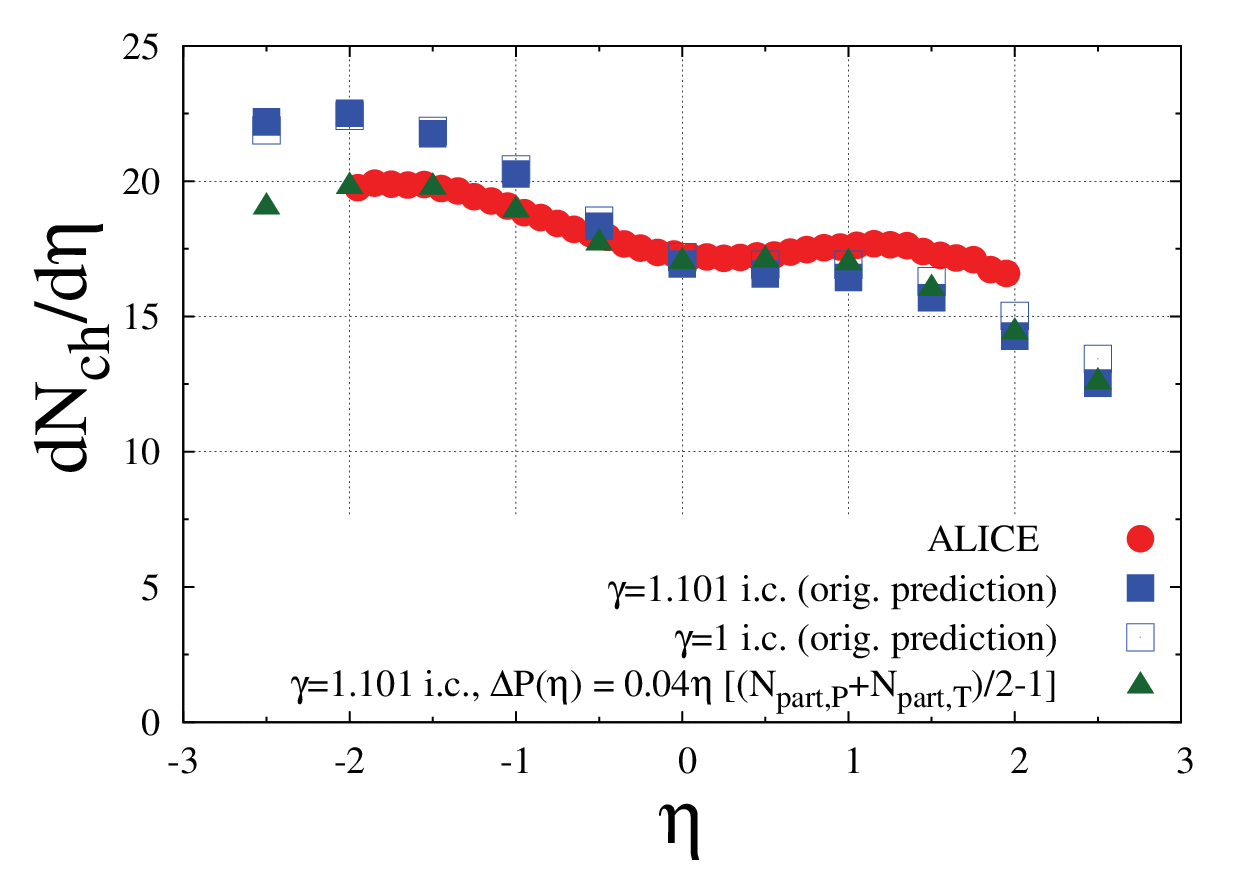}
\end{center}
\caption[]{(Color online)
Charged particle pseudorapidity distribution at $\sqrt{s_{_{NN}}}= 5.02$ 
TeV as a function of $\eta$ with and without the adjusted Jacobian, calculated
by Albacete and Dumitru.
}
\label{fig:dndeta_AD}
\end{figure}

Albacete and Dumitru also show that $dN_{\rm ch}/d\eta$
depends strongly on the $y \rightarrow \eta$ transformation.
The rcBK calculation depends on the Jacobian of this transformation which is not
uniquely defined in the CGC framework. It is necessary
to assume a fixed mini-jet mass, related to the pre-hadronization/fragmentation
stage.  In Ref.~[\refcite{Albacete:2013ei}], they assumed the same 
transformation for $p+p$ and $p+$Pb collisions.
The result in Fig.~\ref{fig:dndeta_AD} shows the dependence of 
$dN_{\rm ch}/d\eta$ on the Jacobian transformation.  
The open and filled squares represent the original result
\cite{Albacete:2013ei} while the filled triangles are based on a Jacobian
with the hadron momentum modified by $\Delta P(\eta) = 0.04 \eta 
[(N_{\rm part}^{\rm proj} + N_{\rm part}^{\rm targ})/2 -1]$.  
The results are essentially
identical in the proton direction but differ considerably in the direction of
the lead beam.  The difference shows the sensitivity of this result to the
mean mass and $p_T$ of the unidentified final-state hadrons.

\subsection[Centrality dependence of $dN_{\rm ch}/d\eta$]{Centrality dependence of $dN_{\rm ch}/d\eta$ (Z. Lin and A. Rezaeian)}

A good description of the minimum-bias data alone cannot be 
considered a sufficient test of a particular approach since there are a number
of alternative approaches \cite{ALICE:2012xs} which can describe the same set 
of data.  The charged hadron multiplicity distribution at different 
centralities provides complementary information to discriminate among models. 

In the b-CGC approach, the impact parameter dependence of the saturation scale 
is an important ingredient for the description of the centrality dependence of 
charged particle production. The impact parameter dependence of $Q_{\rm sat}$ 
in the b-CGC model is self-consistently constrained by a fit to the 
$t$-distribution of diffractive vector meson production at HERA 
\cite{Rezaeian:2013tka,Armesto:2014sma}. Therefore, the centrality dependence
of $dN_{\rm ch}/d\eta$ at the LHC for fixed $m_{\rm jet} = 5$~MeV introduces no
new free parameters.

\begin{figure}[t]     
\begin{center}
\includegraphics[width=0.75\textwidth] {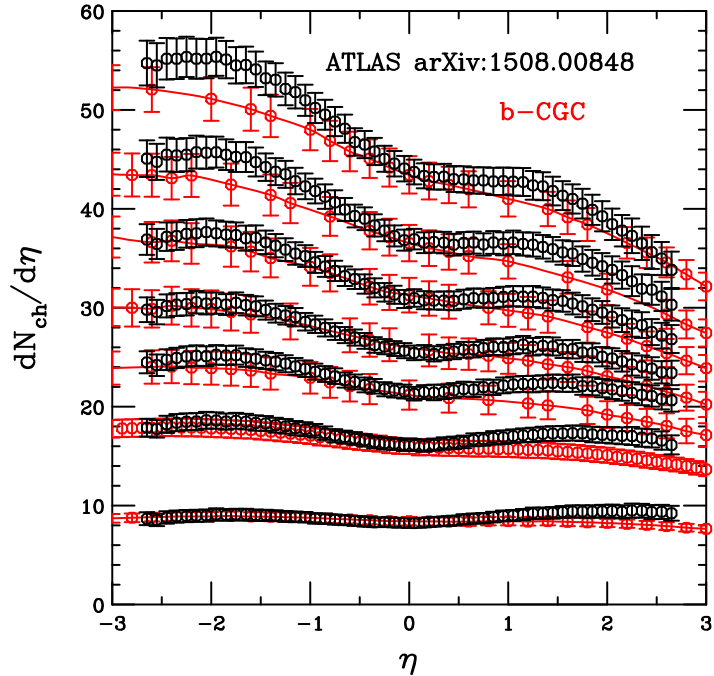}
\end{center}
\caption{(Color online)
The charged particle pseudorapidity distributions in the ATLAS
centrality bins \protect\cite{Aad:2015zza} compared to calculations based 
on leading log $k_t$-factorization in the b-CGC saturation model.  The central
value of the calculation is for $m_{\rm jet} = 5$ MeV.  From top
down the centrality bins are: (1-5)\%, (5-10)\%, (10-20)\%, (20-30)\%, 
(30-40)\%, (40-60)\%, and (60-90)\%.  There is no additional scaling, neither 
on the data nor the calculations.
See Ref.~[\protect\refcite{Rezaeian:2012ye}] for details of the calculation.}
\label{f-alice1b}
\end{figure}

In \fig{f-alice1b}, predictions for the charged hadron multiplicity 
distribution at different centralities are compared to the ATLAS 
data \cite{Aad:2015zza} in $p+$Pb collisions at $5.02$ TeV.  
The theoretical band shown in \fig{f-alice1b} incorporates uncertainties due to 
fixing the $K$ factor and the mini-jet mass by fitting the RHIC minimum-bias
data \cite{Back:2003hx,Arsene:2004cn}.  Figure~\ref{f-alice1b} shows that, 
within theoretical uncertainties, the b-CGC approach generally 
provides a better description of the ATLAS data in the proton region than 
in the nuclear fragmentation region, especially for the more central 
collisions.  The calculations have a rather linear dependence on $\eta$
while the data exhibit more curvature at midrapidity.

The better agreement of the calculations with the results in the proton
direction for more central collisions may be expected since the b-CGC model was 
constrained by small-$x$ data in $e+p$ scattering at HERA. Interestingly, the
b-CGC calculations better reproduce the lead-going multiplicity for the most
peripheral bins, (40-60)\% and (60-90)\%, while underestimating the 
multiplicity in the proton direction.
Future diffractive 
data, including the $t$ distribution of diffractive vector meson production 
in electron-ion collisions, can provide complementary information to constrain 
saturation models, including the impact parameter dependence for nuclear 
targets.  

The recent ALICE data \cite{Adam:2014qja} on the centrality 
dependence of charged hadron production in $p+$Pb collisions are consistent 
with the ATLAS data.  The b-CGC predictions provide a somewhat better 
description of the ALICE data \cite{Adam:2014qja} with the “V0A” centrality 
selection. (The ALICE data are not shown in \fig{f-alice1b}). 

\begin{figure}[htbp]
\begin{center}
\includegraphics[width=0.75\textwidth]{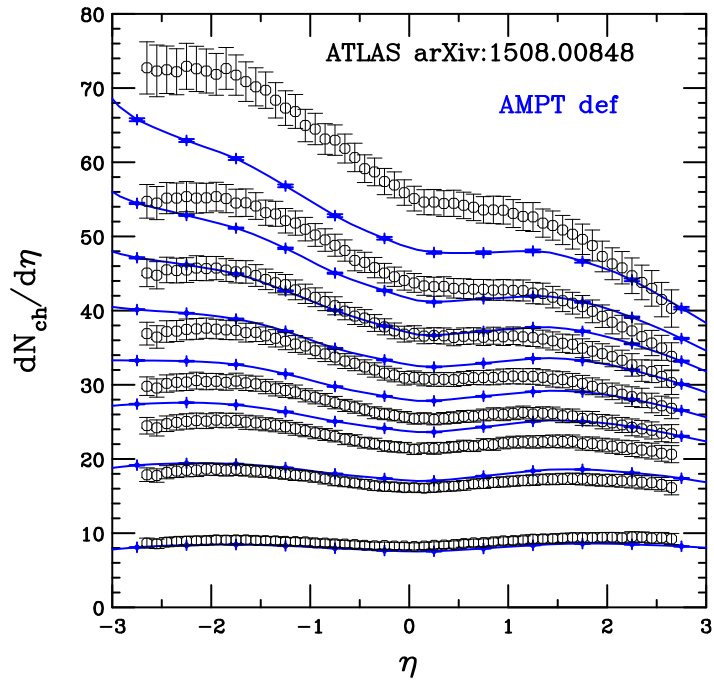}
\caption[]{(Color online)
The charged particle pseudorapidity distributions in the ATLAS
centrality bins \protect\cite{Aad:2015zza} compared to $\mathtt{AMPT}$
calculations.  From top down the centrality bins are: (0-1)\%, (1-5)\%, 
(5-10)\%, (10-20)\%, (20-30)\%, 
(30-40)\%, (40-60)\%, and (60-90)\%.  There is no additional scaling, neither 
on the data nor the calculations.
}
\label{fig5}
\end{center}
\end{figure}

Figure~\ref{fig5} compares the results from the $\mathtt{AMPT}$ event generator
to the same ATLAS data, including also the 0-1\% centrality bin.  In
the $\mathtt{AMPT}$ results, shown in the lab frame, the centrality in 
$p+$Pb collisions is defined according to the total transverse energy 
within the $\eta$ range of the ATLAS forward calorimeter in the lead-going
direction.   The uncertainty on the calculations are statistical.  
The $\mathtt{AMPT}$ results have
an inflection point near midrapidity, similar to the data.  The most 
central results, especially
the 0-1\% bin, underestimate the multiplicity in the lead-going direction
significantly.  However, the (1-5)\%, (5-10)\% and (10-20)\% calculations
reproduce the lead-going direction results relatively well.
The semi-central results, (20-30)\% and (30-40)\%, overestimate the measured
multiplicity.  Finally, the multiplicity in the most peripheral bins are
well reproduced.

Neither calculation reproduces all the ATLAS data.  Both do well in some
centrality bins but the regions where the agreement is good differ in the
two approaches.  The minimum bias results are similar to that of the (40-60)\%
centrality bin while the $p+p$ result is similar to the (60-90)\% centrality
bin.

\subsection[Transverse momentum distribution]{Transverse momentum distribution (J. Albacete, G. G. Barnaf\"oldi, J. Barette,  W.-T. Deng, A. Dumitru, H. Fujii, M. Gyulassy, P. Levai, Z. Lin, Y. Nara, M. Petrovici, V. Topor Pop, X.-N. Wang, and R. Xu)}

Here the transverse momentum distriutions for charged particle production at
midrapidity are compared to the ALICE data \cite{ALICE:2012mj} ($|\eta| < 0.8$)
and the CMS data \cite{CMS:2013cka} ($|\eta|<1$).
While the calculations are not necessarily in the identical rapidity range as
the data, the bin width is divided out so that normalizations of all the 
calculations should be compatible with the data.

\begin{figure}[htbp]
\begin{center}
\includegraphics[width=0.495\textwidth]{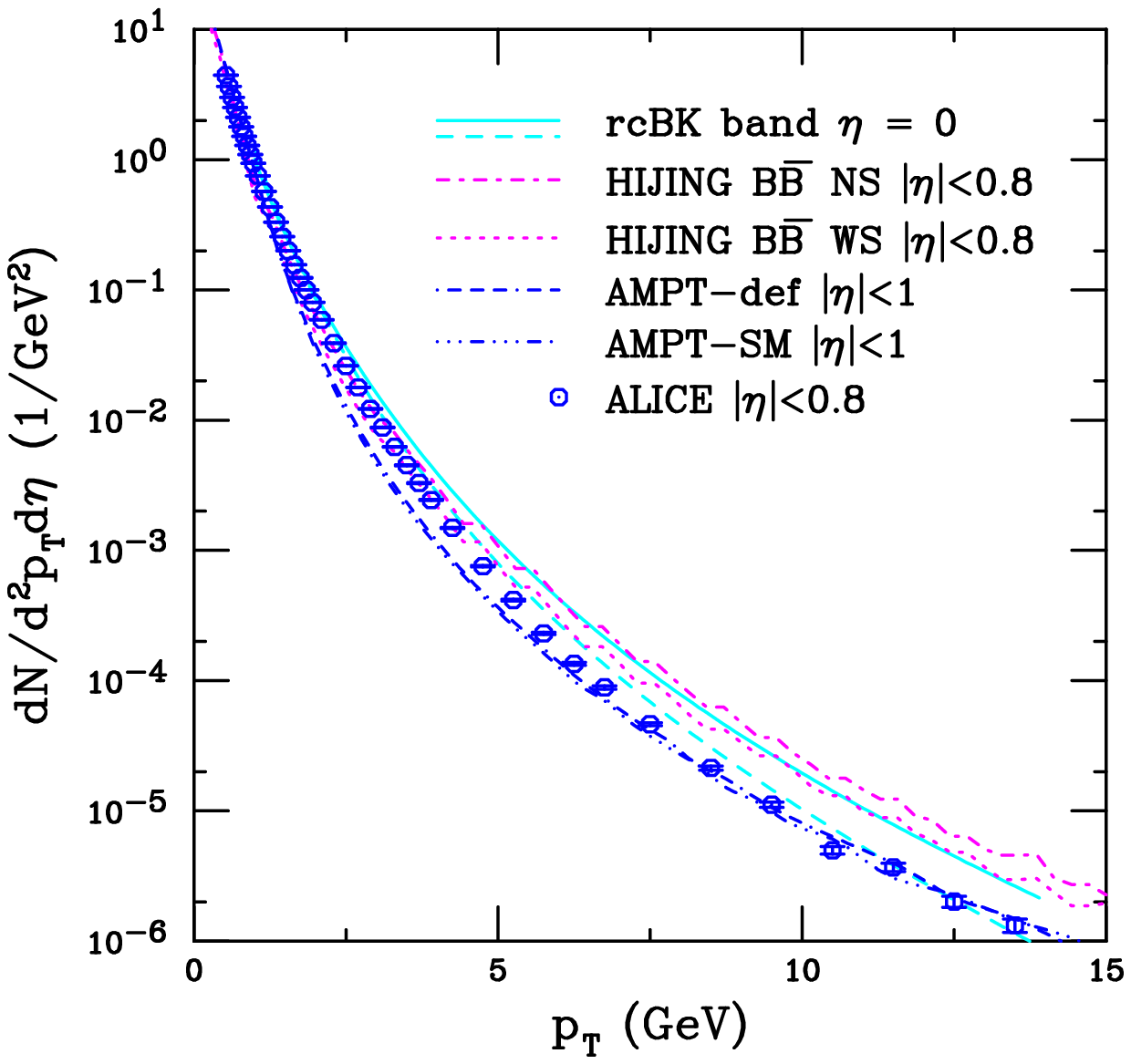}
\includegraphics[width=0.495\textwidth]{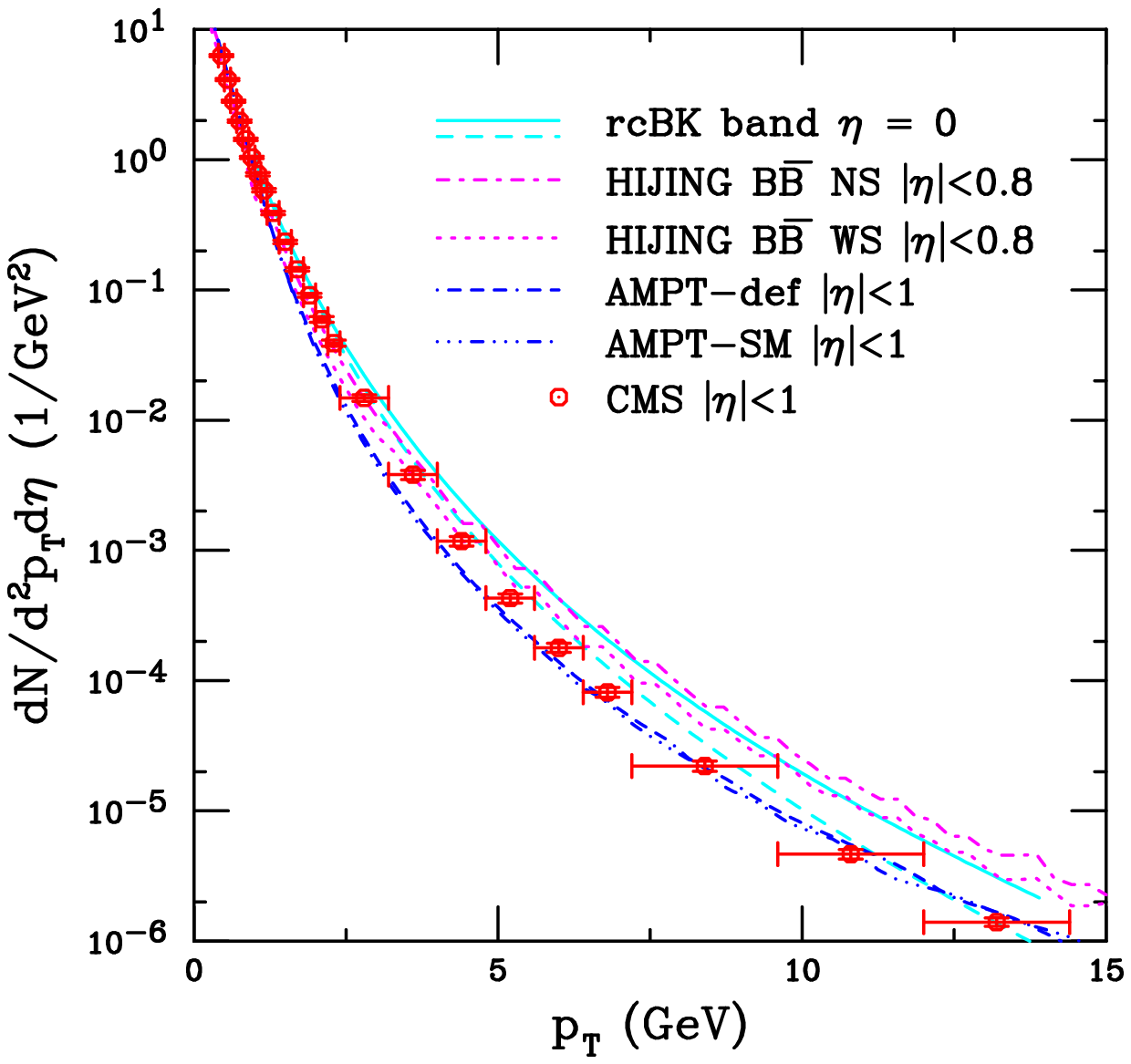}
\caption[]{(Color online) 
Charged particle $p_T$ distributions at $\sqrt{s_{_{NN}}} = 5.02$ TeV.  
The solid and dashed cyan curves outline the rcBK band
\protect\cite{Albacete:2012xq}.
The magenta curves, calculated with $\mathtt{HIJINGB\overline{B}2.0}$ 
are presented
without (dot-dashed) and with (dotted) shadowing. The $\mathtt{AMPT}$ results
are given by the dot-dash-dash-dashed (default) and dot-dot-dot-dashed (SM) blue
curves. The data are from the ALICE \protect\cite{ALICE:2012mj} (left)
and CMS \protect\cite{CMS:2013cka} (right) Collaborations.  
All the calculations were presented in Ref.~[\protect\refcite{Albacete:2013ei}].
}
\label{fig:dnchdpt_comp}
\end{center}
\end{figure}

Figure~\ref{fig:dnchdpt_comp} shows the results for rcBK 
\cite{Albacete:2012xq}, $\mathtt{HIJINGB\overline{B}2.0}$ 
\cite{ToporPop:2011wk,ToporPop:2010qz,Barnafoldi:2011px,Pop:2012ug}, 
and $\mathtt{AMPT}$ \cite{Lin:2004en}.

The rcBK result gives an upper limit (solid curve) and a lower limit (dashed
curve) at $\eta = 0$.  The limits are uncertainty estimates due to
small variations in the scale entering the coupling and fragmentation 
functions.  The results are 
generally in agreement with the data for $p_T < 2$ GeV while, for larger $p_T$,
the results are higher than the data.  This may not be a surprise since, at
sufficiently high $p_T$, the hard scale becomes larger than the saturation scale
and the approach should no longer be valid.

The $\mathtt{HIJINGB\overline{B}2.0}$ distributions 
\cite{ToporPop:2011wk,ToporPop:2010qz,Barnafoldi:2011px,Pop:2012ug} include
strong color fields with $\kappa = 2.1$ GeV/fm and the hard scattering
scale, $p_0$, set to 3.1 GeV.  The results with shadowing (WS) were calculated
with the default $\mathtt{HIJING}$ 
parameterization \cite{Wang:1991hta,Gyulassy:1994ew}.
They are in rather good
agreement with the data up to $p_T \sim 4$ GeV. At higher $p_T$, the results
with and without shadowing bracket the upper limit on the rcBK results.  The
calculation without shadowing is always higher that including shadowing.

On the other hand, the $\mathtt{AMPT}$ distributions, have a rather different
curvature from the ALICE data and the other calculations shown in 
Fig.~\ref{fig:dnchdpt_comp}.  They drop faster at low
$p_T$ than the other results but then become harder at high $p_T$, becoming
similar to the data for $p_T > 5$ GeV, especially for the CMS data.
There is essentially no difference between the default $\mathtt{AMPT}$
results and those with string melting.

\begin{figure}[htbp]
\begin{center}
\includegraphics[width=0.495\textwidth]{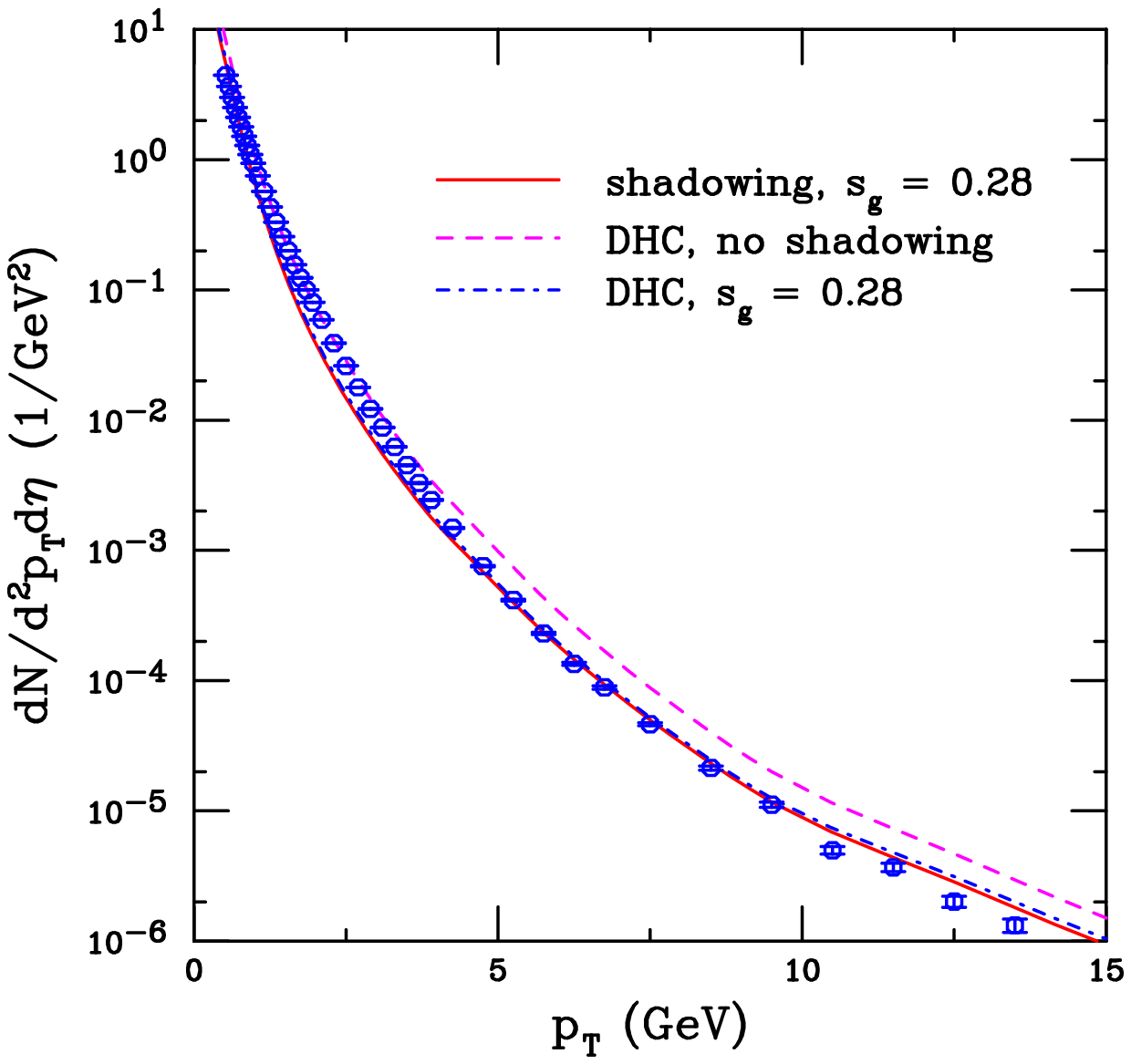}
\includegraphics[width=0.495\textwidth]{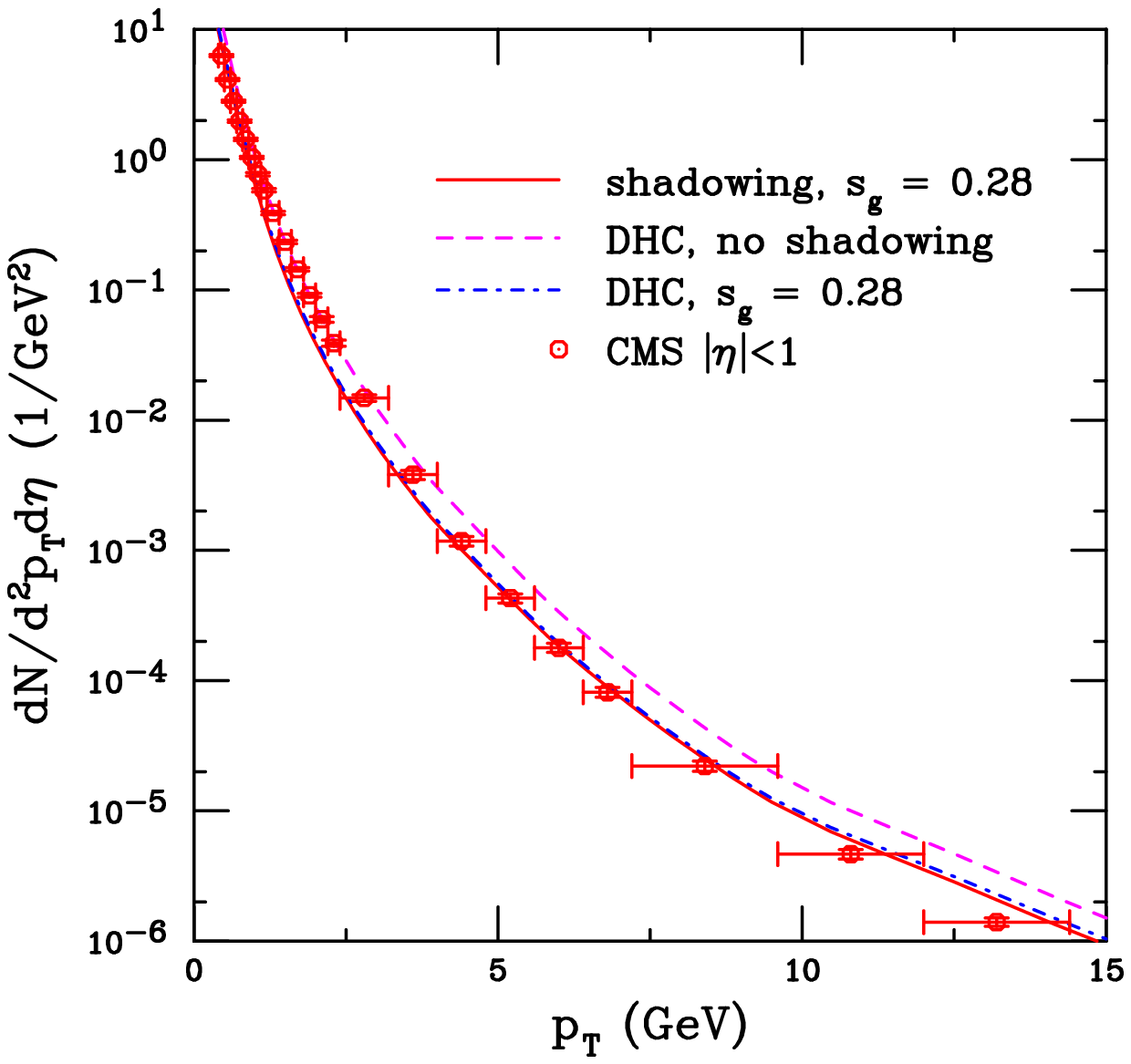}
\caption[]{(Color online)
Charged particle $p_T$ distributions at $\sqrt{s_{_{NN}}} = 5.02$ TeV.  
The charged hadron $p_T$ distribution in $p+$Pb collisions 
with different $\mathtt{HIJING2.1}$ options \protect\cite{Xu:2012au}
is also compared to data.
The data are from the ALICE \protect\cite{ALICE:2012mj} (left)
and CMS \protect\cite{CMS:2013cka} (right) Collaborations.  
All the calculations were presented in Ref.~[\protect\refcite{Albacete:2013ei}].
}
\label{fig:dnchdpt_compXN}
\end{center}
\end{figure}

Figure~\ref{fig:dnchdpt_compXN} shows several options for
cold matter effects in $\mathtt{HIJING2.1}$ \cite{Xu:2012au}.  
The solid red curves labeled ``shadowing,
$s_g = 0.28$'' treats hard scatterings as in default $\mathtt{HIJING}$ but
includes a stronger gluon shadowing than quark scattering, consistent with
the Pb+Pb data at $\sqrt{s_{_{NN}}} = 2.76$~TeV.  The other two results, the
dashed magenta and dot-dashed blue curves, labeled ``DHC'' changes the
order of the scattering processes in $\mathtt{HIJING2.1}$ so that hard
scatterings are simulated first, followed by the soft scatterings so as to
not limit the hard scatterings.

The results including shadowing are very similar for all $p_T$, only the
calculation without shadowing has a different $p_T$ dependence.   The
distributions including shadowing agree well with the ALICE data for
$4 < p_T < 12$ GeV but are somewhat above the data for higher $p_T$.  On
the low end of the $p_T$ range, the calculation without shadowing is in
better agreement with the data.  The same trend is clearly seen for the CMS
data except, at the highest $p_T$, the wide bins can accommodate the
calculations with shadowing.

None of the calculations can describe the entire $p_T$ range of either data
set.  The treatment of hard scatterings in $\mathtt{AMPT}$ and 
$\mathtt{HIJING2.1}$ reproduce the $p_T$ distributions best at high $p_T$.
However, they gave a rather poor description of the ALICE $R_{p{\rm Pb}}(p_T)$ 
from the test beam, see Ref.~[\refcite{Albacete:2013ei}].  Of the
calculated $p_T$ distribution shown here, only the rcBK result gave
a rather good description of $R_{p{\rm Pb}}(p_T)$ for all $p_T$, albeit with
a wide uncertainty band.

\subsection[Average transverse momentum]{Average transverse momentum (A. Rezaeian)}

Within the $k_T$-factorization formalism, supplemented by the b-CGC 
saturation model, it is possible to compute the average transverse momentum  
of charged particles in $p+p$, $p+$Pb and Pb+Pb collisions 
\cite{Rezaeian:2013woa}. 

\begin{figure}[t]                            
\begin{center}
\includegraphics[width=0.6\textwidth] {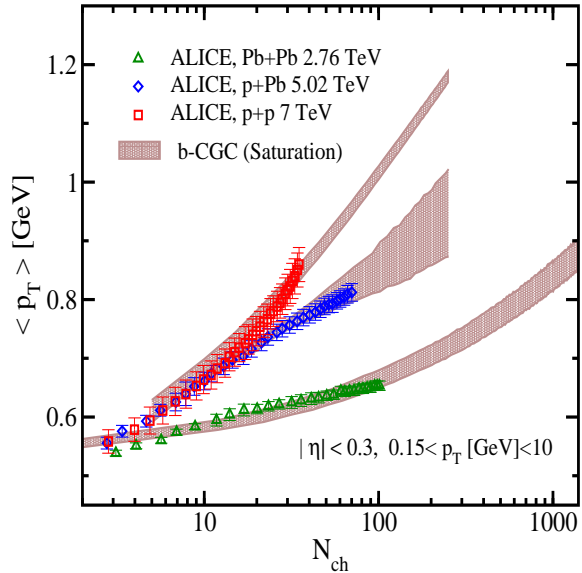}                 
\end{center}
\caption{(Color online) The average transverse momentum,  $\langle p_T\rangle$, 
of charged particles in the range $0.15<p_T<10$ GeV as a function of 
charged particle multiplicity, $N_{\rm ch}$, in $p+p$, $p+$Pb, and Pb+Pb 
collisions at $\sqrt{s_{_{NN}}} = 7$, 5.02 and 2.76 TeV respectively for 
$|\eta|<0.3$.  The b-CGC band includes the theoretical uncertainties 
\protect\cite{Rezaeian:2013woa}.  
The ALICE data \protect\cite{Abelev:2013bla} are also shown. 
(Taken from Ref.~[\protect\refcite{Rezaeian:2013woa}].)}
\label{f-avp}
\end{figure}

In \fig{f-avp}, the average transverse momentum  
of charged particles, $\langle p_T \rangle$, is shown as a function of the
charged particle multiplicity, $N_{\rm ch}$, in $p+p$, $p+$Pb and Pb+Pb 
collisions at $\sqrt{s_{_{NN}}} =7$, 5.02 and 2.76 TeV respectively, in the 
$p_T$ range $0.15<p_T<10$ GeV at midrapidity, $|\eta|<0.3$. The theoretical 
uncertainties are also shown \cite{Rezaeian:2013woa}.  Note that, for large
$N_{\rm ch}$, $\langle p_T\rangle$ is smaller in Pb+Pb collisions than in $p+$Pb 
and $p+p$ for the same value of $N_{\rm ch}$.  This is because the effective 
area of the interaction region is different in Pb+Pb collisions compared to 
the smaller systems.  While the trends of the calculations are similar to the
data and the magnitudes are well matches, the curvature of the $p+p$ and Pb+Pb 
calculations is slightly different than the data.

Events with $N_{\rm ch}<\langle N_{\rm ch} \rangle$ are more perpherial and thus
less dense compared to minimum-bias collisions.  Note that the average 
charged particle multiplicity reported by ALICE \cite{Abelev:2013bla} is 
$\langle N_{\rm ch} \rangle\approx 259.9$, 11.9 and 4.42 in Pb+Pb, $p+$Pb and 
$p+p$ collisions respectively in the kinematics of \fig{f-avp}.  Therefore, 
at moderate $N_{\rm ch}$, $N_{\rm ch} < 150$ in \fig{f-avp}, the Pb+Pb system
is dilute since $\langle N_{\rm ch} \rangle \approx 259.9$ while, in $p+p$
collisions, the same multiplicity selection criteria corresponds to a very 
rare, high-density event.  

Neither final-state hadronization 
nor collective hydrodynamics effects are required in this approach to describe 
the main features of the data shown in \fig{f-avp}. The logarithmic 
rise of $\langle p_T\rangle$  with the density or charged hadron multiplicity  
is directly related to the rise of the saturation scale with density in the 
CGC approach \cite{Rezaeian:2011ia,Rezaeian:2012ye,Levin:2010dw,Levin:2010br,Levin:2011hr,Rezaeian:2013woa}. 

\subsection[Nuclear modification factor]{Nuclear modification factor (K. J. Eskola, I. Helenius, H. Paukkunen and A. Rezaeian}

The original calculations of the nuclear modification factor were compared
to the already available ALICE test beam data \cite{ALICE:2012mj} in 
Ref.~[\refcite{Albacete:2013ei}].  
Those results showed that, with wide uncertainty bands, the CGC-type approaches
agreed with the ALICE data \cite{ALICE:2012mj}.  Perturbative QCD approaches
generally agreed with the trends of the data but underestimated the rise of
$R_{p{\rm Pb}}(p_T)$ at low $p_T$.  Event generators typically did not agree well
with the data.  These results are not reproduced, see 
Ref.~[\refcite{Albacete:2013ei}] for details.  
Only updates are presented here.

\subsubsection[rcBK]{rcBK (A. Rezaeian)}

As emphasized in Ref.~[\refcite{Albacete:2013ei}], the uncertainty on the b-CGC
calculation was rather large because, for fixed scale, $\alpha_s$, the value
of $N$ varied between 1 and 5,  resulted in a wide range of predictions.
Here, further constraints on the allowed value of $N$ gives $4 < N < 6$,
consistent also with the value of the saturation scale extracted from other
observables.


\begin{figure}[t]                                                            
\begin{center}
\includegraphics[width=7 cm] {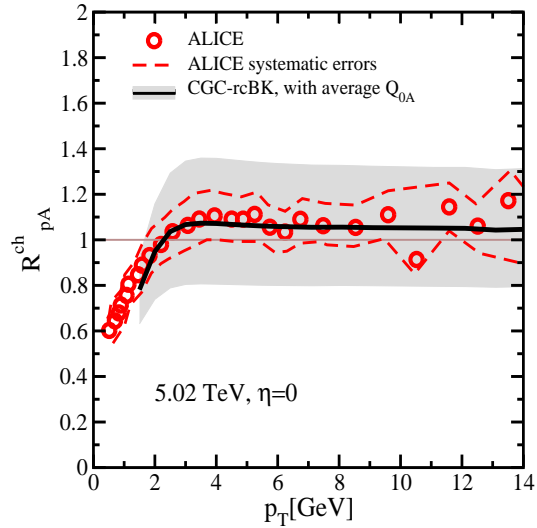}
\end{center}
\caption{(Color online) The nuclear modification factor $R_{pA}(p_T)$ for 
inclusive charged hadron production in minimum-bias $p+$Pb collisions 
at $\sqrt{s_{_{NN}}}=5.02$ TeV and $\eta=0$. 
The calculations are solutions of the 
rcBK equation with $Q_{0A}^2=0.168\, N$ GeV$^2$ for $4 < N < 6$ (grey area).
The black curve shows $N=5$.  The calculations are detailed in 
Ref.~[\protect\refcite{Rezaeian:2012ye}]. 
The points are the ALICE data while the
dashed red lines delineate the experimental systematic uncertainties 
\protect\cite{ALICE:2012mj}. (The plot is taken 
from Ref.~[\protect\refcite{Rezaeian:2012ye}]).}
\label{f-alice2}
\end{figure}

In \fig{f-alice2}, the updated predictions are compared with the ALICE minimum
bias, midrapidity data \cite{ALICE:2012mj}. The ALICE data are in good 
agreement with the predictions shown in Fig.~2 of 
Ref.~[\refcite{Rezaeian:2012ye}].
These solutions of the rcBK evolution equation with average initial nuclear
saturation scale of $Q_{0A}^2 = 0.168 \, N$ GeV$^2$ employed $N\approx 5$, 
constrained in Ref.~[\refcite{Rezaeian:2012ye}].  
It is remarkable that the preferred
value of $N$ corresponds to the average value of $Q_{0A}$ extracted by other 
means, see Eq.~(19) in Ref.~[\refcite{Rezaeian:2012ye}]. 

While the ALICE data have rather large systematic uncertainties, they can
nevertheless impose a strong additional constraint on the initial nuclear 
saturation scale.  They prefer $4 <N < 6$ with effectively zero strong 
coupling to the inelastic terms, $\alpha_s^{\rm in}\approx 0$, as shown in 
the gray region of \fig{f-alice2}. However, a larger $N$ with a finite 
$\alpha_s^{\rm in}$ cannot currently be ruled out. The scale employed in 
$\alpha_s^{\rm in}$ cannot be determined within the current approximation, a 
full NNLO calculation, as yet unavailable, is required. Therefore, the freedom 
to choose $\alpha_s^{\rm in}$ in the hybrid factorization formalism introduces 
rather large uncertainties \cite{JalilianMarian:2011dt}.  

A remarkable feature of the ALICE $R_{p{\rm Pb}}$ data is that the data show no 
evidence of any Cronin-type enhancement.  While the experimental 
uncertainites are too large to draw any firm conclusion, if this feature
persists in more precise data, it can be considered evidence of small-$x$ 
evolution effects at the LHC.  A measurement of $R_{p{\rm Pb}}$ at 
forward rapidities at the LHC could provide an additional crucial test 
of the CGC approach with valuable information about the saturation dynamics 
\cite{Rezaeian:2012ye}.


\subsubsection[Minimum-bias charged-hadron production in $p+$Pb collisions: 
Collinear factorization]{Minimum-bias charged-hadron production in $p+$Pb collisions: Collinear factorization (K. J. Eskola, I. Helenius and H. Paukkunen)}

\begin{figure}[htb!]
\center
\includegraphics[scale=0.75]{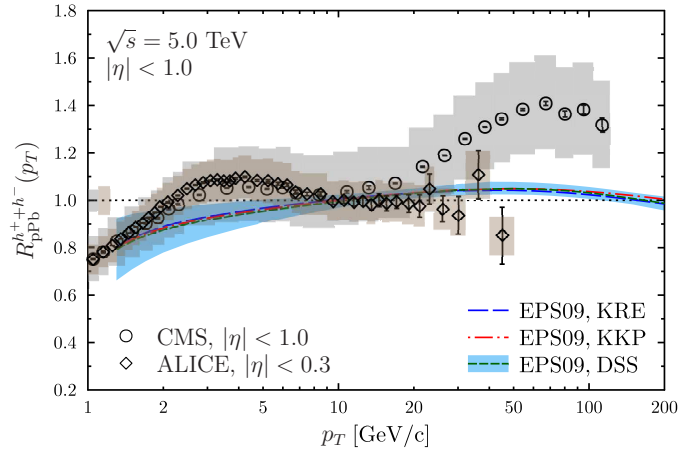}
\caption{(Color online) The charged-hadron nuclear modification factor 
measured in $p+$Pb collisions by the ALICE \protect\cite{Abelev:2014dsa} 
(diamonds) and CMS \protect\cite{Khachatryan:2015xaa} (circles) 
collaborations. The data are compared to NLO calculations 
\protect\cite{Helenius:2012wd} which use the CT10 free proton PDFs 
\protect\cite{Lai:2010vv}, EPS09 NLO nuclear modifications 
\protect\cite{Eskola:2009uj}, and three different set of fragmentation
functions (Kretzer \protect\cite{Kretzer:2000yf}, KKP 
\protect\cite{Kniehl:2000fe} and DSS \protect\cite{deFlorian:2007hc}). 
The EPS09 uncertainty range is shown as a sky blue band and is calculated 
using the DSS fragmentation function.}
\label{fig2}
\end{figure}

The inclusive production of charged hadrons and jets (see Sec.~\ref{Kari-jets})
are intimately related. Thus, based on the excellent description of jet
production by the EPS09 nPDFs seen in Fig.~\ref{fig1} of Sec.~\ref{Kari-jets},
fair agreement of the calculations with charged-hadron production could have 
been expected. As shown in Fig.~\ref{fig2}, there is indeed agreement between 
the EPS09-based predictions \cite{Helenius:2012wd} and the ALICE data 
\cite{Abelev:2014dsa} for the nuclear modification factor $R_{p{\rm Pb}}$ for 
$p_{T} > 10$ GeV\footnote{For identified pions, agreement can be expected to
extend to lower $p_T$ since the enhancement around $p_T \sim 3$ GeV is absent 
in the preliminary ALICE data \cite{Knichel:2014yaa}.}.  However, the CMS 
measurement \cite{Khachatryan:2015xaa} of $R_{p{\rm Pb}}(p_T)$ at 
$p_T \gtrsim 20$ GeV increases by some 40\%, clearly beyond the expectations 
of the EPS09 NLO nPDFs.

Similar indications of a large $p_T$ enhancement are also present in the 
preliminary ATLAS data \cite{Balek:2014uha}, but since these data have been 
centrality selected, no direct comparison is made due to the biases that the 
centrality classification in $p+$Pb collisions is known to pose 
\cite{Armesto:2015kwa}. On the contrary, the ALICE data \cite{Abelev:2014dsa} 
for the same observable shows no sign of such an increase.  However, the 
$p_{T}$ range is more limited. 

\begin{figure}[htb!]
\center
\includegraphics[width=\textwidth]{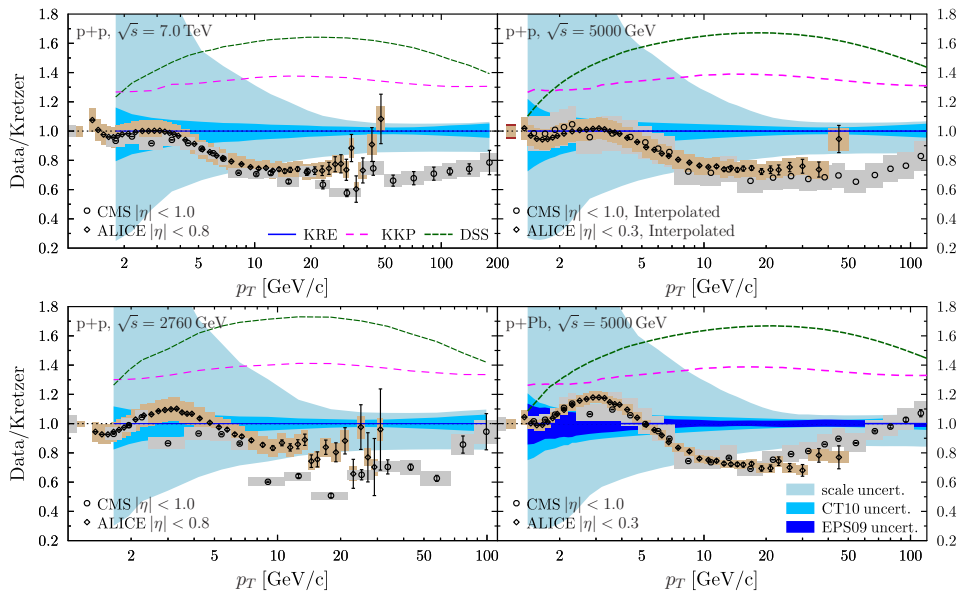}
\caption{(Color online) Ratios between the ALICE 
\protect\cite{Abelev:2014dsa, Abelev:2013ala} (diamonds) and
CMS \protect\cite{Khachatryan:2015xaa, Chatrchyan:2011av, CMS:2012aa} (circles) 
and the NLO calculations employing the CT10 proton PDFs with EPS09 NLO 
(in $p+$Pb collisions), and Kretzer FFs. In all panels, the light blue bands 
quantify the QCD scale uncertainty while the darker blue bands indicate the 
CT10 error range in $p+p$ collisions and the EPS09 NLO error range in $p+$Pb 
collisions.  The purple and green curves are calculations in the same framework
but employing the  KKP \protect\cite{Kniehl:2000fe} and DSS 
\protect\cite{deFlorian:2007hc} FFs respectively, normalized to the calculation 
with the Kretzer FFs. The $\sqrt{s}=7$~TeV and $\sqrt{s}=2.76$~TeV panels are 
from Ref.~[\protect\refcite{d'Enterria:2013vba}].
}
\label{fig3}
\end{figure}

More light can be shed on this mystery by considering the absolute $p_{ T}$ 
spectra measured in $p+$Pb collisions and the baseline $p+p$ data used in 
forming $R_{p{\rm Pb}}$. Since no direct $p+p$ data were available at 
$\sqrt{s_{_{NN}}}=5.02$ TeV, the experimental collaborations have
constructed these baseline data from measurements at other nearby 
center-of-mass energies, generally $\sqrt{s}=2.76$ and 7 TeV. 
In Fig.~\ref{fig2}, the measured $p_T$ spectra are contrasted with the NLO 
calculations. In addition to the nuclear modifications, these calculations 
also depend on the parton-to-hadron fragmentation functions (FFs). While it is 
known \cite{d'Enterria:2013vba} that none of the currently available sets of 
fragmentation functions can optimally reproduce the LHC data\footnote{As 
conjectured in Ref.~[\refcite{d'Enterria:2013vba}], 
this could be related to the 
lack of appropriate constraints (FFs like Kretzer \cite{Kretzer:2000yf} or KKP 
\cite{Kniehl:2000fe} only employ ${\rm e}^+{\rm e}^-$ data) or an inadequate
$p_T$ range (e.g. DSS \cite{deFlorian:2007hc} uses only low $p_T$ data).},
if the same set of FFs is used when comparing calculations to the independent 
$p+p$ and $p+$Pb data sets, it is possible to draw conclusions regarding the 
mutual agreement/disagreement of the calculations with the data. 
Such a comparison 
is presented in Fig.~\ref{fig3} which shows ratios of the CMS and ALICE data 
to NLO calculations using Kretzer FFs \cite{Kretzer:2000yf} in $p+p$ and $p+$Pb 
collisions at different center-of-mass energies. 

The behaviour of the ALICE and CMS data relative to the NLO calculations
in all panels of Fig.~\ref{fig3} is very similar.  They more or less agree 
within the uncertainties, apart, perhaps, from the $\sqrt{s}=2.76$~TeV result. 
In the case of the ALICE and CMS $\sqrt{s}=5$~TeV $p+p$ baseline, the 
data-to-theory ratios are approximately flat for $p_{T} \gtrsim 10$~GeV. 
The same is true for the ALICE $p+$Pb data. However, the CMS $p+$Pb data show 
a distinct upward slope from $p_{T} \approx 20$~GeV onward. It thus appears 
that the origin of the differences in $R_{p{\rm Pb}}$ comes from the $p+$Pb data 
and not the $p+p$ baseline. In any case, gluon antishadowing large enough
to accommodate the 40\% rise of $R_{p{\rm Pb}}$ at high $p_{T}$ would not be 
compatible with the dijet measurements shown in Fig.~\ref{fig1} where the 
required antishadowing is only $\sim 5$\%. 

\subsubsection[Forward-backward asymmetry]{Forward-backward asymmetry (G. G. Barnaf\"oldi, J.Barette, S. M. Harangoz\'o, M. Gyulassy, P. Levai, Z. Lin, G. Papp and V. Topor Pop)}
 
CMS has recently provided data on the forward-backward asymmetry of
charged hadron production \cite{CMS:2013cka,Khachatryan:2015xaa}.
They calculate the asymmetry as
\begin{eqnarray}
Y_{\rm asym}^{h}(p_T) = 
\frac{E_h d^3\sigma^h_{p\rm Pb}/d^2p_T d\eta |_{\eta_{\rm cm+}}}
{E_h d^3\sigma^h_{p {\rm Pb}}/d^2p_T d\eta|_{\eta_{\rm cm-}}} 
= \frac{R^h_{p {\rm Pb}}(p_T,\eta_{\rm cm+})}
{R^h_{p{\rm Pb}}(p_T,\eta_{\rm cm-})} 
\, \, .
\label{yasym}
\end{eqnarray}
where the lead beam is assumed to move toward positive rapidity in the
center of mass frame, $\eta_{\rm CM+}$, while the proton beam is assumed to move
in the direction of negative rapidity in the center of mass, $\eta_{\rm CM-}$.
This is the same convention as assumed in Ref.~[\refcite{Albacete:2013ei}], 
compatible with the 2012 $p+$Pb test run.

\begin{figure}[htpb]
\begin{center}
\includegraphics[height=3in]{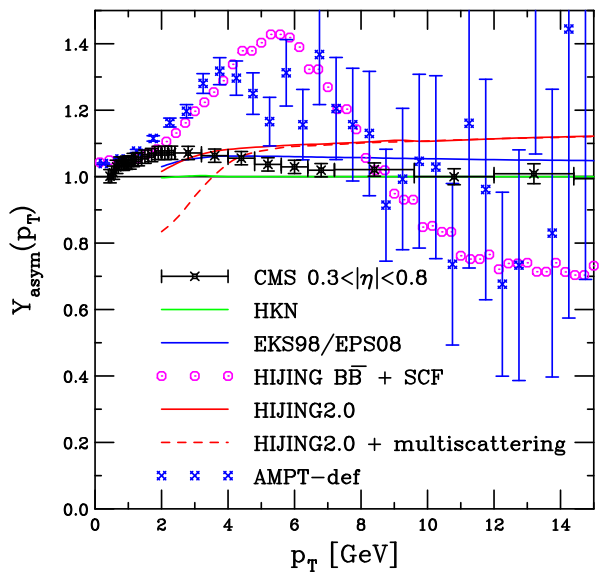}
\end{center}
\caption[]{(Color online)
Predictions for the forward-backward asymmetry, $Y_{\rm asym}^{h}(p_T)$,
from Refs.~\protect\cite{Levai:2011qm,Barnafoldi:2008ec}.  
Centrality-independent results are shown
for the HKN \protect\cite{Hirai:2001np}, EKS98 \protect\cite{Eskola:1998df}
and EPS08 \protect\cite{Eskola:2008ca} parameterizations.  
Minimum bias results obtained by integrating over centrality
are also shown for $\mathtt{HIJINGB\overline{B}}$ and 
$\mathtt{HIJING2.0}$ with and without multiple scattering.  
The blue points are the $\mathtt{AMPT-def}$ results.
The CMS data \protect\cite{CMS:2013cka,Khachatryan:2015xaa} are shown in black.
}
\label{fig:ggb:2}
\end{figure}

The asymmetries, calculated in the center of mass frame
in the range $0.3<|\eta|<0.8$ \cite{Adeluyi:2008gj}, 
both for shadowing in collinear factorization calculations and 
the $\mathtt{HIJINGB\overline{B}}$ and $\mathtt{AMPT}$ event generators, 
are shown in Fig.~\ref{fig:ggb:2}.  These calculations, also shown in
Ref.~[\refcite{Albacete:2013ei}], 
are described there.  All the calculations are
minimum bias (MB).   The results with HKN \cite{Hirai:2001np}, EKS98
\cite{Eskola:1998df}, and EPS08 \cite{Eskola:2008ca}, calculated with the
$\mathtt{kTpQCD\_v2O}$ code\cite{Papp:2002ub}, are MB by
default because they do not include any impact parameter dependence.  The
results labeled $\mathtt{HIJING2.0}$ use the $\mathtt{HIJING2.0}$ shadowing
parameterization and, in one case, the multiple scattering prescriptions,
$\mathtt{HIJINGB\overline{B}}$ and $\mathtt{AMPT-def}$ are integrated over
impact parameter and are thus MB.
One calculation, for the 20\% most central
collisions, shown in Ref.~[\refcite{Albacete:2013ei}], has been removed because
the data are only for minimum bias collisions.
The results for central collisions are greater than unity over all $p_T$ 
because shadowing effects are
expected to be enhanced at more central impact parameters.

In Ref.~[\refcite{Khachatryan:2015xaa}], EPS09 NLO calculations of $Y_{\rm asym}$
with the Kretzer fragmentation functions calculations 
\cite{Helenius:2012wd,Paukkunen:2014vha}, as in Fig.~\ref{fig2},
are shown compared to the data, not only for $0.3 < |\eta_{\rm cm}|<0.8$ but
also for $0.8 < | \eta_{\rm cm} | < 1.3$ and $1.3 < | \eta_{\rm cm}| < 1.8$,
all for $p_T > 5$ GeV.  The central EPS09 NLO results in 
Ref.~[\refcite{Khachatryan:2015xaa}]
are similar to those shown in Fig.~\ref{fig:ggb:2} for EKS98 and EPS08.  The
EPS09 NLO modifications \cite{Khachatryan:2015xaa}) also
but also include the EPS09 NLO uncertainties.  

The data show a mild enhancement on either side of midrapidity in the center
of mass at low $p_T$ ($p_T  <5$ GeV).  
When higher rapidities are studied, the low $p_T$ 
enhancement in the data increases, from $Y_{\rm asym} < 1.1$ for 
$0.3 < |\eta_{\rm cm}| < 0.8$, to a peak of $\sim 1.2$ at 
$0.8 < |\eta_{\rm cm}| < 1.3$
and $\sim 1.3$ at $1.3 < |\eta_{\rm cm}| < 1.8$ \cite{Khachatryan:2015xaa}.
At higher $p_T$, the data at all rapidities are consistent with unity. 

A low $p_T$ enhancement that increases with rapidity is not surprising since
there is an enhancement (antishadowing) in the lead direction (high $x$ in 
lead) and a depletion (shadowing) in the proton direction (low $x$ in lead). 
The antishadowing and shadowing effects both increase at higher rapidity
where $x$ increases in the lead direction and decreases in the proton direction.
The ratio of the two gives an enhancement in $Y_{\rm asym}$ for low $p_T$.
Note that, even for $\eta_{\rm cm} = \pm 1.8$, $x$ is not very large for 
$p_T < 5$ GeV.  At low $p_T$, the effect is particularly enhanced because of
gluon contributions.

As $p_T$ increases, the overall effect of modification of the parton densities
decreases substantially, especially for gluons at low $p_T$ because
the evolution of the gluon nPDFs is large.  For $p_T > 10$ GeV, quark 
dominated processes, like quark-gluon interactions, become more important.
The combination of these effects reduces the calculated $Y_{\rm asym}$ at high
$p_T$, as seen in the calculations in Ref.~[\refcite{Khachatryan:2015xaa}].

The HKN calculation has no discernible asymmetry.  The EKS98 and EPS08 results 
are indistinguishable and are thus labeled as EKS98/EPS08.  These calculations 
are in reasonably good agreement with the CMS data.  The shadowing
parameterization in $\mathtt{HIJING2.0}$, a rather simple $Q^2$-independent
model, agrees relatively well at low $p_T$ but is higher than the data for
$p_T > 6$ GeV.  As mentioned before, the EPS09 NLO result, including 
uncertainties, gives good agreement with the high $p_T$ data at the most central
rapidities but shows a depletion at high $p_T$ relative to the data further
away from midrapidity \cite{Khachatryan:2015xaa}.  The 
$\mathtt{HIJINGB\overline{B}}$ and $\mathtt{AMPT}$ results predict a larger 
enhancement at higher $p_T$ than supported by the data and give an asymmetry
less than unity at higher $p_T$.

\subsection[Flow]{Flow (Z. Lin)}

In the previous compilation \cite{Albacete:2013ei}
$\mathtt{AMPT}$ was used to calculate the
yields, $p_T$ spectra, and flow coefficients of particles produced in
$p+p$ and $p+$Pb collisions  at $\sqrt {s_{_{NN}}}=5$ TeV. 
The same $\mathtt{AMPT}$ event data is used to calculate the $p_T$ dependence
of the anisotropy harmonics $v_n (n=2,3,4)$ from the string melting
$\mathtt{AMPT-SM}$ version of $\mathtt{AMPT}$ \cite{Lin:2004en}. 
$\mathtt{AMPT-SM}$ has been previously used to study these observables 
and direct comparison to the $p+$Pb $v_2$ and $v_3$ data have shown
generally good agreement \cite{Bzdak:2014dia}.
The string-melting mechanism in $\mathtt{AMPT}$ converts
traditional hadronic strings in the initial state to partonic matter
when the energy density in the overlap volume of the collision is
expected to be higher than that of the QCD phase transition. 
It also includes a quark coalescence model to
describe the bulk hadronization of the resultant partonic matter.

$\mathtt{AMPT}$ version 2.26t1 \cite{ampt} was used to generate
the results in Ref.~[\refcite{Albacete:2013ei}] as well as the $\mathtt{AMPT}$
results here.
Following Ref.~[\refcite{Xu:2011fi}], the default $\mathtt{HIJING}$ parameters 
($a=0.5$ and $b=0.9$ GeV$^{-2}$) were used for the Lund symmetric
splitting function.  The same values of the strong coupling constant
and parton cross sections were used as in Ref.~[\refcite{Xu:2011fi}]. 
In these simulations, minimus-bias $p+$Pb events were calculated
with no restrictions on the impact parameter and with the proton beam
moving toward positive rapidity. The $p+$Pb centrality for the flow analysis 
here was defined according to the number of charged hadrons
within $|\eta|<1$ in the laboratory frame.  
Table~\ref{table1} shows the relevant conditions for several 
$p+$Pb centrality classes in the laboratory frame 
including the average, minimum and maximum values of impact parameter;
the total number of participant nucleons in the lead nucleus,
$N_{\rm part}^{\rm Pb}$; the number of participant nucleons in the Pb nucleus
that undergo inelastic scattering,
$N_{\rm part-in}^{\rm Pb}$; and the average number of charged particles
within $|\eta|<1$ calculated with $\mathtt{AMPT-SM}$.

\begin{table}
\caption{Characteristics of several $p+$Pb centrality classes in 
$\mathtt{AMPT-SM}$ where centrality is determined from the number of
charged hadrons within $|\eta|<1$ in the laboratory frame.}
\begin{tabular}{cccccccc}
\hline
Centrality & $\left <b \right >$ (fm) & $b_{\rm min}$ (fm) & 
$b_{\rm max}$ (fm) & $N_{\rm part}^{\rm Pb}$  & $N_{\rm part-in}^{\rm Pb}$ &
$\langle N_{\rm ch}(|\eta|<1) \rangle$ \\
\hline
MB  & 5.84 & 0.0 & 13.2 & 7.51 & 5.37 & 36.8 \\
0-5\% & 3.48 & 0.0 & 8.8 & 15.87 & 12.26 & 102.5 \\
5-10\% & 3.74 & 0.0 & 8.9 & 14.28 & 10.80 & 81.6 \\
10-20\% & 3.97 & 0.0 & 9.8 & 13.00 & 9.64 & 68.1 \\
\hline
\end{tabular}\\[2pt]
\label{table1} 
\end{table}

There are CMS data available on $v_n \{2,|\Delta \eta| > 2\}(p_T)$ 
\cite{Li:2014zza,Chatrchyan:2013nka}.
Their analysis method, with 
$v_n \{2,|\Delta \eta|>2\}(p_T) = v_{n \Delta}(p_T,p_T^{\rm ref})/
\sqrt {v_{n \Delta} (p_T^{\rm ref},p_T^{\rm ref}})$, is used. 
Here $v_{n \Delta}(p_T,p_T^{\rm ref})$ is calculated 
as $\langle \langle \cos(n \Delta \phi) \rangle \rangle $ 
\cite{CMS:2013bza},
where $\langle \langle ... \rangle \rangle $ 
denotes averaging over different charged hadron pairs in each 
event and then averaging over those events. 
The two particles in each pair need to both be
within $|\eta|<2.4$ and have a minimum separation 
$|\Delta \eta|$ of 2. In addition, the reference particle must be within 
$0.3 < p_T^{\rm ref} < 3.0$ GeV. 

\begin{figure}[htb]
\centerline{\includegraphics[height=3in]{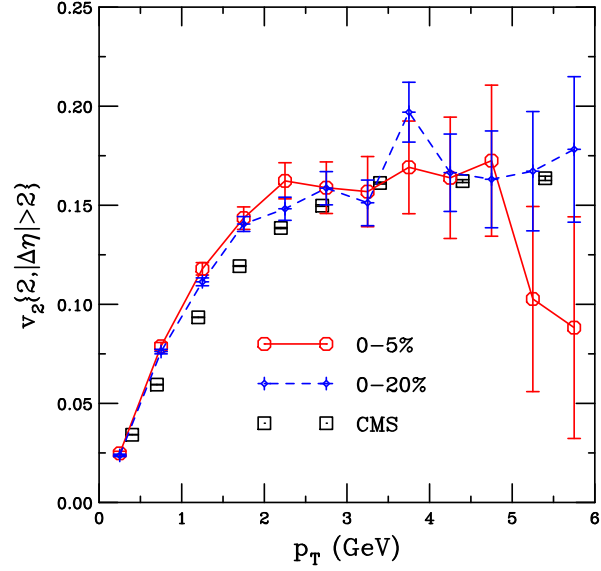}}
\caption{(Color online)
The $v_2\{2,|\Delta \eta|>2\}(p_T)$ calculated
for $p+$Pb collisions with $\mathtt{AMPT-SM}$.  
The data are from Ref.~[\protect\refcite{Chatrchyan:2013nka}].
The uncertainties shown on the data are statistical only.} 
\label{figv2}
\end{figure}

\begin{figure}[htb]
\centerline{\includegraphics[height=3in]{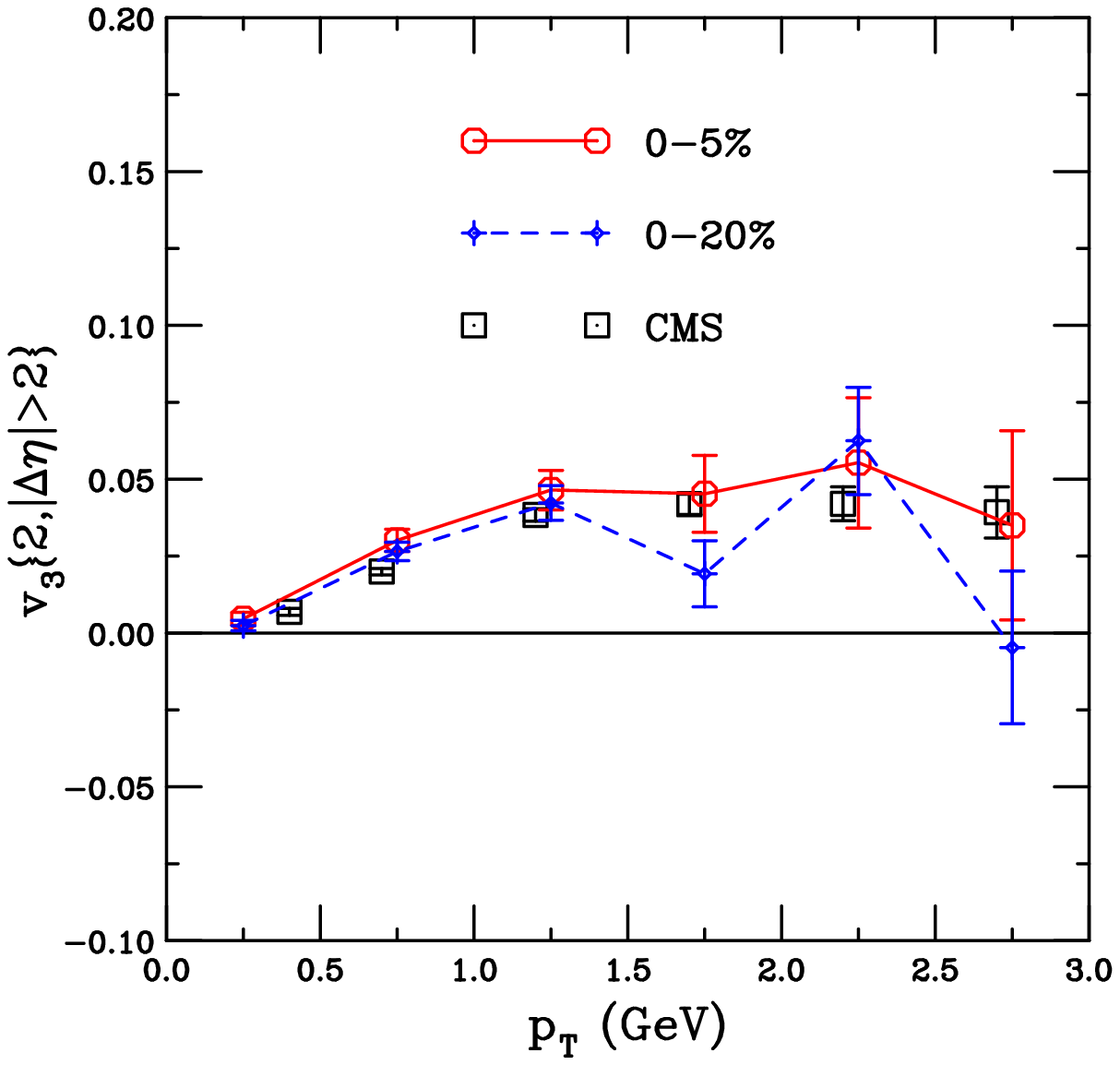}}
\caption{(Color online)
The calculated $v_3\{2,|\Delta \eta|>2\}(p_T)$ 
for $p+$Pb collisions with $\mathtt{AMPT-SM}$.
The data are from Ref.~[\protect\refcite{Chatrchyan:2013nka}].  
The uncertainties shown on the data are statistical only.} 
\label{figv3}
\end{figure}

\begin{figure}[htb]
\centerline{\includegraphics[height=3in]{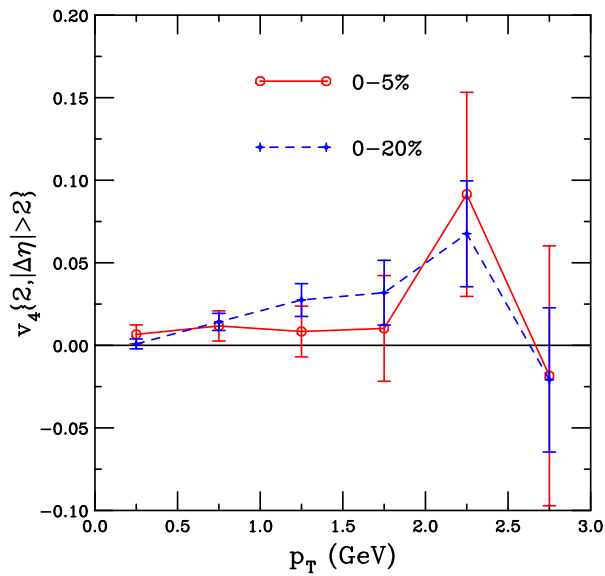}}
\caption{(Color online)
The value of $v_4\{2,|\Delta \eta|>2\}(p_T)$ calculated with $\mathtt{AMPT-SM}$
for $p+$Pb collisions.}  
\label{figv4}
\end{figure}

Figures \ref{figv2}-\ref{figv4} show the anisotropy harmonics $v_n \{2,|\Delta
\eta|>2\}(p_T)$ for $n=2,3,4$ calculated with the two-particle
correlation method just described.  The solid and dashed curves represent the
$\mathtt{AMPT-SM}$ top 5\% and top 20\%-central results, respectively. 
The value of $v_2\{2,|\Delta \eta|>2\}(p_T)$ for the top 5\% centrality 
is close to, but slightly higher than that for the top 20\% centrality. 
The same is observed for $v_3\{2,|\Delta \eta|>2\}(p_T)$. 
On the other hand, $v_4\{2,|\Delta \eta|>2\}(p_T)$ for the top 20\% centrality
seems to be higher than that for the top 5\% centrality, 
although there are large statistical uncertainties on the $v_4$ results. 
Note also that the magnitude of $v_2\{2,|\Delta \eta|>2\}(p_T)$ 
is generally much higher than those of $v_3\{2,|\Delta \eta|>2\}(p_T)$ and
$v_4\{2,|\Delta \eta|>2\}(p_T)$ at the same $p_T$. 

In Figs.~\ref{figv2} and \ref{figv3}, the CMS $p+$Pb data on 
$v_2\{2,|\Delta \eta|>2\}(p_T)$ and $v_3\{2,|\Delta \eta|>2\}(p_T)$ are also 
shown.  The data are
given for the CMS centrality cut $120 < N_{\rm trk} < 150$  
Ref.~[\refcite{Chatrchyan:2013nka}].  The $\mathtt{AMPT}$ results
in these figures employ the same centrality 
definition as in the original $p+$Pb predictions paper
Ref.~[\refcite{Albacete:2013ei}].  
However, this definition is not identical to that of CMS in 
Ref.~[\refcite{Chatrchyan:2013nka}].  
Note that  $120 < N_{\rm trk} < 150$ roughly 
corresponds to $0.5-2.5$\% centrality while the $\mathtt{AMPT}$ 
results are for $0-5$\% and $0-20$\% centrality.   Thus, the comparison to 
data here is inexact.

\section{Jets}


\subsection[Minimum-bias dijets in $p+$Pb collisions]{Minimum-bias dijets in $p+$Pb collisions (K. J. Eskola, I. Helenius and H. Paukkunen)}
\label{Kari-jets}

The jet production data from the first $p+$Pb run 
\cite{Chatrchyan:2014hqa,CMS:2014qca,ATLAS:2014cpa} have proven to provide a 
precise test of the nuclear parton distribution
functions (nPDFs). Here, the normalized 
distribution of dijets measured by the CMS collaboration 
\cite{Chatrchyan:2014hqa}, the first publicly available jet data from the LHC 
$p+$Pb run, are discussed. In this measurement, the jets were binned in dijet 
pseudorapidity, $\eta_{\rm dijet}$, defined as the average pseudorapidity of the 
two hardest (largest $p_T$) jets in the event, 
$\eta_{\rm dijet} \equiv 0.5 (\eta_{\rm leading} + \eta_{\rm subleading})$ (in the 
laboratory frame). The momentum fractions probed on the nucleus side is 
approximately $x_2 \approx (2p_{T}^{\rm leading}/\sqrt{s})e^{-\eta_{\rm dijet}+0.465}$ 
[\refcite{Eskola:2013aya}], such that the $\eta_{\rm dijet}$ dependence rather 
straightforwardly tracks the $x$ dependence of the nuclear PDFs. 

\begin{figure}[htb!]
\center
\includegraphics[scale=0.5]{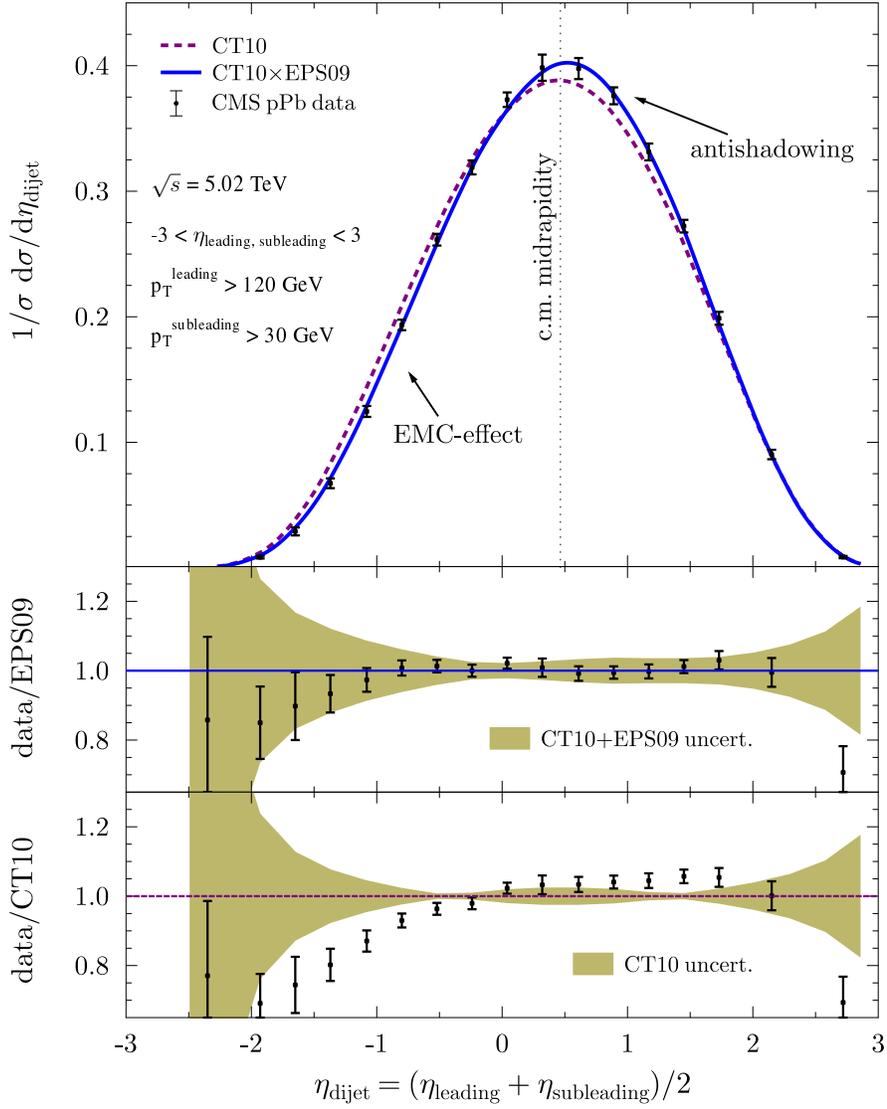}
\caption{(Color online)
The CMS dijet measurements \protect\cite{Chatrchyan:2014hqa} are 
compared to NLO theory calculations \protect\cite{Eskola:2013aya}. 
The predictions employ the CT10 free proton PDFs \protect\cite{Lai:2010vv} 
(purple dashed curve) and the CT10 PDFs modified by the EPS09 NLO 
\protect\cite{Eskola:2009uj} nuclear modifications (blue curve). The upper 
panel shows the normalized cross section as a function of $\eta_{\rm dijet}$.
The lower two panels display the ratio of the data to the CT10+EPS09 and 
CT10 calculations respectively, including the PDF and nPDF uncertainty bands.
The data are given for $p_{T,1} > 120$ GeV, $p_{T,2} > 30$ GeV, and azimuthal
separation $\Delta \phi_{1,2} > 2\pi/3$.  The pseudorapidity interval is chosen
so that there is a gap in psuedorapidity between the dijets and the forward
transverse energy deposited in the CMS HF detectors at $4 < |\eta| < 5.2$.
See Ref.~[\protect\refcite{Chatrchyan:2014hqa}] for more experimental details.
}
\label{fig1}
\end{figure}

The data are 
contrasted with NLO predictions \cite{Eskola:2013aya} 
in Fig.~\ref{fig1}. The plot shows the expectations using both the CT10 free 
proton PDFs \cite{Lai:2010vv} (dashed purple curve) and nPDFs constructed from 
the CT10 free proton PDFs with the EPS09 NLO nuclear modifications 
\cite{Eskola:2009uj} (blue curve). The data clearly favor the EPS09 nuclear 
PDFs and, in practice, rules out the predictions using only free proton PDFs. 
In comparison to the predictions with free proton PDFs, the data show an 
enhancement at $\eta_{\rm dijet} \gtrsim 0$, and a depletion at 
$\eta_{\rm dijet} \lesssim 0$. These are explained by EPS09 in terms of gluon 
antishadowing and the EMC effect at large $x$.  These effects were, in turn, 
both predicted based on inclusive pion production measured by the PHENIX 
Collaboration at RHIC \cite{Adler:2006wg}. 

In the calculations shown in Figure~\ref{fig1} the renormalization scale 
$\mu_R$ and factorization scale $\mu_F$ were both fixed as 
$\mu_F=\mu_R= 0.5 p_T^{\rm leading}$.  The normalized spectrum considered here is 
quite stable against the choice of scale in the central region, 
$-1 \lesssim \eta_{\rm dijet} \lesssim 2$.  Indeed, the variation is less than 
the CT10 uncertainty \cite{Eskola:2013aya}. Thus, the use of nuclear PDFs is 
essential in order to properly describe the data.

\subsection[Single inclusive jet production]{Single inclusive jet production (Z.-B. Kang, I. Vitev and H. Xing)}

The measurements of the centrality and rapidity dependence of single 
inclusive jet production in $p+$Pb collisions show important nuclear 
modifications of the production cross section.  In this section, 
based on Ref.~[\refcite{Kang:2015mta}], these measurements are
studied in the framework of standard cold nuclear matter effects with an
emphasis on initial-state cold matter energy loss.  

The final-state energy loss, $\Delta E$, can be generalized to initial-state 
energy loss.  Using the differential distribution of radiated gluons 
$dN^g/d\omega$, the probability density $P_{q,g}(\epsilon)$ for quarks and 
gluons to lose a fraction $\epsilon=\sum_i \Delta E_i/E$ of their energy due 
to multiple gluon emission, can be calculated in the Poisson approximation.
The mean energy loss fraction is
\bea
\langle\epsilon_{q,g} \rangle=\left\langle\frac{\Delta E_{q,g~ \rm initial-state}}{E} 
\right\rangle = \int_0^1 d\epsilon \,\epsilon\, P_{q,g}(\epsilon), 
\quad { \rm where} \quad \int_0^1d\epsilon \, P_{q,g}(\epsilon)=1 \; .
\label{eq:fraction}
\eea
Note that the subscripts $q$ and $g$ in Eq.~(\ref{eq:fraction}) indicate that 
quarks and gluons radiate different numbers of gluons and thus lose a 
different fraction of their energy.

If the incident parton $a$ loses a fractional energy $\epsilon$, it must 
have originally carried a larger momentum fraction $x_a$ to satisfy the
final-state kinematics.  The energy loss can be included by a modification
of the parton densities in the calculation of the cross section, see
Ref.~[\refcite{Kang:2012kc}].
Since accounting for the fluctuations in the cold nuclear matter energy loss 
by directly calculating $P_{q,g}(\epsilon)$ can be computationally demanding, the
effect is implemented as a shift of momentum fraction in the PDFs,
\bea
f_{q/p}(x_a,\mu) \rightarrow 
f_{q/p}\left(\frac{x_a}{1-\epsilon_{q,\rm eff}},\mu\right),
\qquad
f_{g/p}(x_a,\mu) \rightarrow 
f_{g/p}\left(\frac{x_a}{1-\epsilon_{g,\rm eff}},\mu\right),
\label{eq-mPDFs}
\eea
where $\epsilon_{q,g,\rm eff} = 0.7\,\langle\epsilon_{q,g}\rangle$ with  
$\langle\epsilon_{q,g}\rangle$ given by Eq.~(\ref{eq:fraction})
\cite{Sharma:2009hn,Kang:2012kc}. 
Thus the nuclear modification of single inclusive jet production in $p+$Pb 
collisions depends not only on the magnitude of initial-state cold nuclear 
matter energy loss, but also on the slope of the parton distribution 
functions. In particular, large suppression can be expected for jet production 
at forward rapidity and large $p_T$ where the proton parton momentum fraction 
$x_a$ is large and $f_{q,g/p}(x_a,\mu)$ is steeply falling. 

Since the energy loss calculation is at leading order, the CTEQ6L1 parton 
distribution functions~\cite{Pumplin:2002vw} are used in both $p+p$ and
$p+$Pb collisions.  The factorization and renormalization scales are equal and
fixed to be $\mu=p_T$.  The gluon mean-free path is taken to be 
$\lambda_g \sim 1$~fm and the
interaction strength between the propagating jet and the QCD medium is varied
by changing the typical momentum transfers, $\xi$, over the range 
$0.175 < \xi < 0.7$~GeV, extracted from comparisons to RHIC data. 

\begin{figure}[htb]
\centerline{\includegraphics[angle = 270,width=0.995\textwidth]{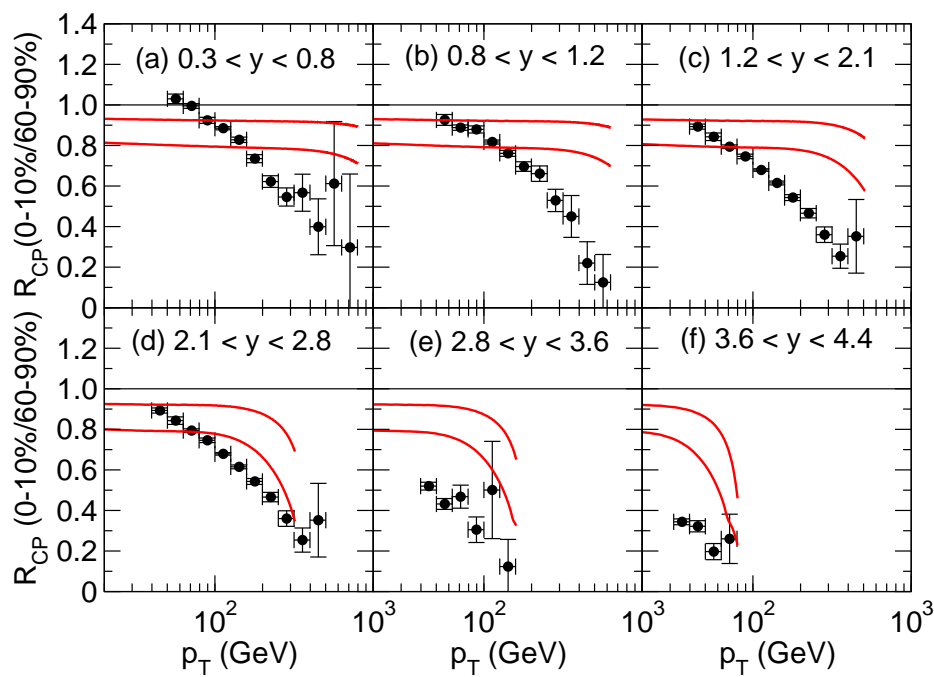}}
\caption[]{(Color online)
Comparison of the calculated $R_{CP}$ with the ATLAS data 
\protect\cite{ATLAS:2014cpa} (points) as a function of $p_T$.  
The upper and lower limits
of the calculated results are given as red solid curves for (a) $0.3 < y < 0.8$,
(b) $0.8 < y < 1.2$, (c) $1.2 < y < 2.1$, (d) $2.1 < y < 2.8$, 
(e) $2.8 < y < 3.6$, and (f) $3.6 < y < 4.4$.} 
\label{fig:Ivan1}
\end{figure}

The $R_{p{\rm Pb}}$ data were presented in four centrality bins: 0-10\%, 20-30\%,
40-60\% and 60-90\% over a range of rapidities from forward to backward.  
At forward rapidities, $R_{p{\rm Pb}}$ decreases with increasing 
$p_T$ for the most central impact parameters while it increases with $p_T$ for 
peripheral collisions.  The value of $p_T$ at which the ratio deviates from
unity increases as the rapidity range moves from forward to backward rapidity.
In the mid-central collisions, the ratio is consistent with unity at all $p_T$.
At backward rapidities, the ratio is also equivaent to unity at all $p_T$.
Note also that $R_{p{\rm Pb}}$ extends to higher $p_T$ at more backward 
rapidities.

The values of the central-to-perpherial ratio, $R_{CP}$, defined as
\begin{eqnarray}
R_{CP} = \frac{\langle N_{\rm coll}^{\rm per}\rangle}{\langle N_{\rm coll}^{\rm cent}
\rangle} \frac{d\sigma_{\rm jet}^{p{\rm Pb}}/dy d^2p_T|_{\rm cent}}{d\sigma_{\rm jet}^{pp}
/dy d^2p_T|_{\rm per}} \, \, , \label{def:rcp}
\end{eqnarray}
where $N_{\rm coll}$ is the number of binary nucleon-nucleon interactions,
are also calculated
as a function of $p_T$ in the same centrality regions, with $R_{CP}$ calculated
for the 0-10\%, 20-30\%, and 40-60\% most central collisions relative to the 
60-90\% centrality bin.  The ratio of the most central to most peripheral
impact parameters, 0-10\%/60-90\% shows the strongest suppresssion with $p_T$
while the suppression is the weakest for the semi-central to most peripheral
collisions, 40-60\%/60-90\%.  At negative rapidities there is even an 
enhancement for $p_T < 100$ GeV.  The greatest separation between the
most central, mid-central and semi-central to peripheral collisions is at
the most forward rapidities.  

When the $R_{CP}$ results for the most central relative to the most peripheral
collisions are plotted as a function of the jet energy,
$p_T \cosh y$, for rapidities greater than $-0.3 < y < 0.3$ approximately
scale with $p_T \cosh y$.  At more backward rapidities, there is no
scaling and the data ratios are clearly separated.  
In addition, for $0.8 < y$, $R_{CP}$ is always less than unity while there is 
an increasing enhancement for the lowest values of $p_T \cosh y$ at 
backward rapidity.

\begin{figure}[htb]
\begin{center}
\includegraphics[angle=270,width=0.495\textwidth]{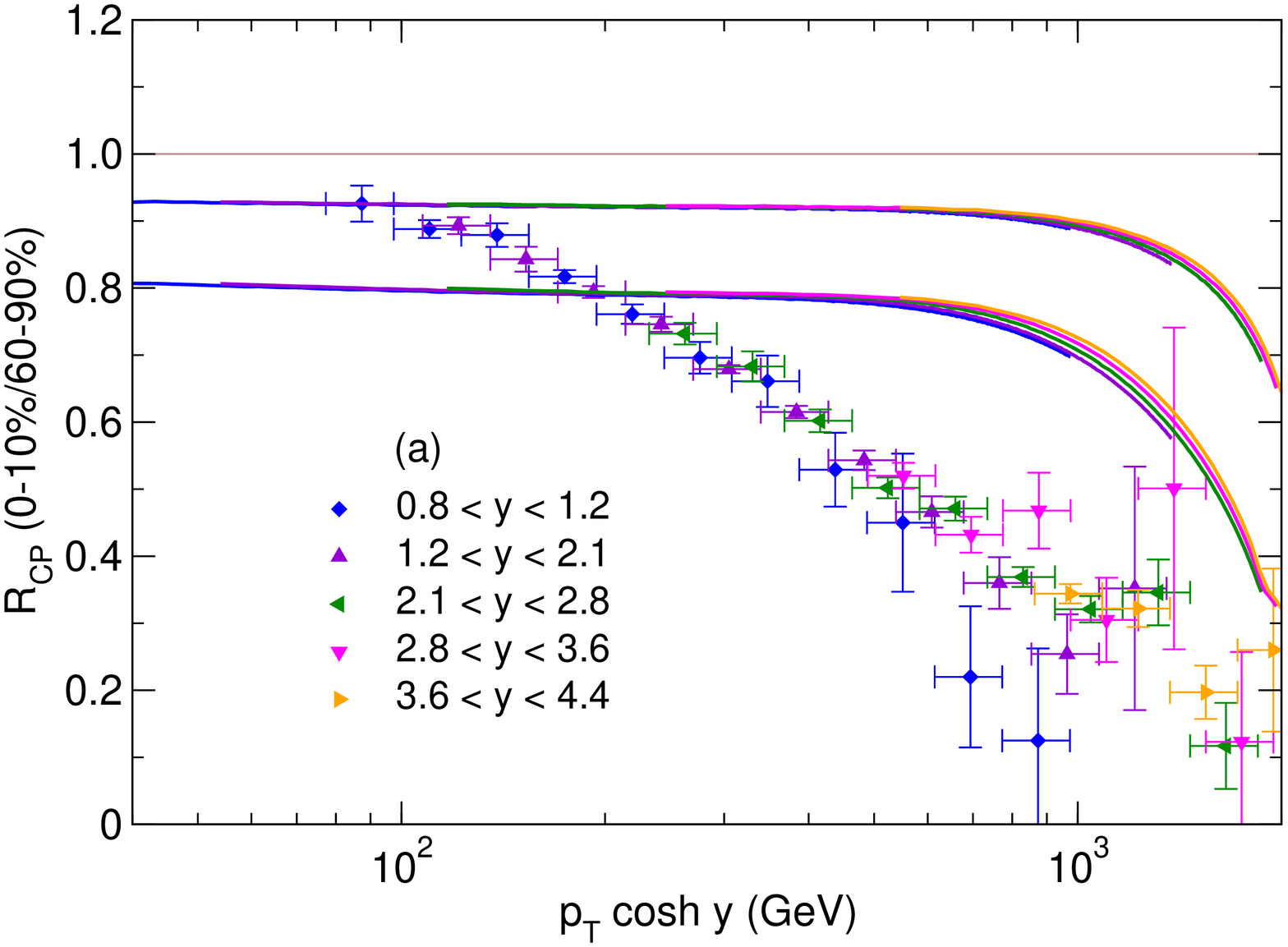}
\includegraphics[angle=270,width=0.495\textwidth]{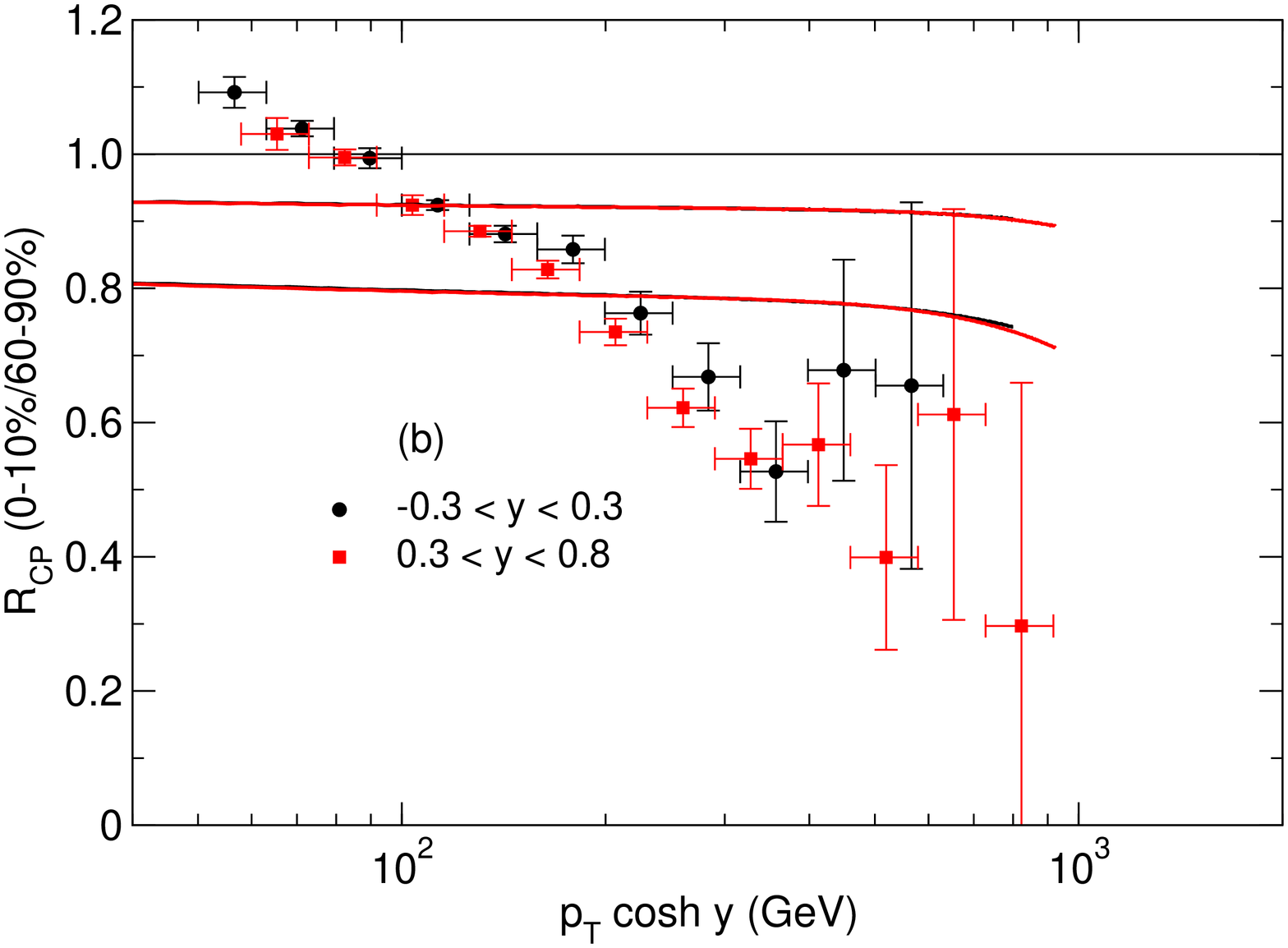}
\end{center}
\caption{(Color online)
Comparison of the calculated $R_{CP}$ with the ATLAS 
data \protect\cite{ATLAS:2014cpa} as a function of $p_T \cosh y$.
In (a), the results at forward rapidities ($0.8 < y < 1.2$ (blue
diamonds), $1.2 < y < 2.1$ (maroon upward-pointing triangles), $2.1 < y < 2.8$
(green left-pointing triangles), $2.8 < y < 3.6$ (magenta downward-pointing
triangles), and $3.6 < y < 4.4$ (orange right-pointing triangles) are shown.
In (b), results near midrapidity are shown ($-0.3 < y < 0.3$ (black
circles) and $0.3 < y < 0.8$ (red squares)).  The upper and lower limits of the
calculation for each rapidity region overlap each other.} 
\label{fig:Ivan2}
\end{figure}

In Figs.~\ref{fig:Ivan1} and \ref{fig:Ivan2}, the calculations of $R_{CP}$
with cold matter energy loss described here are compared
to the ATLAS data \cite{ATLAS:2014cpa}.  Only the ratio of
the most central to most peripheral results (0-10\%/60-90\%) are shown.
The centrality dependence of the calculations comes from the average number
of binary collisions in a given centrality bin, taken from the ATLAS
determination, as well as the effective path length, $L$, through the medium 
for the produced jets.  The value of $L$, which is required in the calculation
of $dN^g/d\omega$, is calculated in a Glauber model consistent with the value of
$\langle N_{\rm coll} \rangle$ appropriate for the given centrality bin.

Figure~\ref{fig:Ivan1} shows the results as a function of $p_T$.  The upper
and lower edges of the calculated ratio, corresponding to $\xi = 0.175$ and
0.7 GeV respectively, are given.  The calculations show the same trend as the
data albeit with somewhat different curvature.  When the calculations are
compared to the values of $R_{CP}$ for mid-central and semi-central collisions,
the agreement with the data improves.  The calculations and data both decrease
with increasing $p_T$ and $y$.  This picture is consistent with cold matter
energy loss where the effect is strongly dependent on the parton momentum 
fraction in the projectile proton, $x_a$, assuming that the proton moves toward
positive rapidity. 

Indeed, if cold matter energy loss is the dominant mechanism for the effect
observed by ATLAS, one would expect to see scaling with $x_a$ in the forward
rapidity region but not backward of midrapidity.  This is in fact the case,
as shown in Fig.~\ref{fig:Ivan2}.  The calculations and the data both scale
with $x_a \propto p_T \cosh y$ although, again, the curvature is not quite
the same.  

The calculations of $R_{p {\rm Pb}}$ show a similar agreement with the forward 
rapidity data in central collisions.  However, the minimum bias result,
dominated by non-central events, is almost independent of $p_T$ and even
suggests some small enhancement.  

The calculations from Ref.~[\refcite{Kang:2015mta}]
shown here capture the bulk of the observed modifications for 
the LHC experiments from central to semi-central collisions. The upper edge of 
the band calculated for cold matter energy loss is consistent with the minimum 
bias data if the statistical and systematic uncertainties are taken into 
account.  The encouraging comparison between the calculations and the data 
indicates the significance of cold nuclear matter energy loss for 
understanding particle and jet production in $p+A$ collisions, particularly
at forward rapidities.

The observed scaling of $R_{CP}$ and $R_{\rm pPb}$ as a function of the total 
jet energy, $p_T \cosh y  \propto x_a$, in the forward rapidity region 
in $p+$Pb collisions can be explained naturally in the picture of cold nuclear 
matter energy loss. Further, $x_a \approx x_F$, Feynman $x$, at forward 
rapidity.  Similar scaling at large $x_F$ has indeed been observed for 
different final states~\cite{Leitch:1999ea}.  On the other hand, an nPDF-only 
calculation is consistent with the minimum bias data in inclusive jet 
production but fails to describe central collisions. 

The observed enhancement in peripheral collisions is difficult to understand 
in either an energy loss or strictly nPDF picture. Such an enhancement might 
have a different origin, for example from ``centrality bias'', and needs to be 
explored further. This enhancement also affects the central-to-peripheral 
ratio and is thus partly responsible for the small values of $R_{CP}$.
It is important to understand whether there is a centrality selection bias and,
if it exists, its dynamical origin, for it to be taken into account correctly
in theoretical calculations.

A next step will be to go beyond the soft-gluon energy
loss approximation and obtain the full medium-induced splitting 
kernels~\cite{Ovanesyan:2011kn}. With these cold nuclear matter splitting 
kernels, the vacuum and in-medium parton showers can be treated
on the same footing, following the progress recently made on the 
implementation of final-state QGP effects
\cite{Kang:2014xsa,Chien:2015vja,Chien:2015hda}. 


There is no significant contradiction between the cold matter energy loss 
interpretation of single inclusive jet suppression in this section and the
dijet enhancement due to nuclear shadowing in the previous section.  The
results for ATLAS are generally in a much higher $p_T$ range where the effect
of nuclear modifications of the parton densities are reduced by the scale
evolution relative to those of the dijets.  In addition, the centrality
dependence of EPS09s \cite{Helenius:2012wd} may not have a strong
effect on $R_{CP}$. 


\subsection[Nuclear modification ratio for production of forward-forward jets in  $p+$Pb collisions with Sudakov effects included]{Nuclear modification ratio for production of forward-forward jets in $p+$Pb collisions with Sudakov effects included (P. Kotko, K. Kutak and S. Sapeta)}

Here the predictions \cite{Albacete:2013ei,vanHameren:2014lna} for the 
emergence of saturation \cite{Gribov:1984tu} effects on dijet production
are updated.  The prescription for including the hard scale, $\mu$,
dependence in the small~$x$ gluon evolution equations using the Sudakov form 
factor, proposed in Refs.~[\refcite{Kutak:2014wga,vanHameren:2014ala}], 
is applied. (For other approaches, see Ref.~[\refcite{Mueller:2013wwa}]).  
The high energy factorization formalism~\cite{Catani:1990eg}, which accounts 
for both the high energy scale of the scattering and the hard momentum scale 
$p_T$ of the produced hard system, is employed.  
 
The study presented in Refs.~[\refcite{Albacete:2013ei,vanHameren:2014ala}] 
concentrated on central-forward dijet production. The results obtained for 
the forward-central jet configuration in $p+p$ collisions
\cite{vanHameren:2014ala} are discussed and compared to preliminary CMS data 
\cite{CMS:2014oma} since the corresponding $p+$Pb data are not yet available.  
Predictions for the case in which both jets are produced in the forward region,
in the spirit of Ref.~[\refcite{vanHameren:2014lna}], are also shown. 
%
%


The hybrid high energy factorization formula in the asymmetric configuration
is \cite{Deak:2010gk}
\begin{eqnarray}
\lefteqn{  \frac{d\sigma}{dy_1dy_2dp_{T1}dp_{T2}d\Delta\phi} 
  = } \nonumber \\
& &  \sum_{a,c,d} 
  \frac{p_{T1}p_{T2}}{8\pi^2 (x_1x_2 S)^2}
  {\cal M}_{ag^*\to cd} \,
  x_1 f_{a/A}(x_1,\mu^2)\,
  {\cal F}_{g/B}(x_2,k^2_T,\mu)\frac{1}{1+\delta_{cd}}\, \,\, 
  \label{eq:cs-fac}
\end{eqnarray}
where
\be
  k_T^2 = p_{T1}^2 + p_{T2}^2 + 2p_{T1}p_{T2} \cos\Delta\phi\,\, \, ,
\ee
$x_1\simeq 1$, $x_2 \ll 1$ and $\Delta\phi$ is the azimuthal distance between 
the outgoing partons.  The squared matrix element, ${\cal M}_{ag^*\to cd}$, 
includes  $2\to 2$ processes with one off-shell initial state gluon, $g^*$, 
and three on-shell partons $a,c,d$.
The following partonic subprocesses contribute to dijet production: 
$qg^*  \to  qg$, $gg^*  \to q\overline q$, and $gg^*\to gg$~\cite{Deak:2009xt}.
The off-shell gluon in Eq.~(\ref{eq:cs-fac}) is obtained from the
unintegrated, hard-scale dependent gluon density 
${\cal F}_{g/B}(x_2,k_T^2,\mu^2)$~\cite{Kutak:2003bd,Kutak:2004ym,Kutak:2012rf}, 
a function of the gluon momentum fraction $x_2$,
the transverse momentum of the off-shell gluon $k_T$, and hard scale $\mu$,
chosen, for example, to be the average transverse momentum of the two leading 
jets. 
In the case of the on-shell parton, at high momentum fraction $x_1$, the 
collinear density, $f_{a/A}(x_1,\mu^2)$, is employed. 

\subsubsection{Forward-central dijets}

\begin{figure}[t!]
  \begin{center}
    \includegraphics[width=0.495\textwidth]{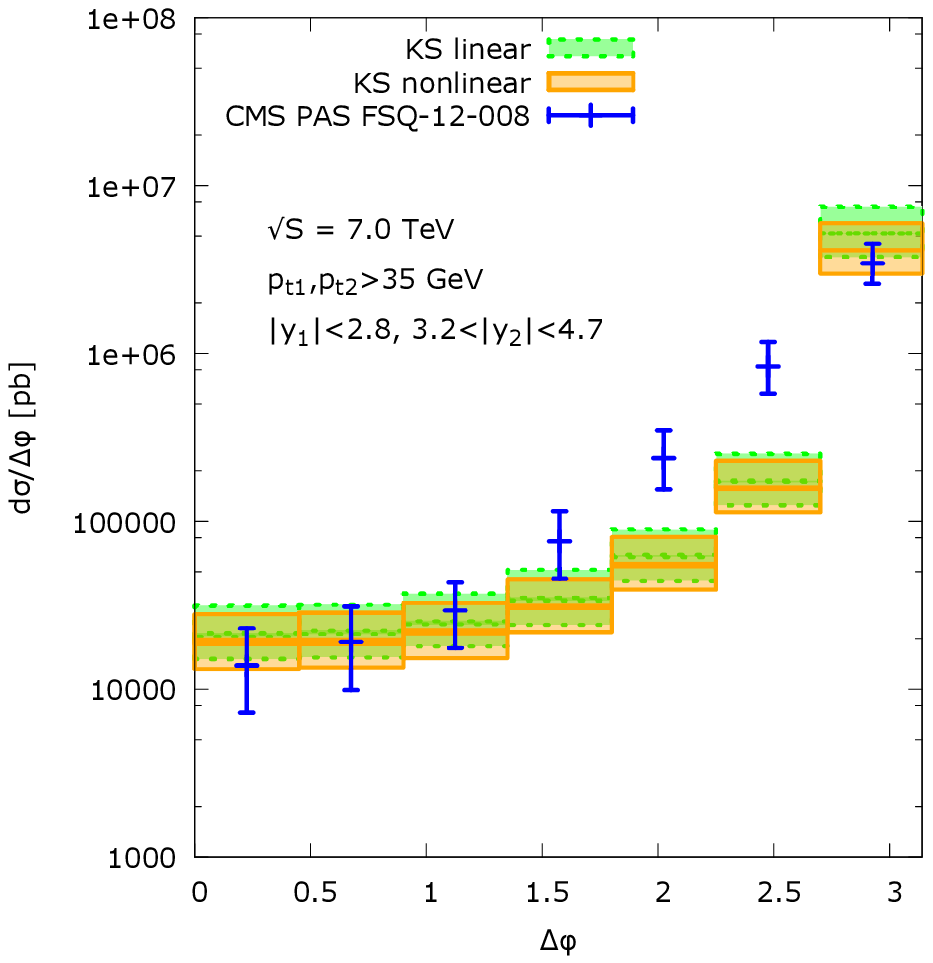}
    \includegraphics[width=0.495\textwidth]{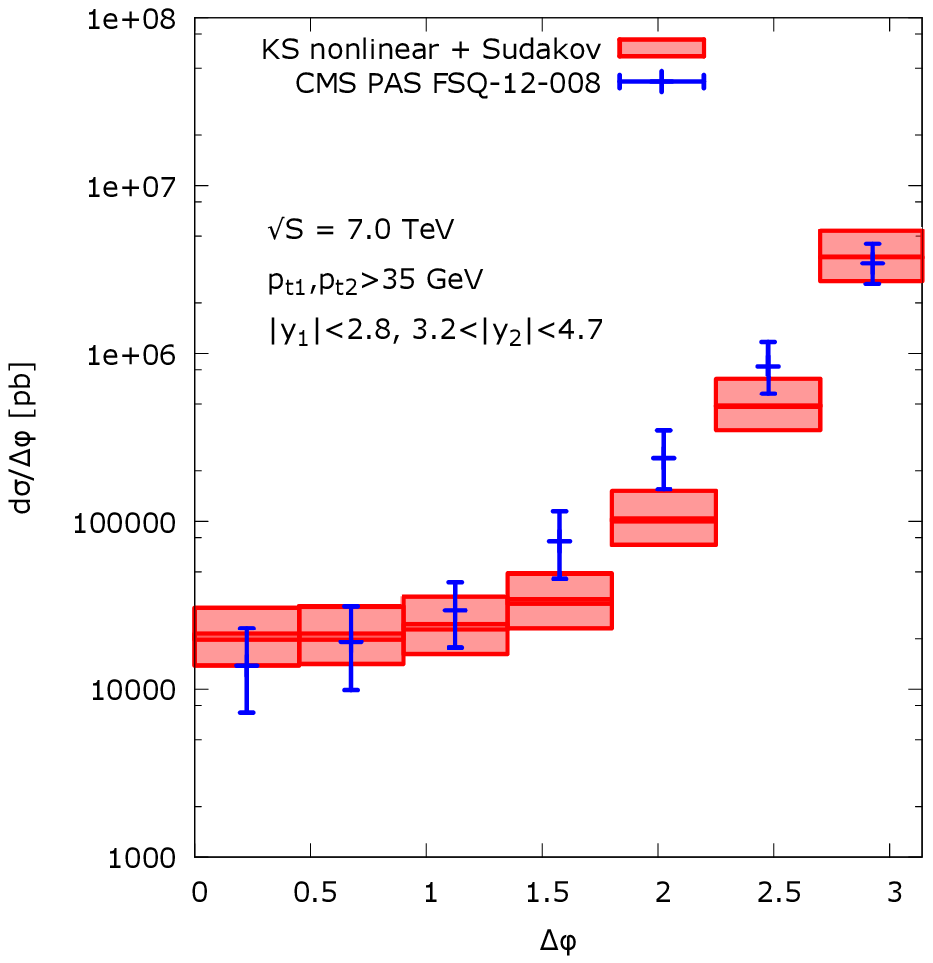}
    \includegraphics[width=0.495\textwidth]{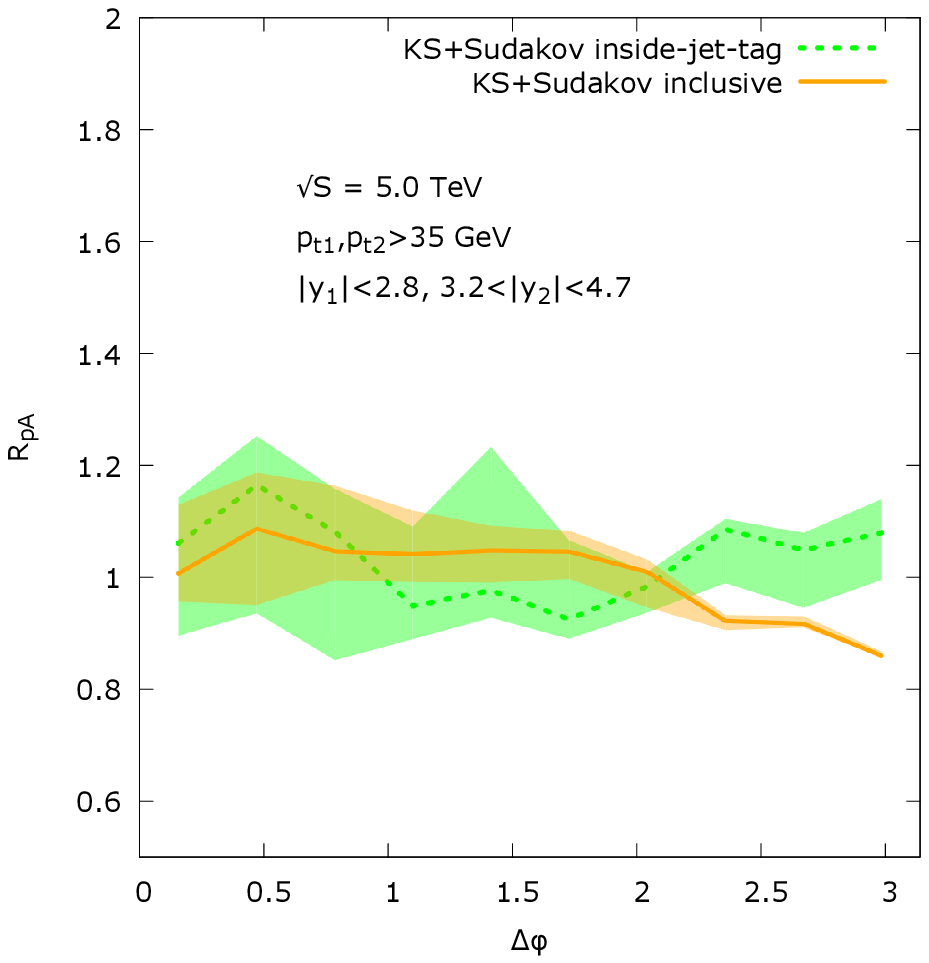}
  \end{center}
  \vspace{-10pt}
  \caption{(Color online)  (Top left) Comparison of the CMS 7 TeV $p+p$ data 
\protect\cite{CMS:2014oma} to predictions in the hybrid high energy 
factorization approach using the KS linear and nonlinear gluon densities. 
(Top right)
The CMS data are compared to the result with the Sudakov-improved KS nonlinear
gluon density.  (Bottom) The prediction 
of the nuclear modification factor for inclusive dijet production as well 
as for the inside-jet tag scenario where there is third jet with 
$p_{T3}>20$ GeV obeying the constraint $y_1>y_3>y_2$.  
From Ref.~[\protect\refcite{vanHameren:2014ala}].
}
\label{fig:decor-p-Pb1}
\end{figure}

Figure~\ref{fig:decor-p-Pb1} shows the results for the azimuthal angle 
decorrelation obtained from the hybrid high-energy factorization formulation
\cite{vanHameren:2014ala} using the KS linear and nonlinear unintegrated 
gluon densities \cite{Kutak:2012rf}. 

The top left part of Fig.~\ref{fig:decor-p-Pb1} shows that, in the range of
available data, there is not much difference between predictions based on 
linear and nonlinear evolution.  Figure~\ref{fig:decor-p-Pb1} (top right) shows
that incorporating the hard scale in the unintegrated gluon density by
including Sudakov effects, the red histograms, 
improves the description of the CMS data \cite{CMS:2014oma}.
The nuclear modification factor, $R_{pA}$, was calculated in two different
scenarios with the CMS cuts: inclusive, with no additional requirement on the 
two leading jets, and the inside-jet tag, with a third jet with $p_T > 20$ GeV
subleading to the dijet intermediate in rapidty.  The results suggest that the
potential saturation signals are rather weak since $R_{pA}$ is consistent with 
unity in both scenarios \cite{vanHameren:2014ala}.
 
\subsubsection{Forward-forward dijets}

In the forward-forward jet configuration the values of $x_2$ are 
approximately an order of magnitude smaller than in the forward-central 
jet configuration. Therefore
this configuration is more sensitive to saturation effects.

\begin{figure}[t!]
  \begin{center}
    \includegraphics[width=0.495\textwidth,angle=-90]{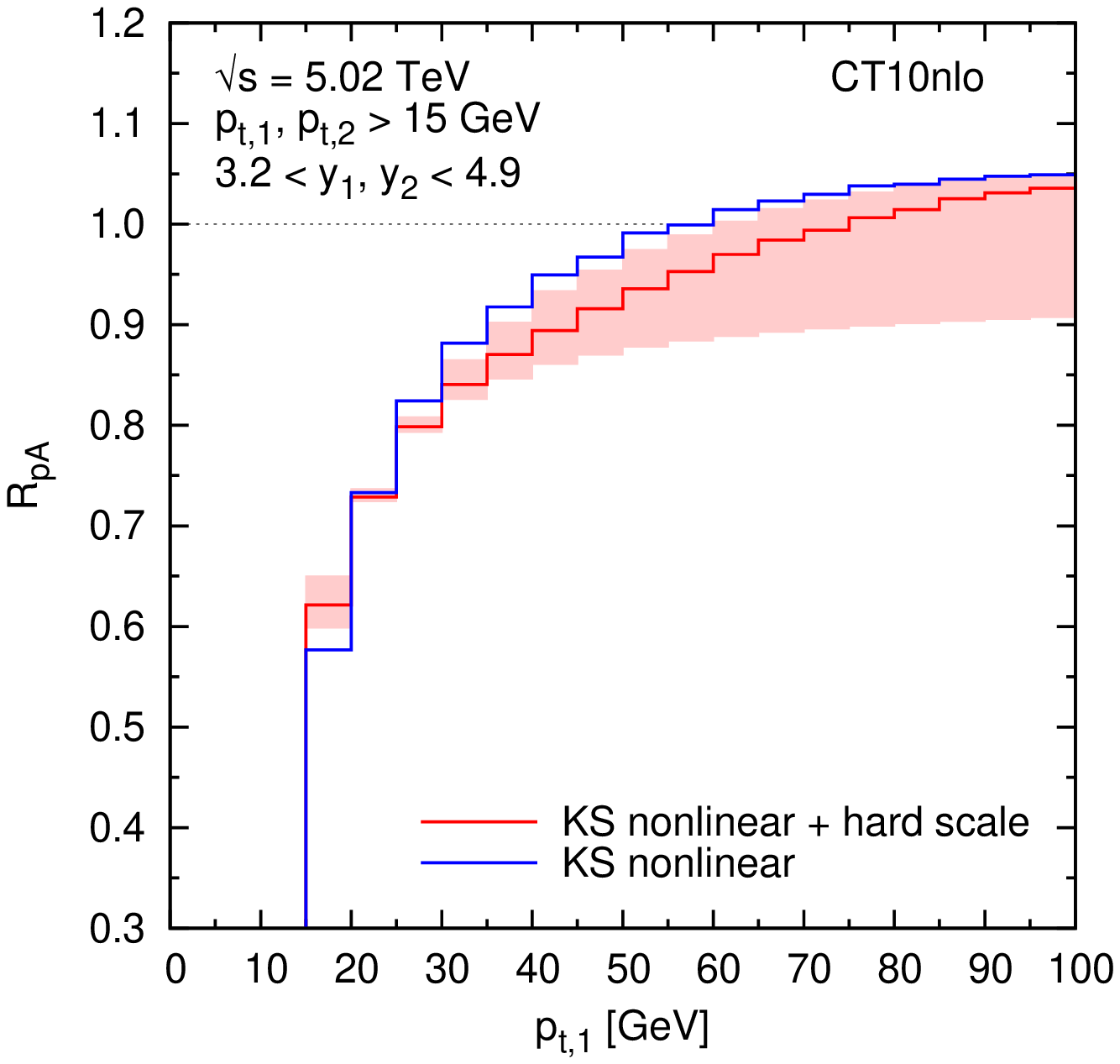}
    \includegraphics[width=0.495\textwidth,angle=-90]{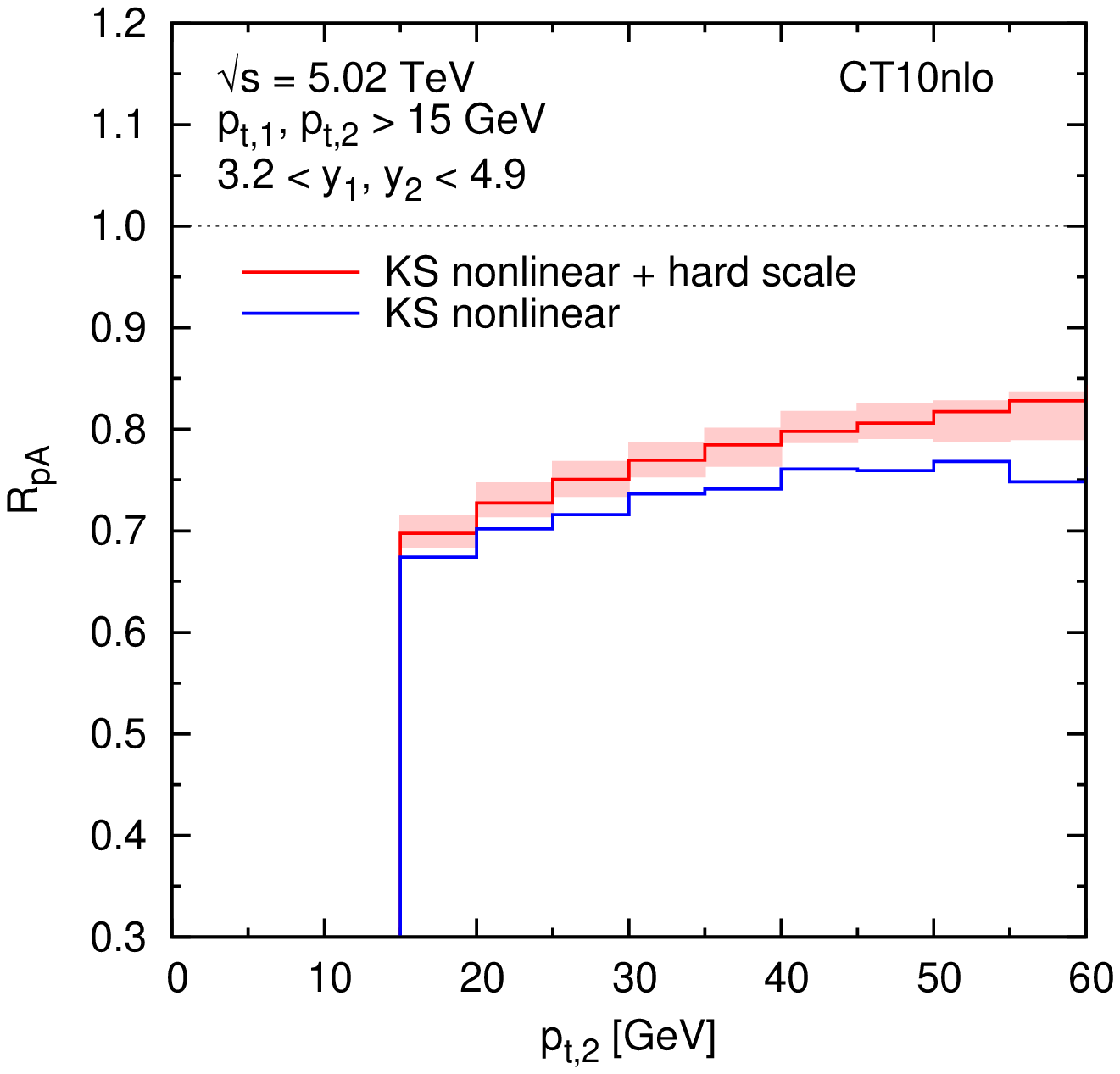}
    \includegraphics[width=0.495\textwidth,angle=-90]{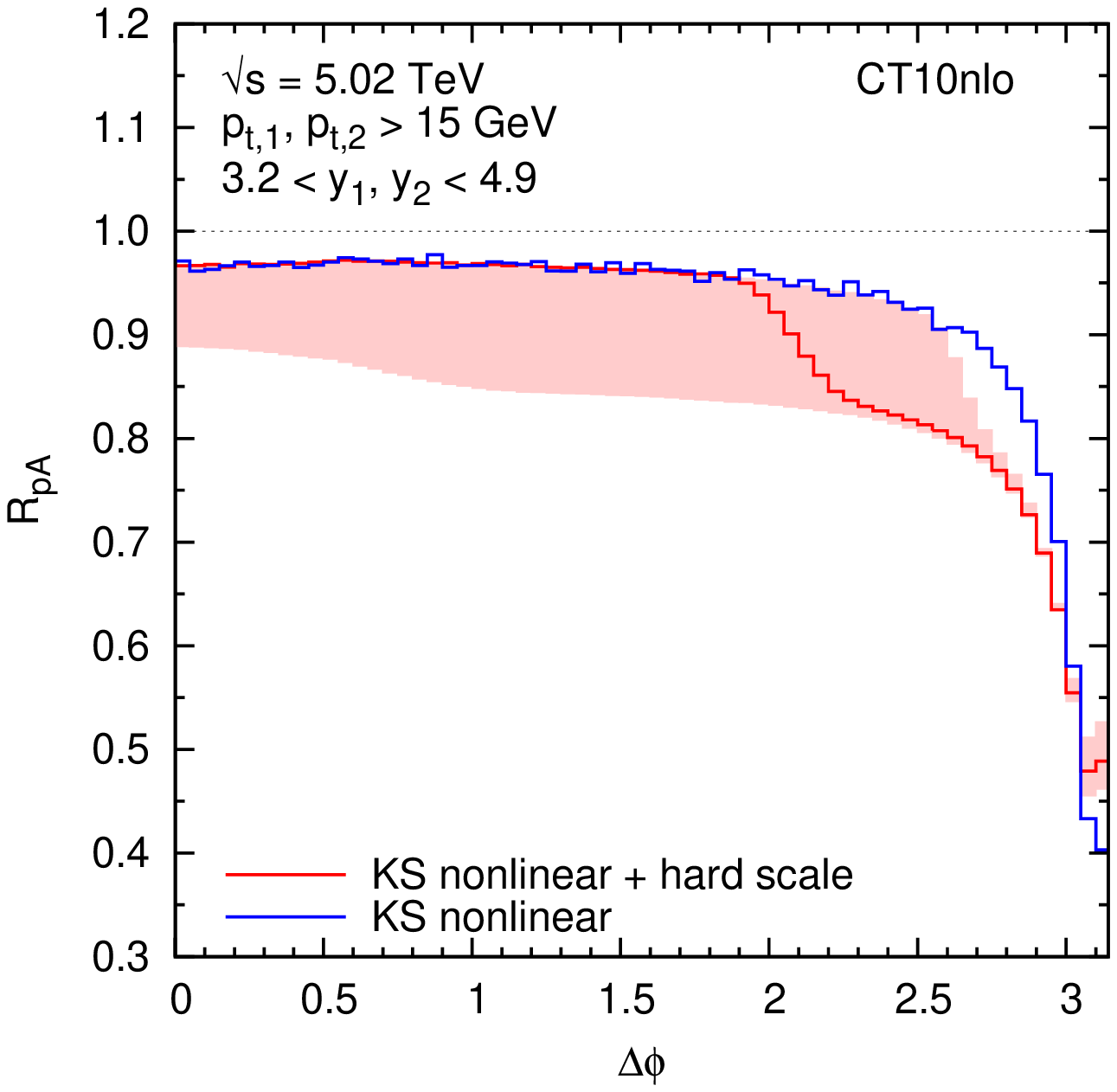}
  \end{center}
  \vspace{-10pt}
  \caption{(Color online) The nuclear modification factor, 
$R_{pA}$, as functions of the
$p_T$ of the harder jet (top left), the $p_T$ of the subleading jet (top
right), and the azimuthal angle between the jets (bottom). 
The blue lines correspond to predictions obtained using the KS gluon density
alone \protect\cite{Kutak:2012rf} while the red lines are predictions including
a gluon density that also depends on the hard scale 
\protect\cite{Kutak:2014wga}.  In both cases, the renormalization and 
factorization scales are $\mu= (p_{T1}+p_{T2})/2$.  The light red bands show 
the effect of varying the scales in the KS nonlinear + hard scale result
from $0.5 \leq \mu \leq 2$.  The analogous variation for the pure KS nonlinear
gluon gives virtually no effect and is therefore not shown.
}
\label{fig:decor-p-Pb}
\end{figure}

In Fig.~\ref{fig:decor-p-Pb}~(top left) the nuclear modification factor 
for forward-forward dijet production is shown as a function of the $p_T$ 
of the hardest jet. 
Figure~\ref{fig:decor-p-Pb}~(top right) shows the corresponding ratio for
the subleading (second hardest) jet.
All results were obtained employing the CT10NLO PDFs~\cite{Lai:2010vv} 
on the side of the projectile (large $x$).
The blue lines correspond to the KS nonlinear gluon 
density~\cite{Kutak:2012rf} while 
the red lines include the Sudakov resummation effects, introducing the hard
scale in the KS nonlinear gluon density~\cite{Kutak:2014wga}.

The motivation to account for the Sudakov effects comes from studies of 
coherence effects which suppress soft gluon emission when the scale of the 
hard process, $\mu$, is larger than the scale $k_T$ of the local gluon density
\cite{Kutak:2014wga}.
The phenomenological significance of these effects has been
demonstrated to improve the
description of decorrelations in forward-central dijet
\cite{vanHameren:2014ala} and $Z+$jet
production \cite{vanHameren:2015uia}.

The results shown in Fig.~\ref{fig:decor-p-Pb} employ a central value of
the hard renormalization and factorization scales of $\mu_0= 0.5(p_{T1}+p_{T2})$.
The hard scale dependence has been investigated by varying $\mu$ between
$0.5\mu_0$ and $2\mu_0$.  The pale red band shows the
result for the KS nonlinear + hard scale gluon incorporating the Sudakov
effects.  The analogous variation for
the pure KS nonlinear gluon gives a negligible effect and is therefore 
not shown.

The bottom panel of Fig.~\ref{fig:decor-p-Pb} gives the predictions for 
azimuthal decorrelations of the forward-forward dijets albeit with a lower
jet $p_T$ cut than in Ref.~[\refcite{Kutak:2014wga}]. The azimuthal
separation, $\Delta\phi \sim \pi$, probes the unintegrated gluon density at
small~$k_T$, where it is strongly suppressed by nonlinear effects.  
As shown in Fig.~\ref{fig:decor-p-Pb}, this observable is a strong signal of
saturation effects and is sensitive to the enhanced saturation going from
a proton to a nuclear target.

The introduction of the hard scale leads to a reduction of the unintegrated
gluon density in lead relative to that of the proton as long as $k_T < \mu$
but, for $k_T \geq \mu$ the hard scale contribution vanishes and $R_{pA}$
transitions to the KS nonlinear result.  The value of $\Delta \phi$ where
this transition takes place depends on the value of $\mu$.  If a lower $\mu$ 
is chosen for the KS nonlinear
+ hard scale calculation, the result would approaches that of the 
KS nonlinear gluon at a higher value of $\Delta \phi$. This
can be expected since lower values of $\mu$ reduce the phase space where the 
Sudakov suppression can have an effect.

\begin{figure}[t!]
  \begin{center}
    \includegraphics[width=0.495\textwidth]{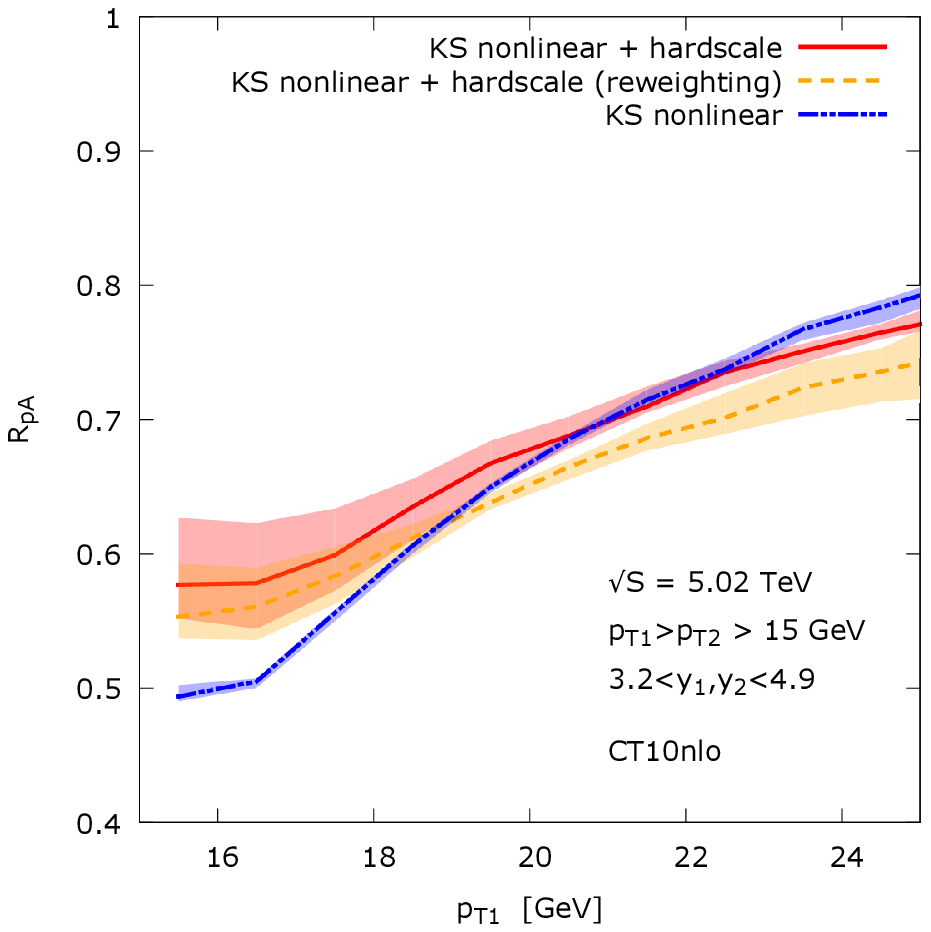}
    \includegraphics[width=0.495\textwidth]{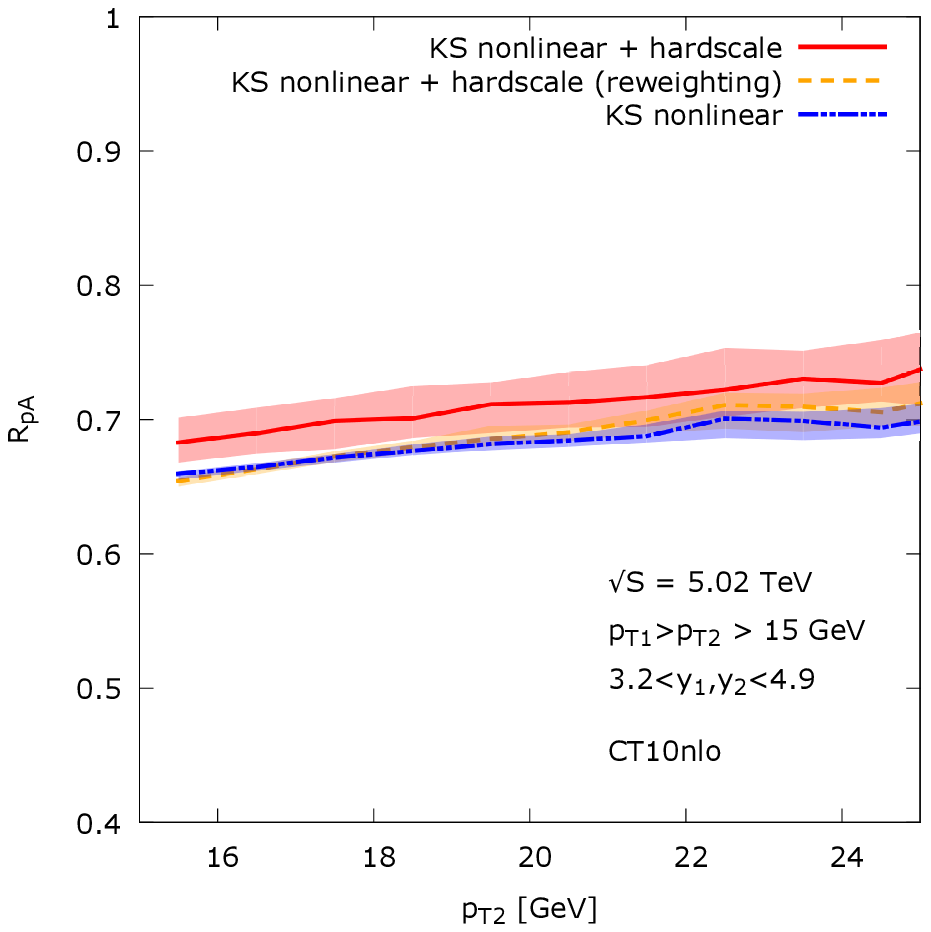}
    \includegraphics[width=0.495\textwidth]{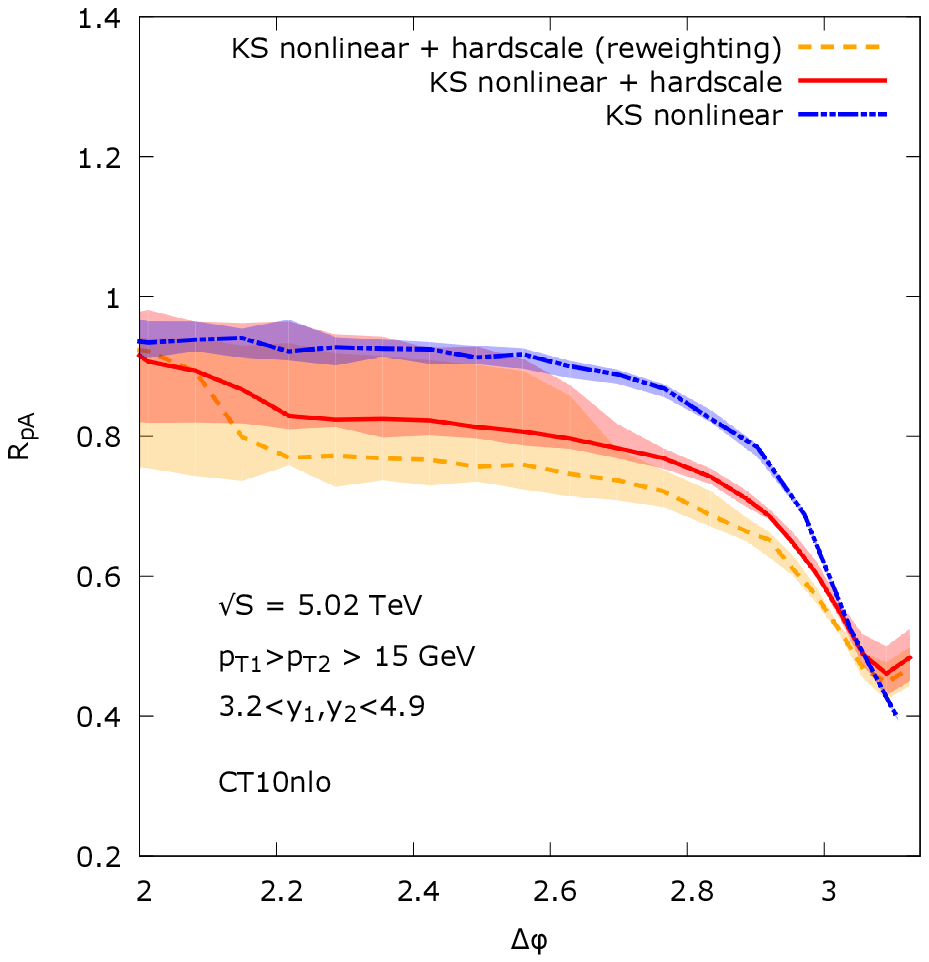}
  \end{center}
    \vspace{-10pt}
\caption{(Color online) Similar to Fig.~\protect\ref{fig:decor-p-Pb} but
expanding the low-$p_T$ ($p_T < 25$ GeV) (top) and high $\Delta \phi$ ($\Delta
\phi > 2$) (bottom) regions.  Calculations using the Sudakov-weighted events 
\protect\cite{vanHameren:2014ala} are given by the orange band. 
The statistical errors on the calculations are included in the 
uncertainty bands.
}
\label{fig:lowpT_Kutak}
\end{figure}

In Fig.~\ref{fig:lowpT_Kutak} the saturation region (low-$p_T$ 
and $\Delta \phi$ near $\pi$) is expanded from Fig.~\ref{fig:decor-p-Pb}.
These results are compared to model predictions 
\cite{Kutak:2014wga} employing the method of reweighting Monte Carlo events 
with $\mu > k_T$ by the Sudakov factor described in
Ref.~[\refcite{vanHameren:2014ala}]. 
Both models give qualitatively similar effects 
for these observables in the low $p_T$ and large $\Delta \phi$ regions.

In addition to the effects included in the calculation, the region
$\Delta\phi \simeq \pi$, for individual processes, is sensitive to corrections
coming from  higher order gluon density correlators, as
discussed in Ref.~[\refcite{Kotko:2015ura}]. Numerical studies of those effects 
are in progress.

The suppression due to the combination of coherence effects and saturation is 
particularly strong for the subleading jet and, in the case of azimuthal
decorrelations, it extends over a significant range of $\Delta\phi$. 

\section{Quarkonium}

\subsection[Ground State Quarkonium]{Ground State Quarkonium (F. Arleo, E. G. Ferreiro, F. Fleuret, H. Fujii, J.-P. Lansberg, A. Rakotozafindrabe, S. Peign\'e and R. Vogt)}

The $J/\psi$ and $\Upsilon$ suppression factors have been measured by ALICE 
\cite{Abelev:2013yxa,Abelev:2014oea} and LHCb \cite{Aaij:2013zxa,Aaij:2014mza} 
in similar rapidity windows, $2.5<y_{\rm cm}<4$ for ALICE and 
$2.5<y_{\rm cm}<5$ for LHCb in
symmetric ($p+p$ and $A+A$) collisions.  Although the muon spectrometers for
both experiments are only on one side of the collision point, results were
obtained forward and backward of midrapidity by switching the beam direction
and running both $p+$Pb and Pb$+p$ collisions.  Due to the rapidity shift in 
asymmetric collisions, the acceptances of the two detectors in the 
collision center of mass was shifted to $2.03< y_{\rm cm} < 3.53$ at
forward rapidity and $-4.46 < y_{\rm cm} < -2.96$ at backward rapidity for
ALICE and $1.5< y_{\rm cm} < 4$ at
forward rapidity and $-5 < y_{\rm cm} < -2.5$ at backward rapidity for
LHCb.  The regions of overlap between the forward and backward rapidity
regions are $2.96 < |y_{\rm cm}| < 3.53$ for ALICE and $2.5 < |y_{\rm cm}| < 4$
for LHCb.  

The results were presented first as $R_{p{\rm Pb}}(y)$ 
with an extrapolated $p+p$ normalization since there
is no $pp$ measurement at $\sqrt{s} = 5$ TeV.  The $p+p$ normalization
is based on an interpolation between the $p+p$ measurements at $\sqrt{s} = 2.76$
and 7 TeV, along with a model-based systematic uncertainty 
\cite{Abelev:2013yxa}.
In addition, to eliminate the dependence on the uncertain $p+p$ 
normalization, a forward-backward production ratio,
$R_{FB}(y,\sqrt{s_{NN}}) = R_{p{\rm Pb}}(+|y|,
\sqrt{s_{_{NN}}})/R_{p{\rm Pb}}(-|y|,\sqrt{s_{_{NN}}})$,
was extracted
where $R_{FB}$ is defined with the proton beam moving toward positive $y$
in the numerator and negative $y$ in the denominator.  Thus cold matter effects
dominant at small $x$ are in the numerator while the denominator probes larger
$x$.  The $p+p$ contributions to $R_{p{\rm Pb}}$ cancel in the ratio because $p+p$
collisions are symmetric around midrapidity.  

In addition to the $J/\psi$ calculations in the color evaporation model (CEM)
at next-to-leading order presented in Ref.~[\refcite{Albacete:2013ei}], 
several other calculations are also shown in 
Figs.~\ref{fig:psiy}-\ref{fig:upsy}.  They are a LO color singlet model
calculation \cite{Ferreiro:2013pua}; a coherent energy loss calculation 
with no nuclear modifications of the parton 
densities \cite{Arleo:2012hn,Arleo:2012rs}; 
and a color glass condensate calculation employing a
color evaporation model for production \cite{Fujii:2013gxa}.
They are briefly discussed here.

The EPS09 NLO CEM band is obtained by calculating the deviations
from the central value for the 15 parameter variations on either side of the 
central set and adding them in quadrature.  
The calculation was done employing the charm production
parameters obtained in Ref.~[\refcite{Nelson:2012bc}]. 
The results shown here and in Ref.~[\refcite{Vogt:2015uba}] 
correct the results in Ref.~[\refcite{Albacete:2013ei}] 
which used incorrect scale inputs.
The EPS09 NLO band is narrower and exhibits less shadowing 
than the corresponding EPS09 LO CEM result, see Ref.~[\refcite{Vogt:2015uba}] 
for more details and a full comparison to all the data.

The EPS09 LO curves \cite{Ferreiro:2013pua} are obtained with the Monte Carlo 
Glauber calculation code JIN \cite{Ferreiro:2008qj} 
that calculates cold nuclear matter 
effects in the exact kinematics of a specific partonic process. In this case, 
generic $2\rightarrow 2$ matrix elements are used. These matrix elements are 
systematically compared to $p+p$ data after convolution with proton PDFs to 
verify that they yield to the correct phase-space 
weighting\footnote{In the $\Upsilon$ case, the LO CSM partonic 
matrix element is used.  However, the physical context is unimportant for 
evaluation of the matrix element since only the kinematics can affect the 
result. In particular, the color state of the $Q \overline Q$ pair is not 
taken into account.}.  The factorization scale employed in the nPDFs when 
calculating the nuclear modification factor was taken to be the transverse 
mass of the observed quarkonium in each event.  To simplify the comparison, 
the central EPS09 LO set is used along with four specific extrema 
(minimal/maximal shadowing and minimal/maximal EMC effect) that reproduce 
the envelope of the gluon nPDF uncertainty encoded in EPS09 LO.

Note that the EPS09 LO CEM ($2 \rightarrow 1$ partonic process)
result in Ref.~[\refcite{Vogt:2015uba}]
is similar to that of the EPS09 LO result described in 
Ref.~[\refcite{Ferreiro:2013pua}] for the generic $2 \rightarrow 2$ matrix
element.  The differences between the two results shown on the
right-hand side of Fig.~\ref{fig:psiy} are due to the production model
are smaller than those due to the mass and scale
parameters employed in the two calculations.

In the coherent energy loss model~\cite{Arleo:2012hn,Arleo:2012rs}, 
the differential $p+A$ production cross section as a function of the 
quarkonium (labelled $\psi$) energy is
\be
\label{eq:xspA0-energy}
\frac{1}{A}\frac{\dd\sigma_{pA}^{\psi}}{\dd \Ea} \left( \Ea \right)  = 
\int_0^{\epsamax} \dd \epsa \,{\cal P}(\epsa, \Ea, \ell_{_A}^2) \, 
\frac{\dd\sigma_{pp}^{\psi}}{\dd \Ea} \left( \Ea+\epsa \right) \, ,
\ee
where $\Ea$ ($\epsa$) is the energy (energy loss) of the $Q \overline{Q}$ pair 
in the rest frame of nucleus $A$. The upper limit on the energy loss is 
$\epsamax=\min(\Ea,\Ep-\Ea)$ where $\Ep$ is the beam energy in that frame.
The energy loss probability distribution, or \emph{quenching weight}, 
${\cal P}$, is related to the medium-induced, coherent radiation spectrum given 
in Refs.~[\refcite{Arleo:2012rs,Arleo:2010rb}].  
This result proved to be an excellent 
approximation of the spectrum computed to all orders in the opacity 
expansion~\cite{Peigne:2014uha}.  It depends on the accumulated transverse 
momentum transfer $\ell_{_{\rm A}} = \sqrt{\qhat \ellb}$ due to soft rescatterings 
in the nucleus where $\ellb$ is the medium path length obtained from a Glauber 
calculation using realistic nuclear densities and $\qhat$ is the transport 
coefficient in cold nuclear matter. The transport coefficient 
is~\cite{Arleo:2012rs}
\be
\label{qhat-model}
\hat{q}(x_2) \equiv \hat{q}_0 \left[ \frac{10^{-2}}{x_2} \right]^{0.3}\ ; \ \ \  
x_2 \equiv  \frac{\mT}{\sqrts} \, e^{-y} \, ,
\ee
at small values of $x_2$, $x_2 < 0.01$, 
and $x_2$ is defined in $2 \rightarrow 1$
kinematics.  Here $y$ is the quarkonium rapidity in 
the center-of-mass frame of an elementary proton-nucleon collision, $\mT$ is 
the transverse mass and $\qzero$ is the only free parameter of the model. 
It is determined by fitting the $\jpsi$ suppression measured by 
the E866 Collaboration~\cite{Leitch:1999ea} in $p+$W relative to $p+$Be 
collisions at $\sqrt{s_{_{NN}}}=38.7$~GeV, see Ref.~[\refcite{Arleo:2012rs}]. 
The fitted value is $\qzero=0.075^{+0.015}_{-0.005}$~\gevsqfm.
The $p+p$ production cross section appearing in \eq{eq:xspA0-energy}
is given by the simple parametrization 
\begin{eqnarray}
\frac{\dd\sigma_{pp}^{\psi}}{ \dd y}  \propto \left(1- \frac{2 m_T}{\sqrt{s}} 
\cosh{y} \right)^{n(\sqrts)}\  \, \, , 
\end{eqnarray}
where the exponent $n$ is obtained from a fit to $p+p$ measurements at
different center-of-mass energies. 

\begin{figure}[htb]
    \begin{center}
      \includegraphics[height=3in, angle = 270]{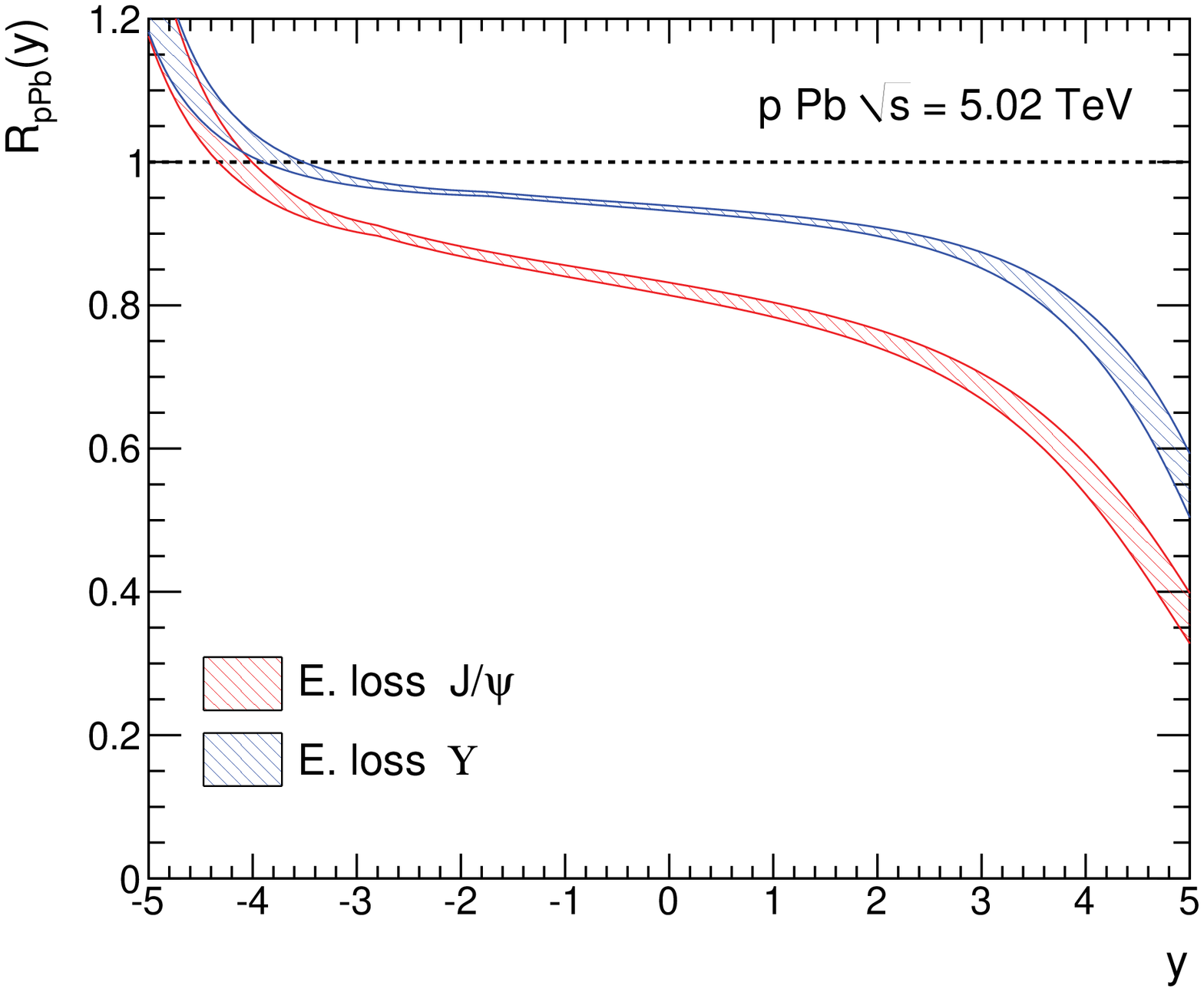}
    \end{center}
    \caption{(Color online)
The $\jpsi$ and $\Upsilon$ suppression factor as a function of
rapidity in $p+$Pb collisions at $\sqrts=5.02$~TeV for cold matter energy loss
alone.}
    \label{fig:lhc}
\end{figure}

\begin{figure}[htbp]
\begin{center}
\includegraphics[width=0.495\textwidth]{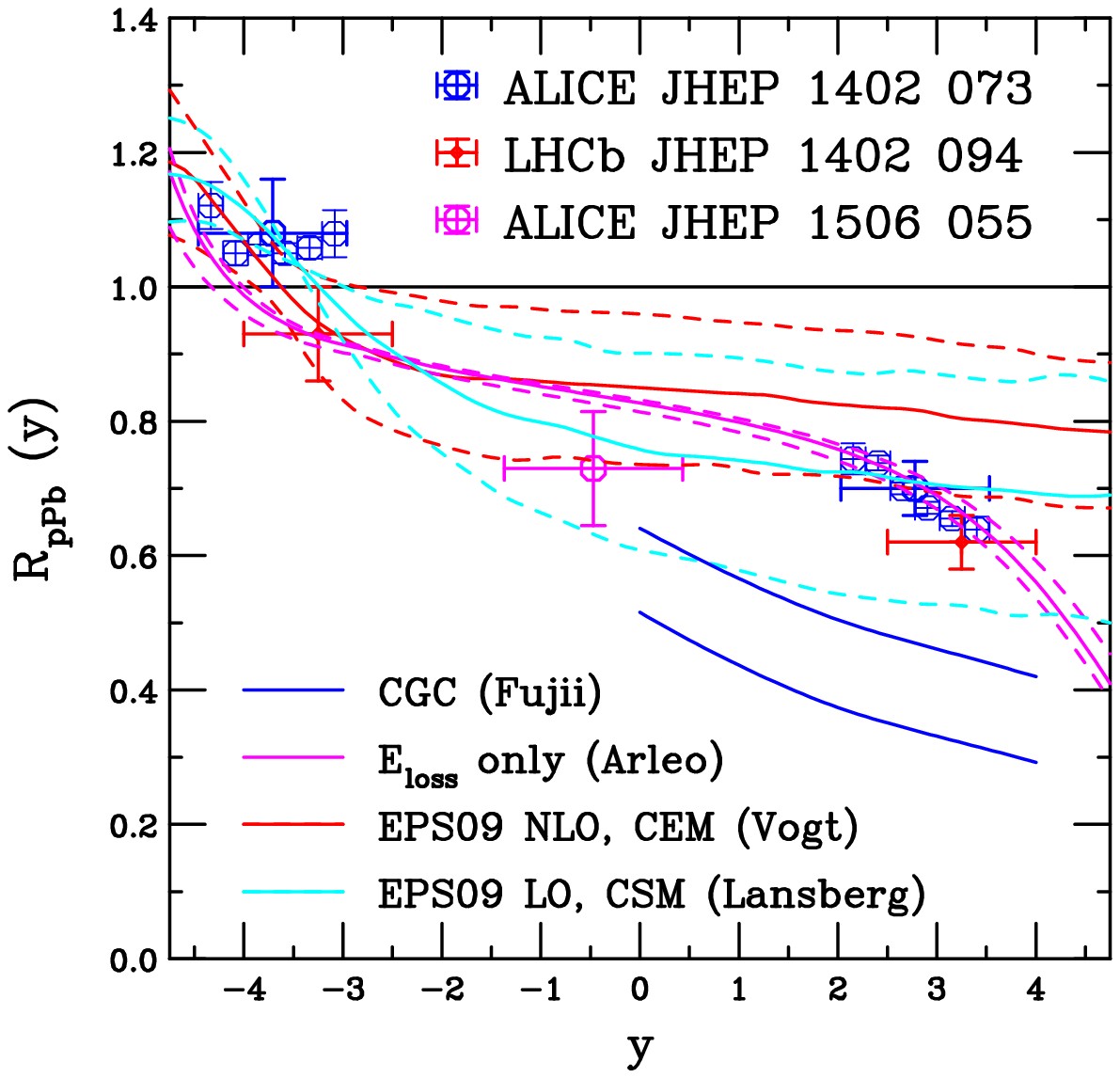} 
\includegraphics[width=0.495\textwidth]{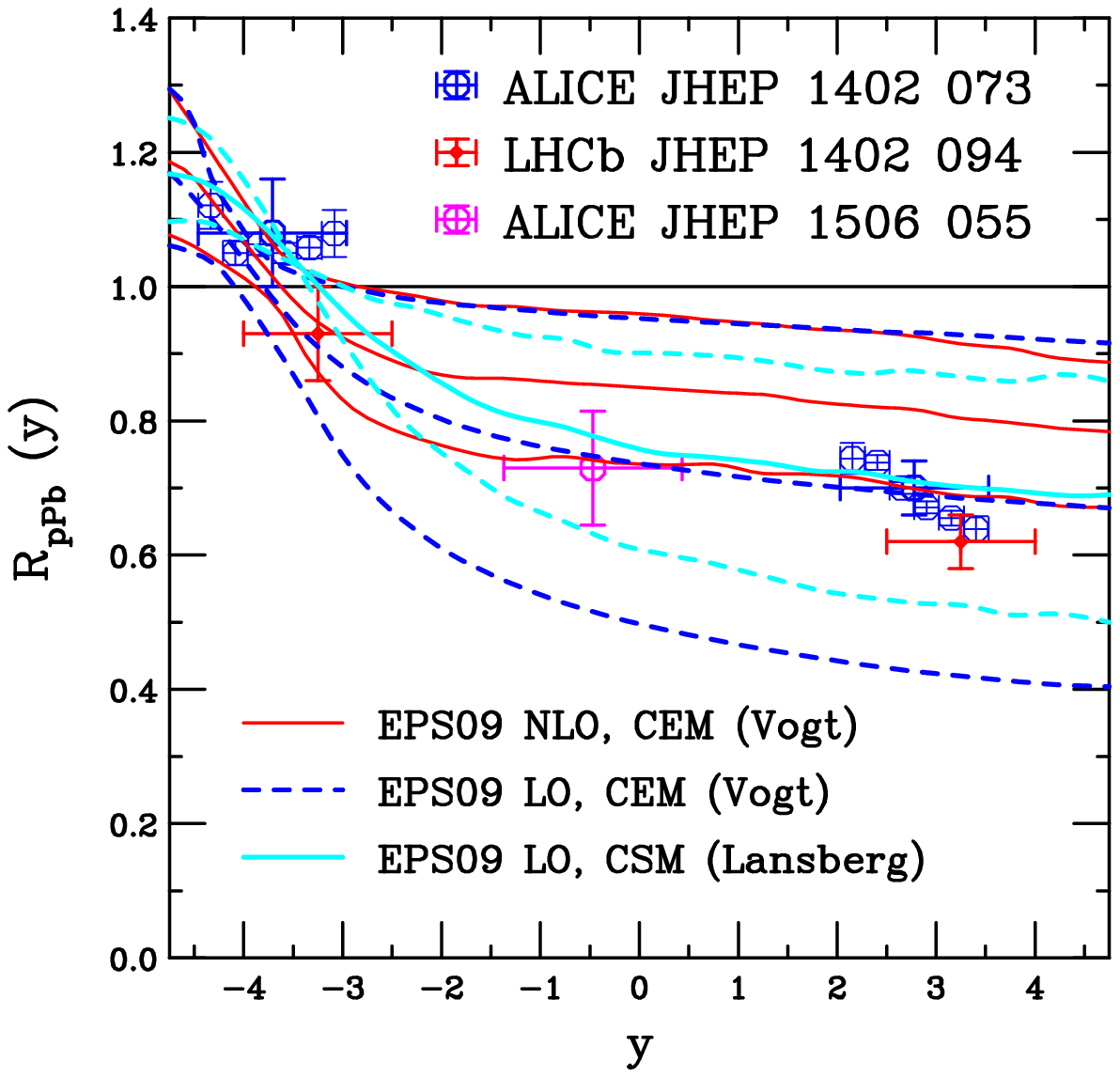} 
\end{center}
\caption[]{(Color online) (Left)
The ratio $R_{p{\rm Pb}}$ for $J/\psi$ as a function of $y$.
The red curves show the EPS09 NLO CEM uncertainties \protect\cite{Vogt:2015uba}.
The EPS09 LO CSM
calculation \protect\cite{Ferreiro:2013pua} is shown in cyan.  The
energy loss only calculations
\protect\cite{Arleo:2012rs,Arleo:2013zua} are shown in magenta. The upper
and lower limits of the CGC calculation \cite{Fujii:2013gxa}
are in blue at forward rapidity.
(Right) The EPS09 LO calculations in the CEM (red) and CSM (cyan) are compared
with each other and with the EPS09 NLO CEM calculation on the left-hand side.
The CEM calculation \protect\cite{Vogt:2015uba}
includes the full EPS09 LO uncertainty added in quadrature
while the CSM calculation  \protect\cite{Ferreiro:2013pua}
includes only the minimum and maximum uncertainty
sets.
The ALICE \protect\cite{Abelev:2013yxa,Abelev:2014oea} and LHCb 
\protect\cite{Aaij:2013zxa,Aaij:2014mza} data are also shown.
}
\label{fig:psiy}
\end{figure}

\begin{figure}[htbp]
\begin{center}
\includegraphics[width=0.495\textwidth]{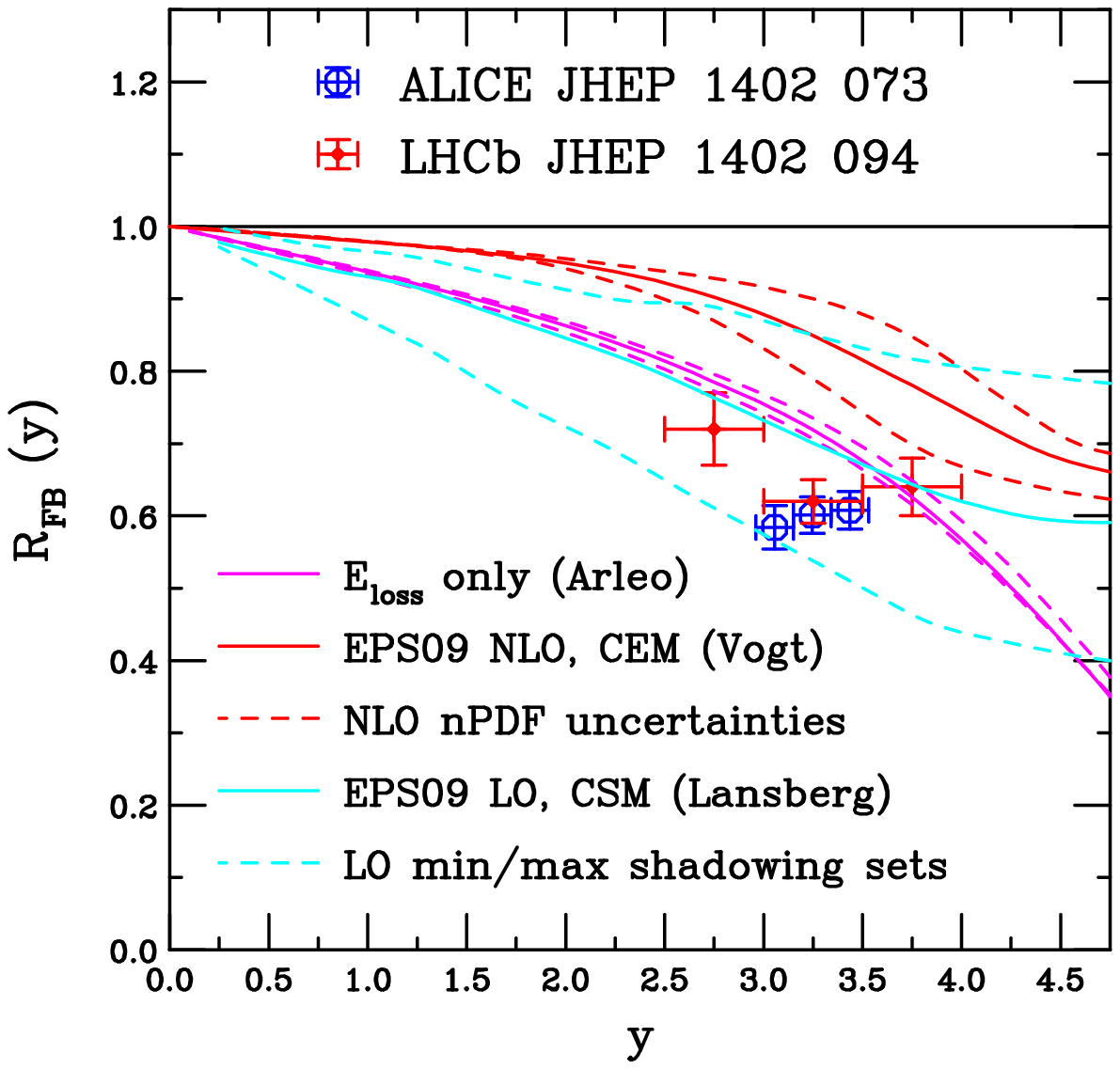} 
\includegraphics[width=0.495\textwidth]{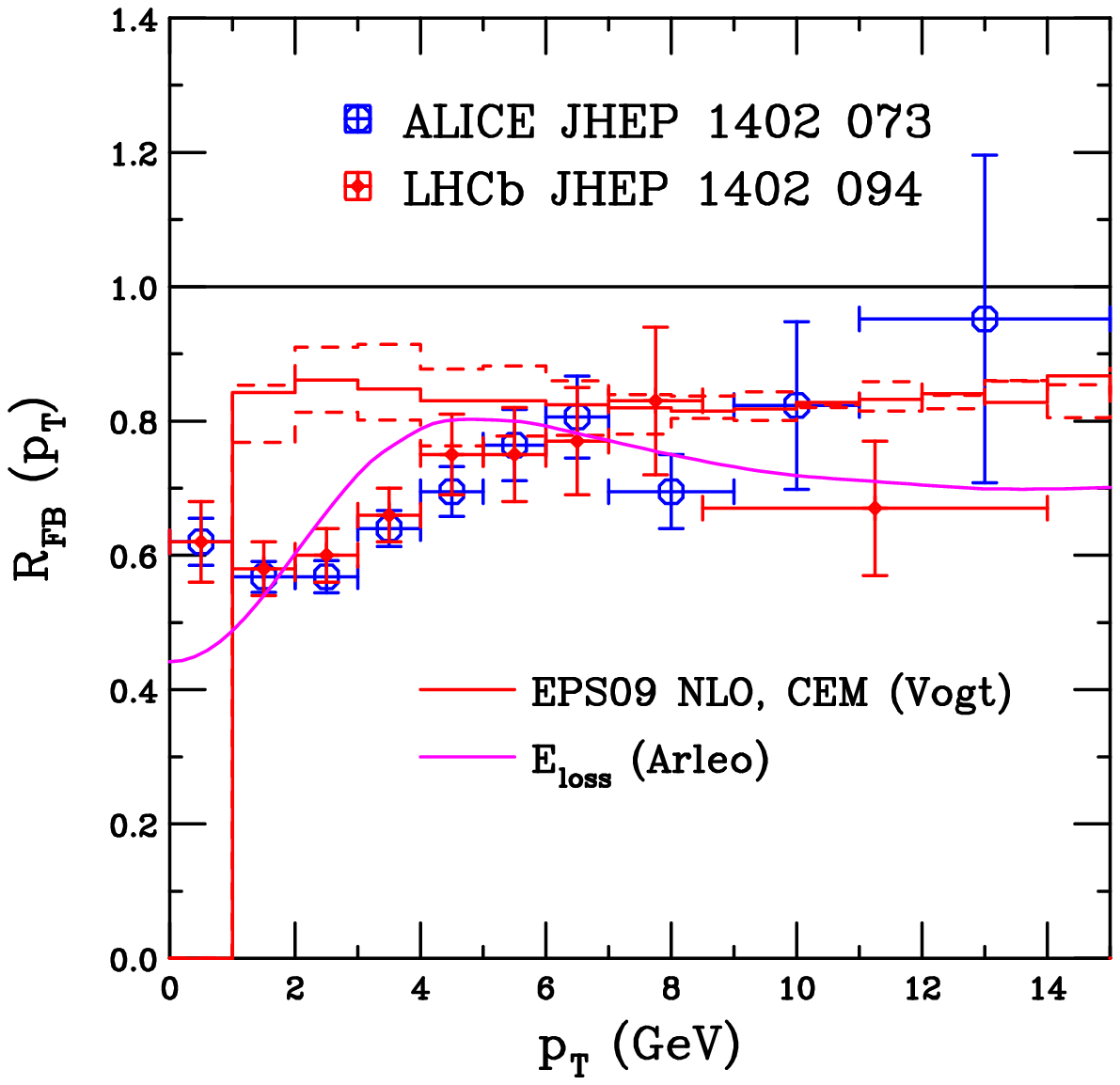} 
\end{center}
\caption[]{(Color online)
The ratio $R_{FB}$ for $J/\psi$ as a function of $y$
(left) and $p_T$ (right).
The red curves show the EPS09 NLO uncertainties \protect\cite{Vogt:2015uba}.  
The result with energy loss alone
\protect\cite{Arleo:2012rs,Arleo:2013zua} are shown in magenta.
The EPS09 LO CSM results \protect\cite{Ferreiro:2013pua} are given by the cyan
curves for the rapidity dependence only.
The ALICE \protect\cite{Abelev:2013yxa,Abelev:2014oea} and LHCb 
\protect\cite{Aaij:2013zxa,Aaij:2014mza} data are also shown.
}
\label{fig:psi_rfb}
\end{figure}

\begin{figure}[htbp]
\begin{center}
\includegraphics[width=0.495\textwidth]{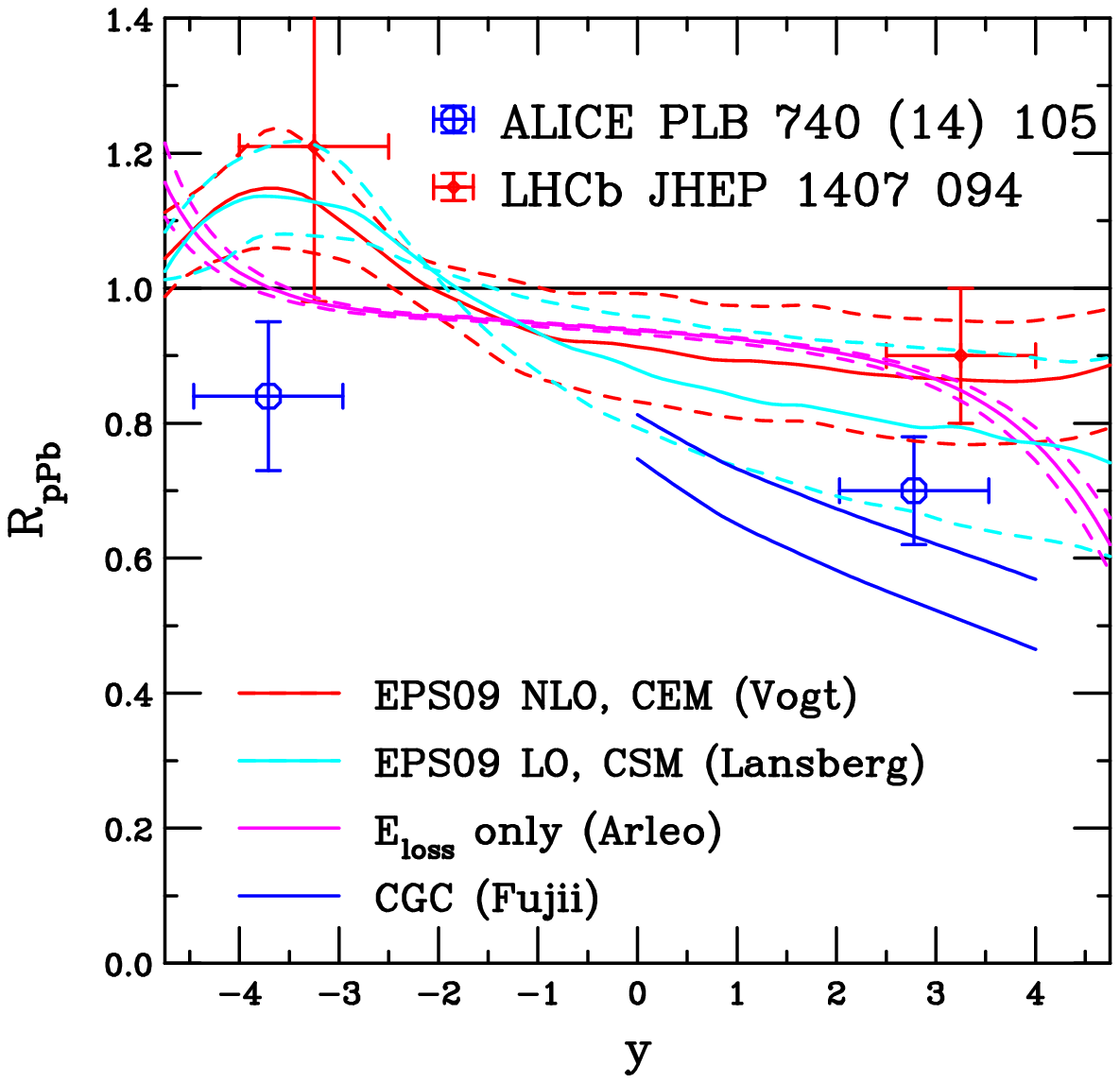} 
\includegraphics[width=0.495\textwidth]{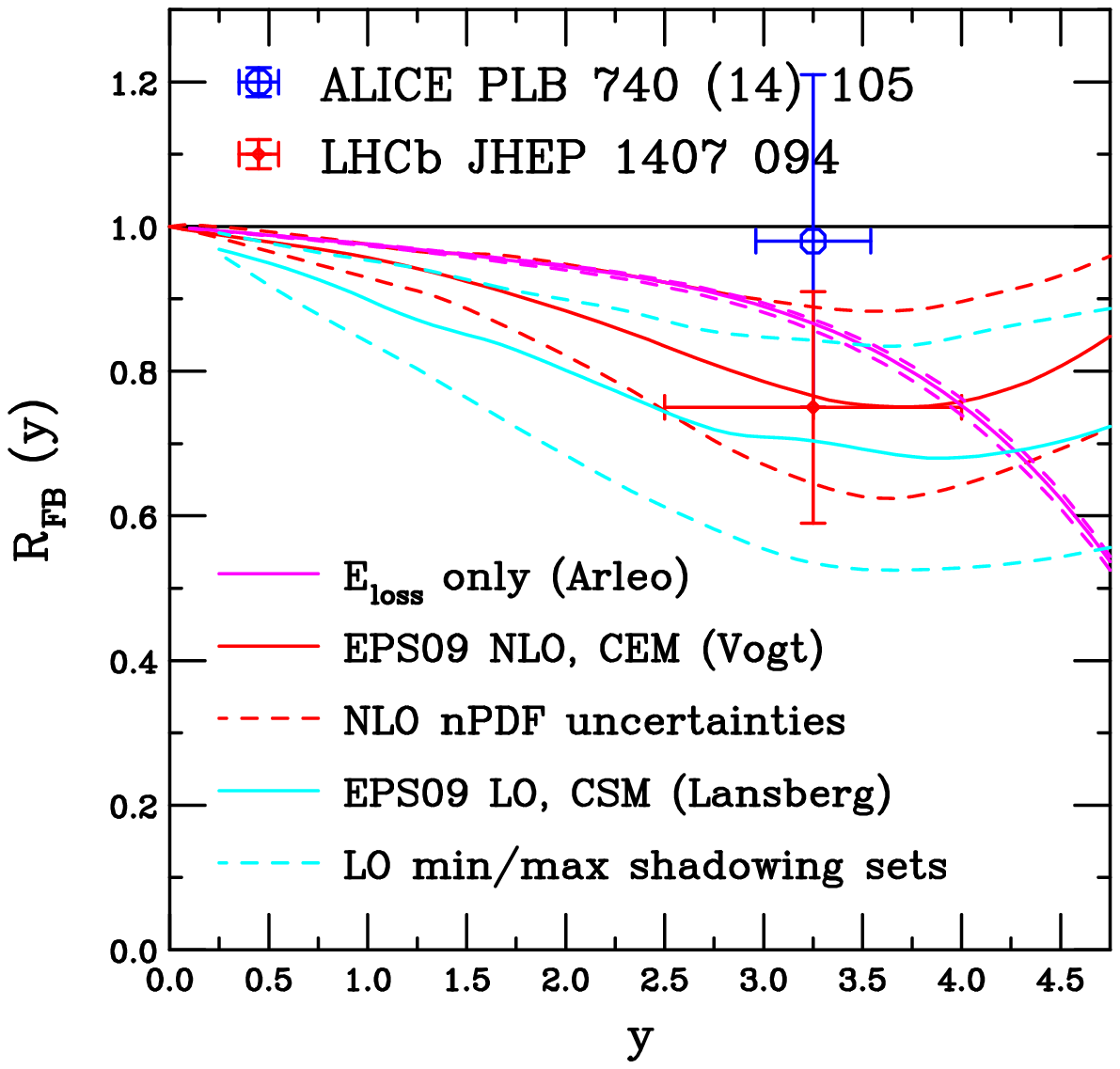} 
\end{center}
\caption[]{(Color online) (Left)
The ratio $R_{p{\rm Pb}}$ for $\Upsilon$ as a function of $y$.
The red curves show the EPS09 NLO CEM uncertainties \protect\cite{Vogt:2015uba}.
The EPS09 LO CSM calculation
\protect\cite{Ferreiro:2013pua} is shown in cyan.  The
energy loss only calculations
\protect\cite{Arleo:2012rs,Arleo:2013zua} are shown in magenta. The upper
and lower limits of the CGC calculation \cite{Fujii:2013gxa}
are in blue at forward rapidity.
(Right) The forward-backward ratio, $R_{FB}$, as a function of rapidity.
The same results as on the left-hand side are given, except for the CGC
result which is not calculable at backward rapidity.
The data from ALICE \protect\cite{Abelev:2014oea} and
LHCb \protect\cite{Aaij:2013zxa} are also shown in both panels.}
\label{fig:upsy}
\end{figure}

The predictions for $\jpsi$ and $\Upsilon$ suppression in $p+$Pb collisions 
at $\sqrt{s_{_{NN}}}=5.02$~TeV are shown in Fig.~\ref{fig:lhc}.  The model predicts 
rather strong $\jpsi$ suppression at forward rapidity, $y\gtrsim3$, and a 
slight enhancement in the most backward rapidity bins, $y<-4$. 
The suppression predicted for the $\Upsilon$ 
shares the same features.  However, the suppression is less pronounced than 
that of the $\jpsi$ since 
the (average) coherent energy loss scales as $\mT^{-1}$~\cite{Arleo:2010rb}.

Finally, CGC calculations from Ref.~[\refcite{Fujii:2013gxa}] are also shown.
The uncertainty comes from varying the saturation scale in the nucleus between
4 and 6 times that of the proton, $Q^2_{0, A} \sim (4-6)Q^2_{0, p}$,
as well as varying the quark mass.  The saturation scale is the biggest source
of uncertainty.  Indeed, more recent calculations suggest a smaller value of
the saturation scale, $Q^2_{0, A} \sim 3 Q^2_{0, p}$, is considered more reasonable
for minimum bias events, bringing the CGC result closer to the data
\cite{Ducloue:2015gfa,Fujii:2015lld}.


\subsection[Charmonium suppression due to comover interaction]{Charmonium suppression due to comover interactions (E. G. Ferreiro)}

Recent results on charmonium production in d+Au and $p+$Pb collisions from the
PHENIX \cite{Adare:2013ezl} and ALICE  \cite{Abelev:2014zpa,Arnaldi:2014kta} 
collaborations have shown an unexpectedly strong suppression of excited 
quarkonium states compared to their ground states.
In particular, stronger suppression of the $\psi(2S)$ relative to the $J/\psi$ 
has been detected.

At lower energies, this difference can been interpreted as the result of 
$c\overline{c}$ breakup in interactions with the primordial nucleons, the 
so-called nuclear absorption.  If the time spent
traversing the nucleus by the $c\overline{c}$ pair is longer than
the charmonium formation time, the larger $\psi(2S)$ meson will
be further suppressed by a stronger nuclear breakup effect.

However, at higher energies, the charmonium formation time is expected to be 
larger than the nucleus radius.  This results in identical breakup 
probabilities for the $\psi(2S)$ and $J/\psi$ since these
states cannot be distinguished during the time they are traversing the 
nucleus.  Moreover, this nuclear absorption is negligible at the LHC 
energies because the $c \overline c$ pair is still small \cite{Ferreiro:2012mm}.

Other usual explanations, such as that based on shadowing due to the 
modification of the nuclear gluon distribution, cannot be invoked since the 
shadowing effects are indistinguishable for the $\psi(2S)$ and $J/\psi$ 
as long as the same mass scale is used in the calculations
\cite{Ferreiro:2012mm}.

However, the difference in the suppression pattern can be easily 
explained by the interactions of the quarkonium states with a comoving 
medium \cite{Ferreiro:2014bia}.  In the comover framework, the suppression 
arises from scattering of the nascent $\psi$ with the produced particles, the
{\em comovers}, that happen to travel along with the $c\overline{c}$ pair 
\cite{Gavin:1996yd,Capella:1996va}.
Comover dissociation affects the $\psi(2S)$ more strongly than the $J/\psi$ 
due to the larger size of the $\psi(2S)$.  This suppression is stronger where 
the comover densities are larger: it increases with centrality and, in
asymmetric proton-nucleus collisions, it is stronger in the nucleus-going 
direction.

In the  comover interaction model, CIM \cite{Capella:1996va,Armesto:1997sa,Armesto:1998rc,Capella:2000zp,Capella:2005cn,Capella:2006mb}, the rate equation 
that governs the charmonium density, $\rho^{\psi}(b,s,y)$, at a given 
transverse coordinate $s$, impact parameter $b$ and rapidity~$y$, obeys the 
simple expression
\beq
\label{eq:comovrateeq}
\tau \frac{\mbox{d} \rho^{\psi}}{\mbox{d} \tau} \, \left( b,s,y \right)
\;=\; -\sigma^{{\rm co}+\psi}\; \rho^{\rm co}(b,s,y)\; \rho^{\psi}(b,s,y) \;
\eeq
where $\sigma^{{\rm co}+\psi}$ is the charmonium dissociation cross section
due to interactions with the comoving medium of transverse density
$\rho^{\rm co}(b,s,y)$.

Assuming that the system becomes more dilute as a function of time due to the
longitudinal motion leads to a $\tau^{-1}$ dependence on
proper time.  The rate equation can be solved analytically. The result depends 
only on the ratio $\tau_f/ \tau_0$ of final over initial times. Using the 
inverse proportionality of proper time to density, 
$\tau_f/ \tau_0= \rho^{\rm co}(b, s, y)/\rho_{pp}(y)$, it is assumed that 
the interactions stop when the comover density has become as dilute as 
the $p+p$ collision density at the same energy. Thus, the 
solution to Eq.~(\ref{eq:comovrateeq}) is
\beq
\label{eq:survivalco}
S^{\rm co}_{\psi}(b,s,y)  \;=\; \exp \left\{-\sigma^{{\rm co}+\psi}
  \, \rho^{\rm co}(b,s,y)\, \ln
\left[\frac{\rho^{\rm co}(b,s,y)}{\rho_{pp} (y)}\right] \right\} \;
\eeq
where the argument of the logarithmic term is the interaction time of the 
$\psi$ with the comovers.

The only adjustable parameter of the comover interaction model is the cross 
section for charmonium dissociation due to interactions with the comoving 
medium, $\sigma^{{\rm co}+\psi}$. It was fixed \cite{Armesto:1997sa} from fits to 
low-energy experimental data to be
$\sigma^{{\rm co}+J/\psi}=0.65$~mb for the $J/\psi$ and 
$\sigma^{{\rm co}+\psi(2S)}=6$~mb for the $\psi(2S)$. 
This value has been also successfully applied at higher energies to reproduce 
the RHIC \cite{Capella:2007jv} and LHC \cite{Ferreiro:2012rq} data on $J/\psi$ 
suppression in nucleus-nucleus collisions.

As mentioned previously, another important effect that should be taken into 
account in quarkonium production in nuclei is shadowing of the gluon 
distribution in the nucleus. This effect is assumed to be identical for the
the $J/\psi$ and the $\psi(2S)$ [\refcite{Ferreiro:2012mm}].
The nuclear modification of the parton distribution functions will result in
a common effect on the $J/\psi$ and the $\psi(2S)$ yields: a decrease in the 
mid and forward rapidity regions at LHC energies and an increase in the 
backward rapidity region.

The nuclear modification factor for comover interactions, together with 
shadowing effects, is
\beq
\label{eq:ratiopsi}
R^{\psi}_{pA}(b) =
\frac{\int\mbox{d}^2s \, 
  \sigma_{pA}(b) \, n(b,s) \,  S_{\psi}^{\rm sh}(b,s) \, S^{\rm co}_{\psi}(b,s)
}{\int \mbox{d}^2 s \, \sigma_{pA} (b) \, n(b,s)} \;, 
\eeq
where $S^{\rm co}_{\psi}$ is the survival probability due to the comover 
interactions and $S_{\psi}^{\rm sh}$ takes
the shadowing of the nuclear parton distribution functions into account. 

\begin{figure}[htbp]
\begin{center}
\vspace*{0.2in}
\includegraphics[width=0.75\textwidth]{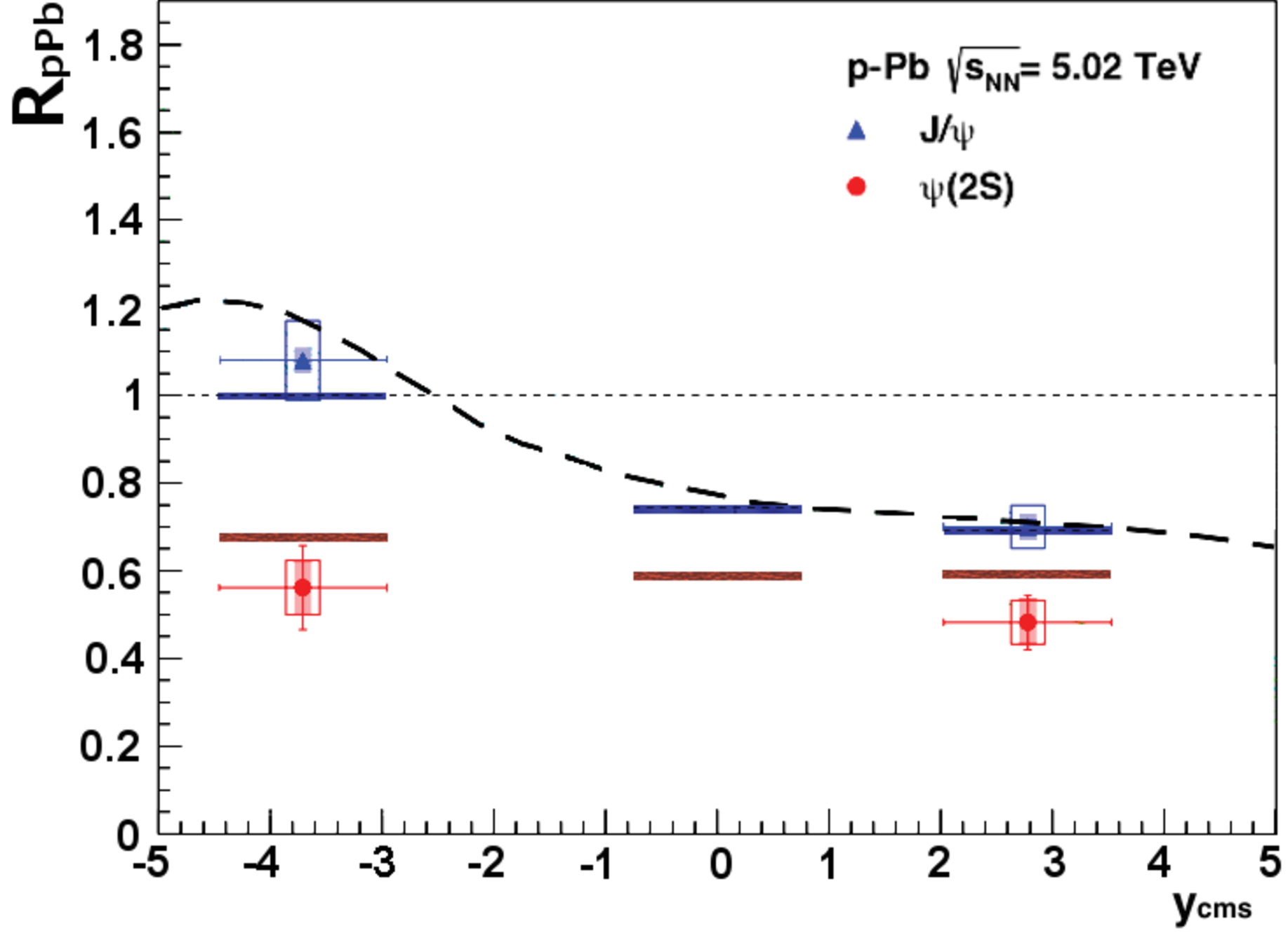}
\end{center}
\caption{\label{fig:figpPby}(Color online) 
The $J/\psi$ (blue line) and $\psi(2S)$ (red line)
nuclear modification factors $R_{p {\rm Pb}}$ as a function of rapidity
compared to the ALICE data \protect\cite{Abelev:2014zpa}. The 
suppression due to the shadowing corrections (dashed line) is also shown.
}
\end{figure}

Figure~\ref{fig:figpPby} shows the nuclear modification factor, $R_{p {\rm Pb}}$ 
as a function of rapidity.  The experimental data \cite{Abelev:2014zpa} on 
$J/\psi$ and $\psi(2S)$ production in $p+$Pb collisions at 
$\sqrt{s_{_{NN}}} = 5.02$~TeV are compared to the CIM results. The EPS09 LO 
shadowing effects are assumed to be identical
\cite{Ferreiro:2013pua,Eskola:2009uj} for both the $J/\psi$ and $\psi(2S)$. 
The effect of the EPS09 shadowing is strongly dependent on the 
rapidity interval considered. While it induces an increase, antishadowing, in 
the backward region, it produces a suppression, shadowing, in 
the forward region.  On the other hand, the interaction with comovers 
introduces a stronger suppression in the backward, lead-going, region due 
to the higher comover density.  The effect will be more important on 
$\psi(2S)$ than on $J/\psi$ production due to the larger
$\sigma^{co-\psi}$ for the $\psi(2S)$.

\begin{figure}[htbp]
\begin{center}
\includegraphics[width=0.75\textwidth]{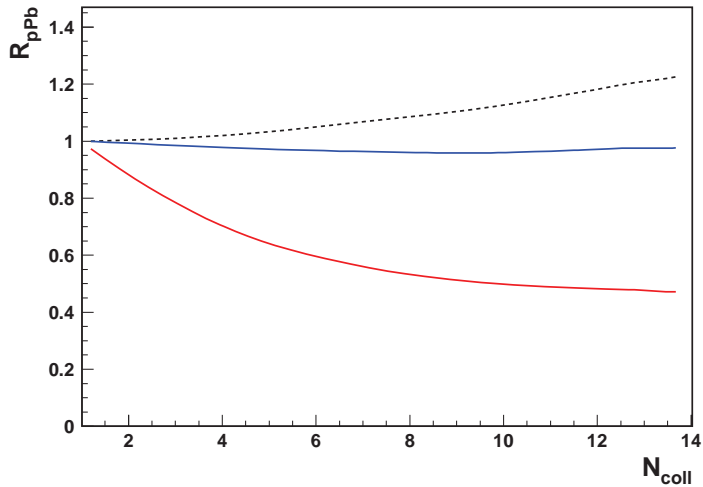}
\end{center}
\caption{\label{fig:figpPbvncoll1}(Color online) 
The nuclear modification factor $R_{p{\rm Pb}}$ as a function of the number of 
binary nucleon-nucleon collisions, $N_{\rm coll}$, in the backward $-4.46<y<-2.96$
rapidity interval for the $J/\psi$ (blue line) and the $\psi(2S)$ (red line). 
The modification due to the shadowing corrections alone (dotted line) is 
also shown.
}
\end{figure}

\begin{figure}[htpb]
\begin{center}
\includegraphics[width=0.75\textwidth]{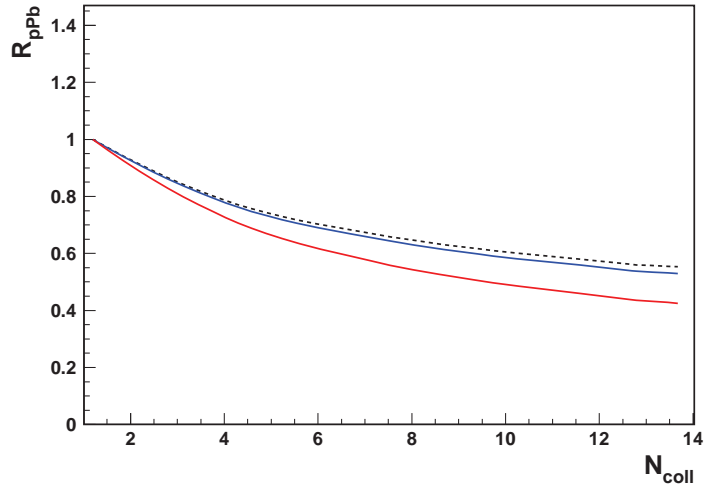}
\end{center}
\caption{\label{fig:figpPbvncoll2}(Color online) 
The nuclear modification factor $R_{p{\rm Pb}}$ as a function of the number of 
binary nucleon-nucleon collisions, $N_{\rm coll}$, in the forward $2.03<y<3.53$
rapidity interval for the $J/\psi$ (blue line) and the $\psi(2S)$ (red line). 
The modification due to the shadowing corrections alone (dotted line) 
is also shown.
}
\end{figure}

In Figs.~\ref{fig:figpPbvncoll1} and \ref{fig:figpPbvncoll2}, the
results for $J/\psi$ and $\psi(2S)$ production are given as a function of 
collision centrality.  Two rapidity intervals are studied: the $p$-going 
direction, $2.03<y<3.53$ and the Pb-going direction, $-4.46<y<-2.96$.
In the backward region, a nuclear modification factor, $R_{p{\rm Pb}}$, 
compatible with unity 
is obtained for the $J/\psi$ due to the combined effect of 
the EPS09 LO antishadowing together with comover suppression.  In the case of 
$\psi(2S)$ production, antishadowing is dominated by the stronger effect of 
comover suppression.  The total $J/\psi$ suppression in the forward region
is almost 50\%, primarily due to shadowing.  On the other hand, for the 
$\psi(2S)$, both shadowing and a limited comover effect contribute to an
overall suppression at forward rapidity.

In summary, a detailed study of $J/\psi$ and $\psi(2S)$ production in $p+$Pb 
collisions at $\sqrt{s_{_{NN}}}=5.02$ TeV has been performed.
The available data are consistent with
the interaction of fully formed physical quarkonia 
with produced particles, the comovers,
that travel along with the $c\overline{c}$ pair.

\section[Gauge boson production]{Gauge boson production (Z.-B. Kang, J.-W. Qiu, P. Ru, E. Wang, B.-W. Zhang and W.-N. Zhang)}

The production of the $Z^0$ and $W^\pm$ gauge bosons in $p+$Pb collisions is
studied here.  The calculations in Sec.~\ref{Zhangetal} are done with 
perturbative QCD up to NLO and are compared to several gauge boson observables.
The effects of shadowing and isospin are studied for two different sets of
underlying proton PDFs.  The calculations of $Z^0$ production in 
Sec.~\ref{KangandQiu} include resummation of large logarithims of $m_Z^2/p_T^2$
and concentrate on the low $p_T$ region of the $p_T$ distribution.

\subsection[$W^\pm$ and $Z^0$ Production to NNLO]{$W^\pm$ and $Z^0$ Production to NNLO (P. Ru, E. Wang, B.-W. Zhang and W.-N. Zhang)}
\label{Zhangetal}

Here the perturbative QCD results are compared to the latest LHC data~(or 
preliminary data) for several observables, including the (pseudo-)rapidity 
dependence, the transverse momentum spectra, the forward-backward asymmetry
of $Z^0$ and leptons from $W^\pm$ decays, 
and the $W^\pm$ charge asymmetry.

The numerical simulations utilize the Monte Carlo program 
DYNNLO~\cite{Catani:2009sm,Catani:2007vq} which was developed to study 
the Drell-Yan process in hadronic collisions at next-to-leading order~(NLO) and 
next-to-next-to-leading order~(NNLO).  Cold nuclear matter (CNM) effects are 
included by incorporating phenomenological parametrizations of the nPDFs.
The CT10~\cite{Gao:2013xoa} and MSTW2008~\cite{Martin:2009iq} 
proton PDFs are employed with the EPS09~\cite{Eskola:2009uj} and 
nCTEQ~\cite{Schienbein:2009kk,Kovarik:2010uv} NLO nPDFs. 
Calculations with the CT10 and MSTW PDFs are shown with three nuclear 
modifications: EPS09 NLO and nCTEQ, including isospin, and isospin alone, 
without shadowing.
The cross sections shown here are obtained by scaling the nucleon-nucleon 
results by the Pb mass number, $A=208$.  More detailed 
discussions can be found in Ref.~[\refcite{Ru:2014yma}].

In the calculations, the renormalization and factorization
scales, $\mu_R$ and $\mu_F$, are set to $\mu_R=\mu_F=m_V$ where 
$m_V$ is the mass of vector boson.

\subsubsection{(pseudo-)rapidity dependence}

\begin{figure}[h]
\includegraphics[width = \textwidth]{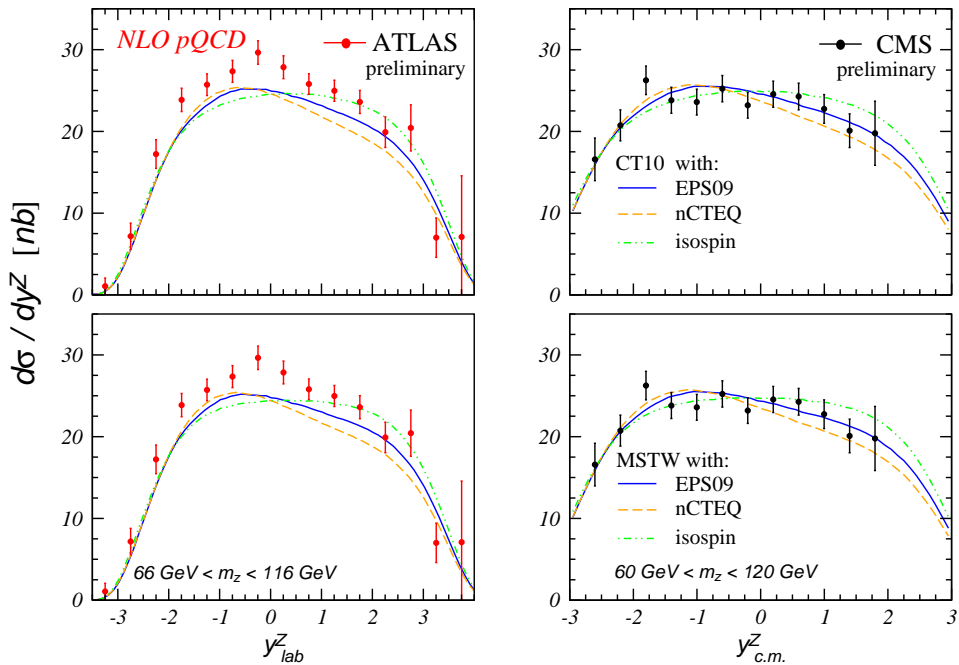}
\caption{(Color online) The $Z^0$ rapidity distribution in $p+$Pb collisions 
at $\sqrt{s_{NN}}=5.02$~TeV.
The left panels show the results compared to the ATLAS 
data~\protect\cite{ZpPbATLAS} while those on the right show comparisons to 
the CMS data~\protect\cite{CMS:2014sca}.
The top panels show calculations with the CT10 PDFs while results with MSTW2008
are shown on the bottom.}
\label{fig:Z-y@pPb}
\end{figure}

The NLO $Z^0$ boson rapidity distributions are compared with the preliminary
ATLAS~($e,\mu$ combined)~\cite{ZpPbATLAS} and CMS~($\mu$ 
only)~\cite{CMS:2014sca} 
data.  The only cut on the final state is the $Z^0$ mass window, 
$66 <m_Z<116$~GeV for ATLAS and $60 <m_Z<120$~GeV for CMS.
The results are shown in Fig.~\ref{fig:Z-y@pPb}. The 
calculations~\cite{Ru:2014yma} agree well with the CMS data,
but not that well with the ATLAS results in the region 
$-2\lesssim y_{l\rm ab}^Z\lesssim1$, where $y_{\rm lab}=y_{\rm cm}+0.465$.
The dependence on free proton PDFs is rather small.  However, differences 
among the three types of nuclear modifications can be observed, especially in 
the forward rapidity region~($y^Z_{\rm lab}>0$).

\begin{figure}[h]
\includegraphics[width = \textwidth]{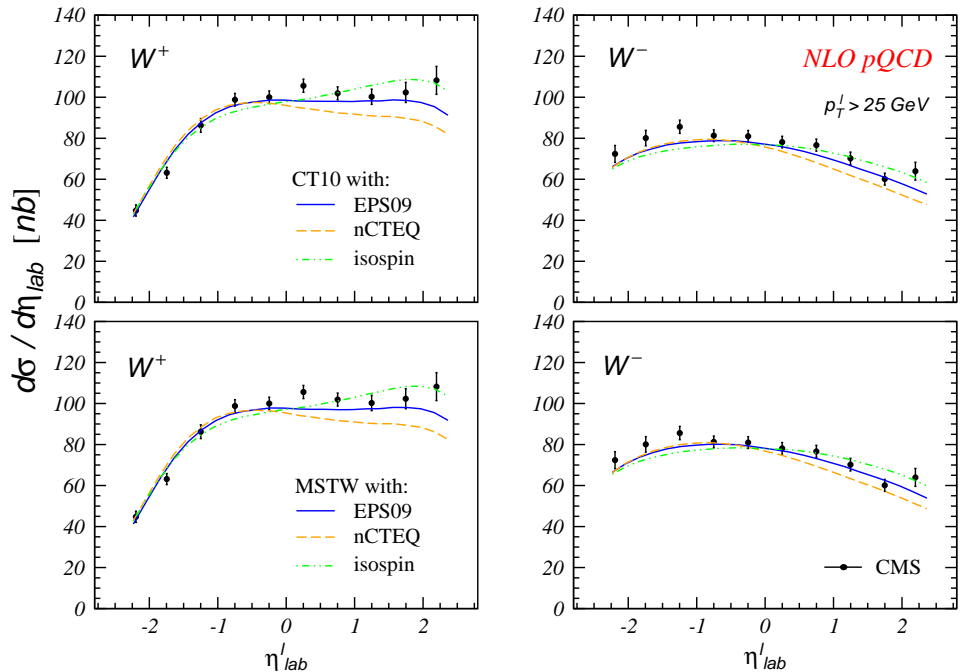}
\caption{(Color online) The charged lepton pseudorapidity distributions for $W$
boson production in $p+$Pb collisions at $\sqrt{s_{NN}}=5.02$~TeV.
The results for $W^+$ are shown on the left-hand side while those for $W^-$ 
are shown on the right.
The results with the CT10~(top panels) and MSTW2008 PDFs~(bottom panels) are 
compared to the CMS data~\protect\cite{Khachatryan:2015hha}.}
\label{fig:W-eta@pPb}
\end{figure}

The NLO charged lepton pseudorapidity distributions for $W^\pm$ boson 
decays are shown in Fig.~\ref{fig:W-eta@pPb}.
The final state cut on the charged lepton transverse momentum is $p_T^l>25$~GeV.
The calculations are in good agreement with the CMS $e,\mu$ combined 
data~\cite{Khachatryan:2015hha}.
There are obvious differences in the nuclear modifications for $W^+$ production,
particularly in the forward region.  Small differences can also be seen for 
$W^-$ production.

\subsubsection{$Z^0$ transverse momentum distribution}

\begin{figure}[h]
\includegraphics[width = \textwidth]{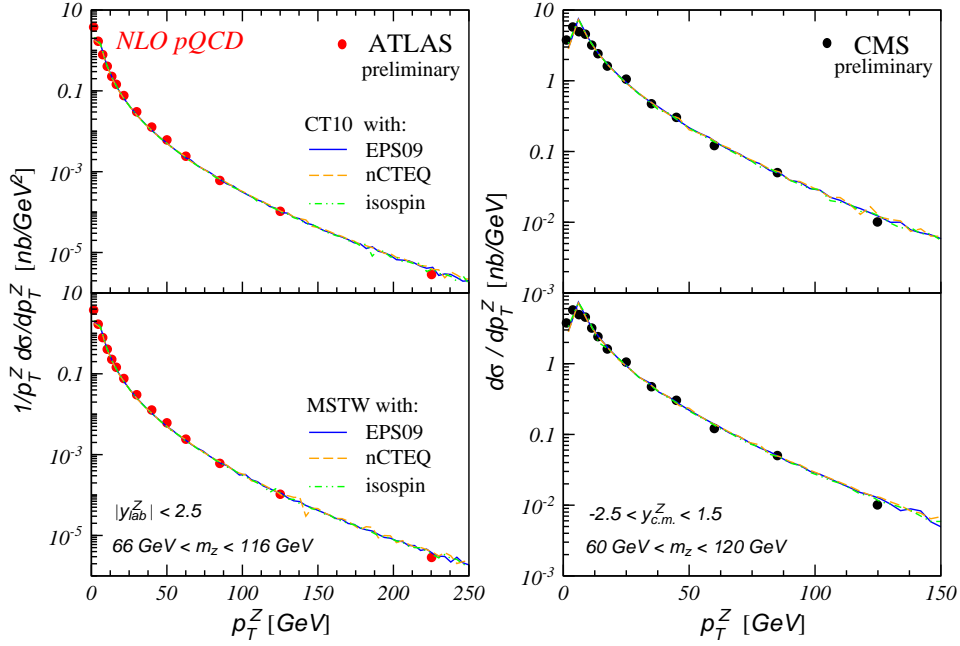}
\caption{(Color online) The $Z^0$ transverse momentum spectra in $p+$Pb 
collisions at $\sqrt{s_{NN}}=5.02$~TeV.
The results for ATLAS~\protect\cite{ZpPbATLAS} and 
CMS~\protect\cite{CMS:2014sca} are shown in the left and right panels, 
respectively.  The top panels show results with the CT10 proton PDFs while
results with the MSTW2008 PDFs are shown in the bottom panels.}
\label{fig:Z-pt@pPb}
\end{figure}

The NLO $Z^0$ transverse momentum distributions are compared to the preliminary
data from ATLAS~($e,\mu$ combined)~\cite{ZpPbATLAS} and 
CMS~($\mu$ only)~\cite{CMS:2014sca}.
In addition to the different mass windows, the $Z^0$ rapidity regions are also 
different for ATLAS and CMS: $|y_{\rm lab}^Z|<2.5$ in the laboratory frame for 
ATLAS and $-2.5<y_{\rm cm}^Z<1.5$ in the center of mass frame for CMS.
The calculations agree quite well with the data, as shown in 
Fig.~\ref{fig:Z-pt@pPb}.  Note that, on the logarithmic scale of the 
distributions, no difference between the type of nuclear effects included can
be observed.

\subsubsection{Forward-Backward Rapidity Asymmetry}

\begin{figure}[h]
\includegraphics[width = \textwidth]{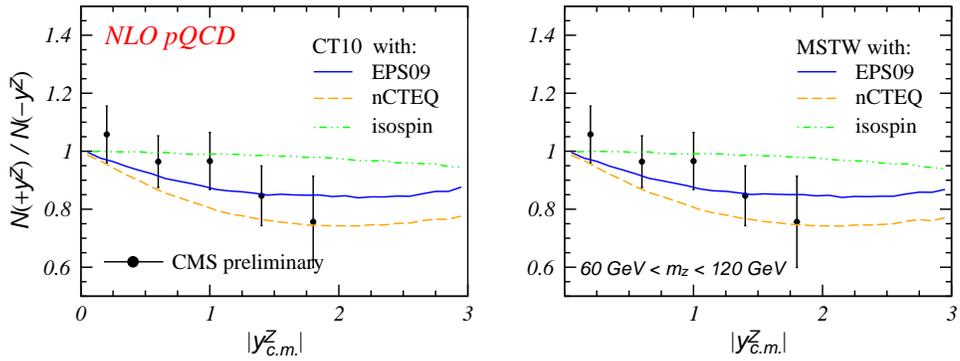}
\caption{(Color online) The forward-backward asymmetry as a function of the 
absolute value of $Z^0$ rapidity in the center of mass frame.
The results with the CT10 proton PDFs are shown on the left and the results with
the MSTW2008 PDFs are shown on the right. The CMS data are from 
Ref.~[\protect\refcite{CMS:2014sca}].}
\label{fig:Z-fbasy@pPb}
\end{figure}

The forward-backward asymmetry for vector boson production can be 
observed in the asymmetric $p+$Pb collision system.  The asymmetry arises from
the cold nuclear matter effects~\cite{Ru:2014yma}.
First, the forward-backward asymmetry is studied as a function of the absolute 
value of the $Z^0$ rapidity in the center of mass frame in the CMS mass window, 
$60 <m_Z<120$~GeV~\cite{CMS:2014sca}.
The NLO results are compared with the CMS muon data in 
Fig.~\ref{fig:Z-fbasy@pPb}.
Differences between the three nuclear modifications are emphasized by the
asymmetry.
Isospin alone gives only a small forward-backward asymmetry while nuclear 
modifications such as antishadowing give a larger symmetry, 20-25\%.
The nCTEQ nuclear modification gives the largest asymmetry.
The calculations agree with the data within the uncertainties.
However, the data favor nuclear modifications with EPS09 NLO and nCTEQ in the 
region $1.2\lesssim|y_{\rm cm}^Z|\lesssim2$.

\begin{figure}[h]
\includegraphics[width = \textwidth]{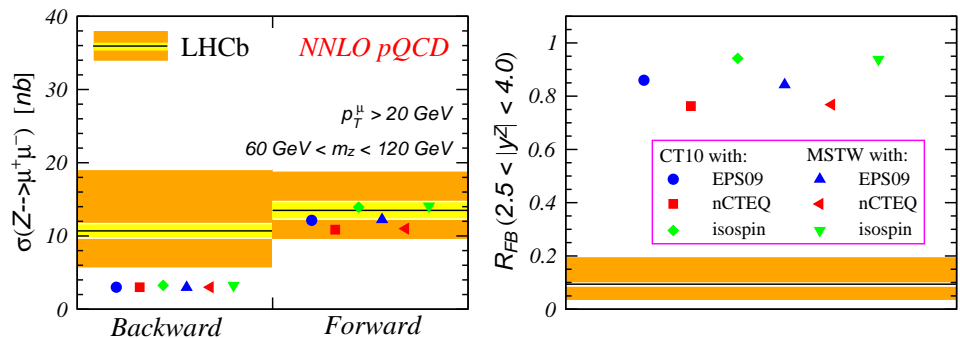}
\caption{(Color online) The forward and backward $Z^0$ cross sections are shown
on the left-hand side while the forward-backward asymmetry is shown on 
the right-hand side.  The results are compared with the LHCb 
data~\protect\cite{Aaij:2014pvu}.}
\label{fig:Z-fbasylhcb@pPb}
\end{figure}

The LHCb collaboration has measured the $Z^0$ forward-bacward asymmetry in a 
different, more forward, rapidity region from CMS~\cite{Aaij:2014pvu}.
The forward and backward cross sections have been measured in the
region $60 <m_Z<120$~GeV, $p_T^{\mu}>20$~GeV, and $2.0<\eta^{\mu}<4.5$ in the 
laboratory frame.  In the center of mass frame, the muon pseudorapidity range 
is $1.53<\eta_{\rm cm}^{\mu}<4.03$ in the forward region and
$-4.97<\eta_{\rm cm}^{\mu}<-2.47$ in the backward region.
The NNLO cross sections are calculated and compared with the LHCb data on the 
left-hand side of Fig.~\ref{fig:Z-fbasylhcb@pPb}.
The calculations in the forward region agree with the data while, in the 
backward region, the results are much smaller than the data,
even though the experimental uncertainty is rather large.

On the right-hand side of Fig.~\ref{fig:Z-fbasylhcb@pPb}, the forward-backward 
ratios $R_{FB}(2.5<|y_{\rm cm}^Z|<4.0)$ are calculated at NNLO and compared to 
the LHCb data.  The ratio is defined as
\begin{eqnarray}
\label{rffbary}
R_{FB}(2.5<|y_{\rm cm}^Z|<4.0)=\frac{\sigma(2.5<y_{\rm cm}^Z<4.0)}
{\sigma(-4.0<y_{\rm cm}^Z<-2.5)} \, \, .
\end{eqnarray}
The calculations considerably
overestimate the data.  In addition, the uncertainty with 
the nuclear PDFs is rather small.  Thus the deviation between theory and data 
is significant.
Considering the results of the individual forward and backward cross sections, 
this overestimate is likely due to a significant (factor of 3-4) underestimate 
of the backward cross section.  

There seems to be an apparent mismatch between the ratio of the forward to
backward cross sections as seen on the left-hand side of 
Fig.~\ref{fig:Z-fbasylhcb@pPb}.  A `by-eye' view might lead one to expect 
$R_{FB} > 1$.
However, in this panel, the forward and backward regions cover the entire
rapidity space of 2.5 units.  When the rapidity range is restricted to 1.5
units the statistical signal is reduced, 
leading to the value of $R_{FB} < 1$ on the right-hand side.

Note that the LHCb results are based on a fairly small sample of $Z^0$ bosons so
that, while the signal is strong, the statistical signifcance is not.  There
are four $Z^0$ candidates in the backward rapidity region and eleven at forward
rapidity over the full phase space.  
When the range is restricted to the overlap of the forward and
backward regions,  only two candidates are left in the forward region while
the four candidates in the backward region are not reduced.  After corections 
for acceptance in the different regions are taken into account the measured
$R_{FB}$ is reduced to $R_{FB}(2.5 < |y| < 4.0) = 0.094^{+0.104}_{-0.062} \, 
({\rm stat.}) \, ^{+0.004}_{-0.007} \, ({\rm syst.})$.

To leading order, the momentum fraction $x$ carried by initial parton in the
nucleus for $p_T \sim 0$ $Z^0$ boson production at rapidity $y^Z$ is
$x= (m_Z/\sqrt{s_{_{NN}}})e^{-y^Z}$.
Thus $x\in(3.32\times 10^{-4},1.49\times 10^{-3})$ in the forward rapidity 
region and $x\in(0.22, 0.989)$ in the backward direction for LHCb.
Also note that forward $Z^0$ production proceeds via nuclear sea quarks, 
$u_s \overline{u}_s$ and $d_s \overline{d}_s$ (with only small higher order 
contributions from gluons), while backward $Z^0$
production is dominated by nuclear valence quarks, $u_v$ and 
$d_v$~\cite{Ru:2015pfa}.
Therefore the forward-backward ratio can schematically be written as
\begin{eqnarray}
\label{rffbarys}
R_{FB}(2.5<|y_{\rm cm}^Z|<4.0)
& \thicksim & \frac{R_{u_s,d_s}(x\in[3.32\times 10^{-4},1.49\times 10^{-3}])}
{R_{u_v,d_v}(x\in[0.22, 0.989])} \nonumber \\
& \thicksim & \frac{u_s ,d_s \, {\rm shadowing}}{u_v, d_v \, {\rm EMC}} \, \, ,
\end{eqnarray}
where $R_f(x)$ is the flavor-dependent nuclear modification factor. The 
nuclear effect in the forward region is thus predominantly related to sea 
quark shadowing while that in the backward direction is mainly due to EMC 
effects on the valence quarks~\cite{Ru:2014yma, Ru:2015pfa}.

Because the forward and backward yields in the rapidity region covered by LHCb 
are rather small~\cite{Aaij:2014pvu},
the experimental precision should be improved for more robust 
comparisons of the calculations to data.  
Unfortunately the rapidity regions of CMS and LHCb do not overlap.
However, it is clear that the forward-backward ratio in the rapidity range
measured by LHCb is significantly lower than one would expect from extrapolating
the CMS data to higher rapidity.

\begin{figure}[h]
\includegraphics[width = \textwidth]{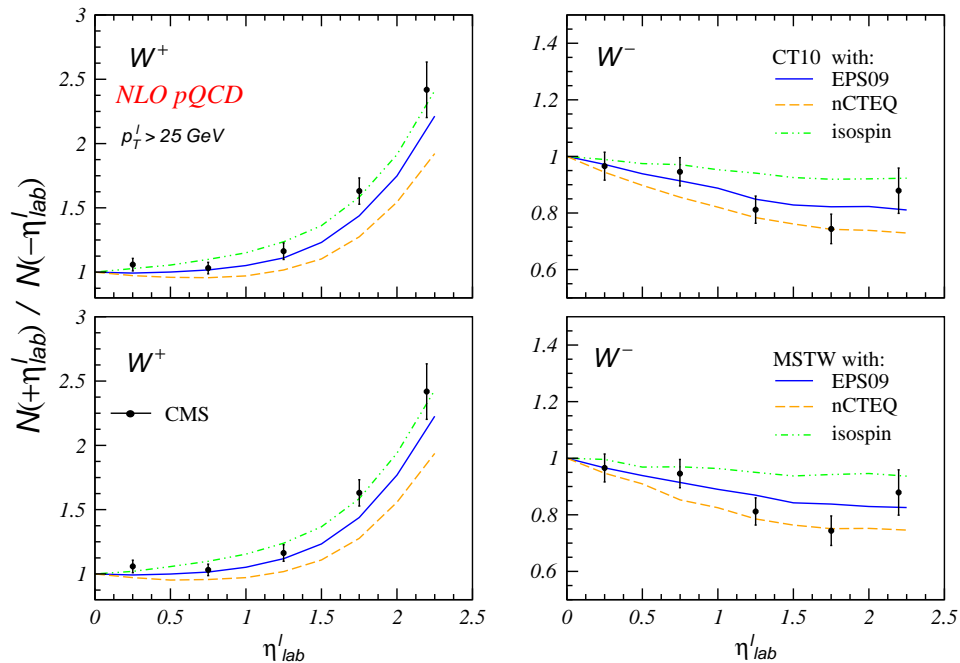}
\caption{(Color online) The froward-backward asymmetry as a function of
the charged lepton pseudorapidity in the laboratory frame
for $W^\pm$ production.  The results for $W^+$ are shown on the left-hand side
while those for $W^-$ are shown on the right-hand side.  The results with the 
CT10 proton PDFs are shown on top while those with MSTW2008 are shown on the
bottom.  The calculations are compared to the CMS 
data~\cite{Khachatryan:2015hha}.}
\label{fig:W-fbasy@pPb}
\end{figure}

The forward-backward asymmetry for $W$ boson production has also been 
measured by CMS~\cite{Khachatryan:2015hha}.
The forward-backward asymmetry is calculated to NLO as a function of the 
charged lepton pseudorapidity and compared to the CMS data.
Figure~\ref{fig:W-fbasy@pPb} shows that the calculations can describe the 
CMS data rather well.
The $W^+$ data seem to favor the isospin alone, without any addition nuclear
PDF modifications in the region $1.5\lesssim\eta^l\lesssim2.5$.  However, the 
$W^-$ data are in better agreement with the EPS09 and nCTEQ modifications.
Note also that the asymmetry is large and greater than unity for the $W^+$ while
the $W^-$ asymmetry is small and less than unity, similar to that of the $Z^0$.
Indeed, the asymmetry for $W^-$, on the right-hand side of 
Fig.~\ref{fig:W-fbasy@pPb}, is quite similar to those for $Z^0$ 
production shown in Fig.~\ref{fig:Z-fbasy@pPb}.

\subsubsection{$W^+/W^-$ charge asymmetry}

\begin{figure}[h]
\includegraphics[width = \textwidth]{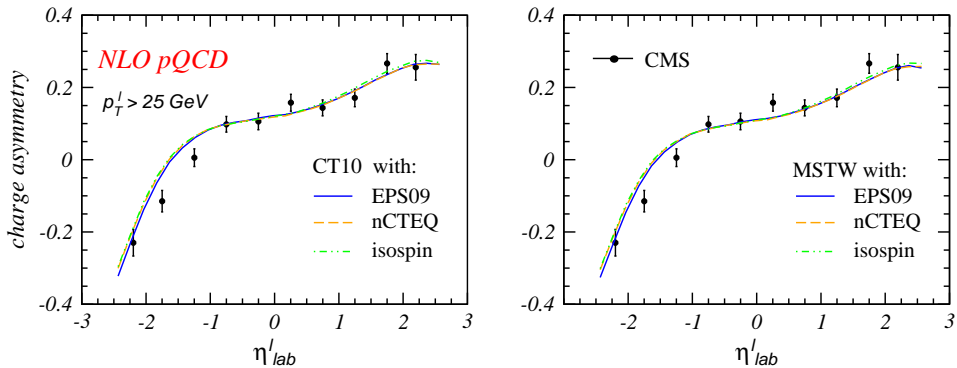}
\caption{(Color online)  The $W^+/W^-$ charge asymmetry as a function of
the charged lepton pseudorapidity calculated with the CT10 proton PDFs is
shown on the left while results with MSTW2008 are shown on the right.
The CMS data are from Ref.~[\protect\refcite{Khachatryan:2015hha}].}
\label{fig:W-casy@pPb}
\end{figure}

The charge asymmetry for $W$ boson production, defined as
\begin{eqnarray}
\label{cas}
\mathcal{A}=\frac{N(W^+)-N(W^-)}{N(W^+)+N(W^-)} \, \, ,
\end{eqnarray}
has been measured by CMS~\cite{Khachatryan:2015hha}. 
The data are compared to NLO calculations in Fig.~\ref{fig:W-casy@pPb}.
The calculations agree with the data except for the region $-2<\eta^l<-1$.
The charge asymmetry is not sensitive to the nPDFs with no obvious
flavor dependence, {\it e.g.} between the $u$ and $d$ valence quarks
\cite{Ru:2014yma}.
This is not surprising because, at a given value of $\eta$, the $W^+$ and
$W^-$ probe the same $x$ values and thus nearly the same values of the nPDFs.

\subsubsection{Summary}

The NLO and NNLO perturbative QCD calculations of vector boson production 
have been compared to the LHC data.
The calculations shown generally good agreement with the data except for the 
low statistics forward-backward asymmetry measured by LHCb.
The results are not sensitive to the choice of free proton PDFs.
However, there is a clear distinction between calculations with 
different parametrizations of the nPDFs for observables such as the 
pseudorapidity distribution and the forward-backward asymmetry.

The significant deviation of the calculations from the LHCb data
could arise from a poor understnading of the modifications of the valence
quark distributions at large $x$, $x\in[0.22,0.989]$.  Alternatively it could
be resolved by a higher-statistics measurement.
The first measurements of vector boson production in $p+$Pb collisions at 
the LHC have demonstrated the capability of studying cold nuclear matter 
effects at $x \rightarrow 1$ and high momentum transfers, $Q^2 \sim m_V^2$. 
Further robust theoretical investigations and more precise data are needed 
to place stringent constraints on the nPDFs and thus gain a 
deeper understanding of cold nuclear matter effects in this relatively 
unexplored region.

\subsection[$Z^0$ production at low $p_T$]{$Z^0$ production at low $p_T$ (Z.-B. Kang and J.-W. Qiu)}
\label{KangandQiu}

At low transverse momentum, $p_T\ll m_Z$, the conventional fixed-order 
calculation for the $Z$ boson differential cross section, $d\sigma/dydp_T$, 
includes a large logarithm $\ln(m_Z^2/p_T^2)$. The convergence of a conventional
perturbative expansion is thus impaired and these large logarithms must be 
resummed. The Collins-Soper-Sterman (CSS) formalism is well-known and was
developed for precisely such 
purposes~\cite{Collins:1984kg,Ellis:1997ii,Qiu:2000hf,Landry:2002ix}.

\begin{figure}[h]
\includegraphics[width = 0.495\textwidth]{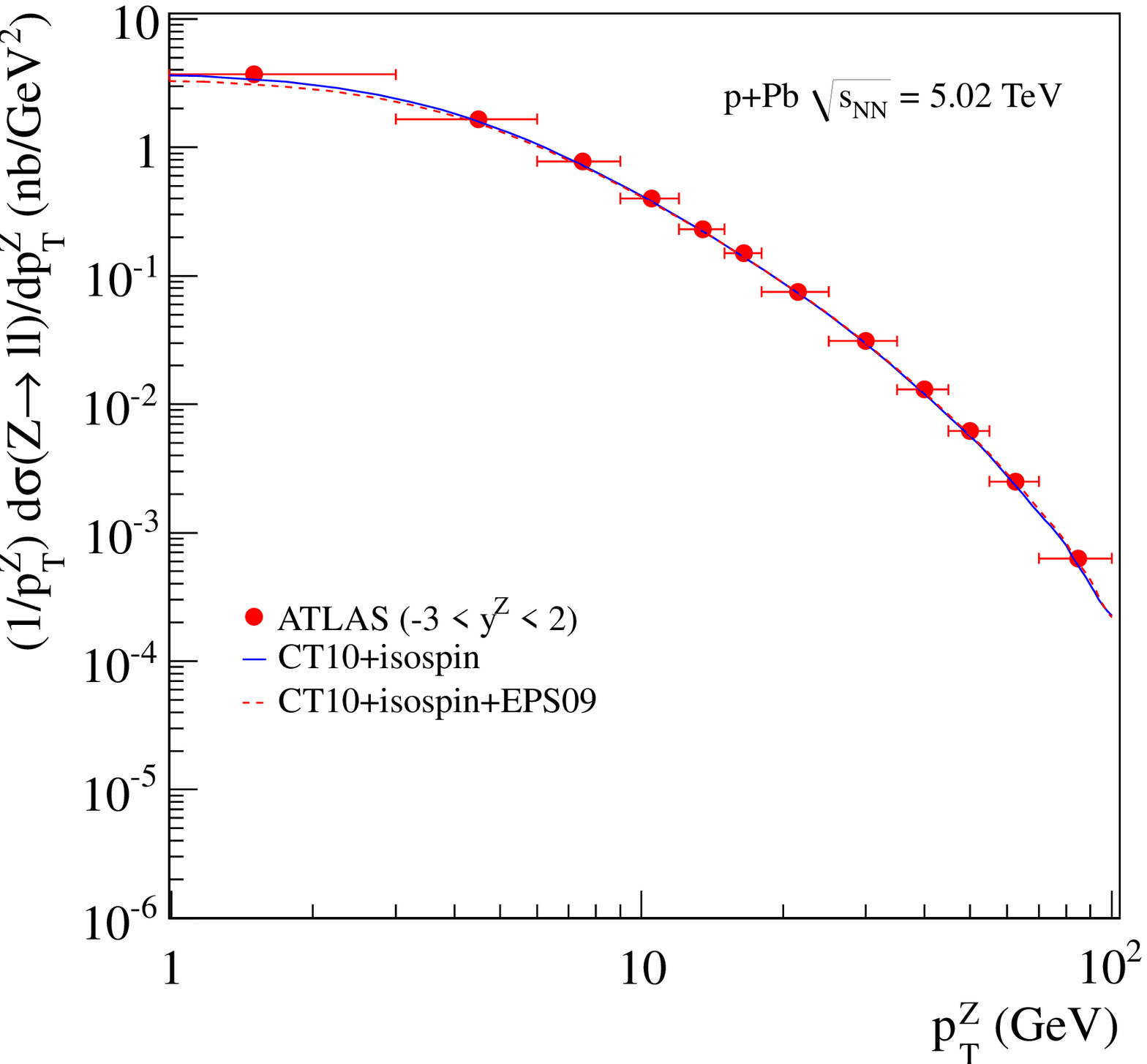}
\includegraphics[width = 0.495\textwidth]{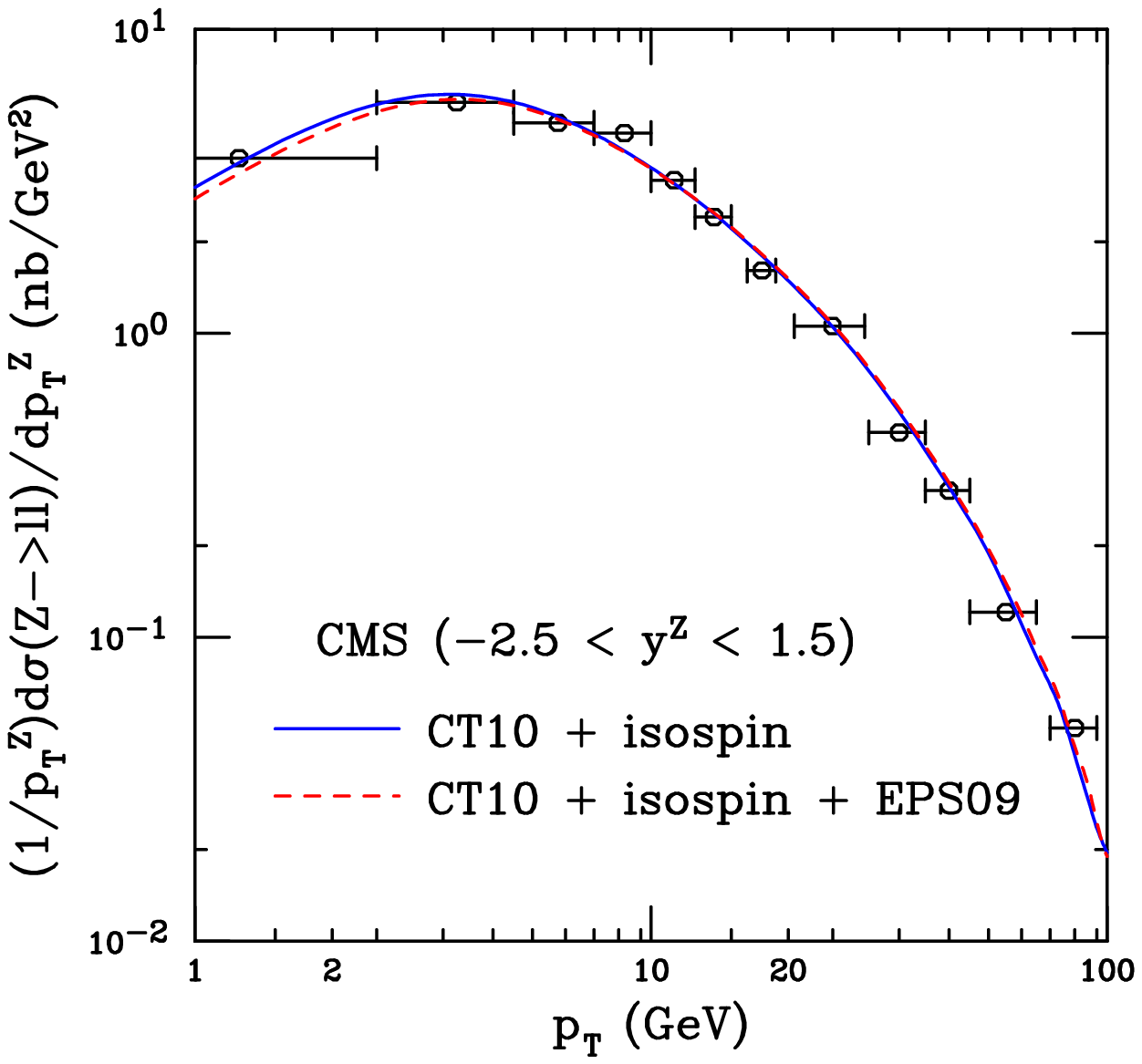}
\caption{(Color online) The $Z^0$ transverse momentum spectra in $p+$Pb 
collisions at $\sqrt{s_{NN}}=5.02$~TeV with the low $p_T$ region emphasized.
The results for ATLAS~\protect\cite{ZpPbATLAS} and 
CMS~\protect\cite{CMS:2014sca} are shown in the left and right panels, 
respectively.}
\label{lowpTZ}
\end{figure}

The calculations are based on the CSS formalism.  The CT10 parameterization is 
employed for the proton PDFs, with EPS09 NLO for the nuclear PDFs.  The 
factorization scale is $\mu = m_T/2 = 0.5 \sqrt{m_Z^2+p_T^2}$.  The nuclear 
size ($\propto A^{1/3}$) enhanced multiple scattering effects are taken into 
account in $p+$Pb collisions, as discussed in detail in 
Ref.~[\refcite{Kang:2012am}]. As can be seen in Fig.~\ref{lowpTZ}, emphasizing 
the low $p_T$ region, this formalism describes the data rather well. It predicts
that $Z^0$ boson production in $p+$Pb collisions is suppressed at low $p_T$ 
due to shadowing, while being slightly enhanced at relatively large $p_T$ due 
to antishadowing in EPS09. However, these modifications are within the current 
experimental uncertainties, and thus no definite conclusions can be made at 
this point. 

\section{Conclusions}

The predictions from Ref.~[\refcite{Albacete:2013ei} have been compared to a
wide range of data from the 2013 $p+$Pb run.  While some results are in good 
agreement with the data, other surprised have been found.  The solution of
some await results either from a higher statistics run, as for the LHCb $Z^0$ 
forward-backward asymmetry, or from a $p+p$ run at a similar 
center-of-mass energy,
as in the case of $R_{p{\rm Pb}}^{\rm ch}(p_T)$ at high $p_T$.  Indeed, LHC Run II
has already made a $p+p$ run at 5 TeV to replace the extrapolated baselines
used in the $R_{p{\rm Pb}}$ results shown here.  A follow up two week $p+$Pb 
run at 5 TeV will come at the end of 2016 to augment the data shown here.  An
additional $p+$Pb run at 7 TeV is also planned that can be compared to the
extensive $p+p$ results at this energy from Run I.  With the higher LHC
luminosity in Run II, good statistics can be expected, even for a shorter run.

\section*{Acknowledgements}


The JET Collaboration is thanked for support for the initiation of 
Ref.~[\refcite{Albacete:2013ei}] and the suggestion of a follow up with this
work.

The work of F. Arleo and S. Peign\'e is funded by ``Agence Nationale de la
Recherche'' under grant ANR-PARTONPROP.

The work of F. Fleuret was supported in part by the French CNRS via the
GDR QCD.

The work of I. Helenius has been supported by the MCnetITN FP7 Marie Curie 
Initial Training Network, contract PITN-GA-2012-315877.

The work of J.-P. Lansberg was supported in part by the French CNRS via the
grants FCPPL-Quarkonium4AFTER \& D\'efi Imphyniti-Th\'eorie LHC France and the
GDR QCD.

The work of Z.~Kang, I.~Vitev and H.~Xing is supported by the U.S. Department 
of Energy under Contract No.~DE-AC52-06NA25396.

The work of K. Kutak has been supported by Narodowe Centrum Nauki with 
Sonata Bis grant DEC-2013/10/E/ST2/00656. P. Kotko acknowledges the
support of DOE grants DE-SC-0002145 and DE-FG02-93ER40771.

The work of J.~Qiu is supported by the U.~S. Department of Energy under 
Contract No.~DE-AC02-98CH10886.

The work of H. Paukkunen was supported by the European Research Council 
grant HotLHC ERC-2011-StG-279579.

The work of A. Rezaeian is supported in part by Fondecyt grant 1110781.

The work of P. Ru, B.-W. Zhang, E. Wang and W.-N. Zhang
is supported in part by the Natural Science Foundation of China with 
Project Nos. 11322546, 11435004, 11221504.

The work of R.~Vogt was performed under the 
auspices of the U.~S. Department of Energy by Lawrence Livermore National 
Laboratory under Contract DE-AC52-07NA27344  and supported by the U.~S. 
Department of Energy, Office of Science, Office of Nuclear Physics (Nuclear 
Theory) under contract number DE-SC-0004014.

The work of X.-N.~Wang was performed under the auspices of the
U.~S. Department of Energy
under Contract No. DE-AC02-05CH11231, by the National Natural Science 
Foundation of China under grant No. 11221504.




\end{document}